\pdfoutput=1
\documentclass[
    reprint,
    superscriptaddress,
    amsmath,amssymb,
    aps,
    nofootinbib,
    prx,longbibliography  
]{revtex4-2}

\usepackage{graphicx}
\usepackage{bm}
\usepackage[svgnames]{xcolor}
\usepackage[colorlinks,citecolor=DarkGreen,urlcolor=DarkBlue,linkcolor=FireBrick,linktocpage,unicode]{hyperref}
\usepackage{mathtools}
\usepackage{array}[=2016-10-06]  
\usepackage{booktabs,tabularx,makecell}
\usepackage{orcidlink}
\usepackage{longtable}

\newcommand{\relmiddle}[1]{\mathrel{}\middle#1\mathrel{}}
\newcommand{\myset}[2]{\left\{ #1 \relmiddle| #2 \right\}}

\hypersetup{pdftitle={Systematic Magnetic Structure Generation Based on Oriented Spin Space Groups: Formulation, Applications, and High-Throughput First-Principles Calculations},
    pdfauthor={Takuya Nomoto, Kohei Shinohara, Hikaru Watanabe, and Ryotaro Arita}}

\begin{document}

\title{Systematic Magnetic Structure Generation Based on Oriented Spin Space Groups: Formulation, Applications, and High-Throughput First-Principles Calculations}%

\author{Takuya Nomoto  \orcidlink{0000-0002-4333-6773}} 
    \email[Contact author: ]{tnomoto@tmu.ac.jp}
    \thanks{These authors contributed equally to this work.}
    \affiliation{Department of Physics, Tokyo Metropolitan University, {Hachioji}, Tokyo 192-0397, Japan}

\author{Kohei Shinohara  \orcidlink{0000-0002-5907-2549}} 
    \email[Contact author: ]{kshinohara0508@gmail.com}
    \thanks{These authors contributed equally to this work.}
    \affiliation{Department of Physics, University of Tokyo, {Bunkyo-ku}, Tokyo 113-0033, Japan}

\author{Hikaru Watanabe  \orcidlink{0000-0001-7329-9638}} 
\affiliation{Graduate School of Engineering, Hokkaido University, Sapporo 060-8628, Japan}

\author{Ryotaro Arita  \orcidlink{0000-0001-5725-072X}} 
    \affiliation{Department of Physics, University of Tokyo, {Bunkyo-ku}, Tokyo 113-0033, Japan}
    \affiliation{RIKEN, Center for Emergent Matter Science, {Wako} Saitama 351-0198, Japan}


\begin{abstract}
We propose a framework for generating magnetic structures, inspired by the concept of oriented spin space groups (SSGs): Magnetic structures are first generated as totally symmetric representations of an SSG and are then rotated such that they belong to the maximal magnetic space group of the SSG, which we term spin-symmetry-adapted (SSA) structures and oriented SSA structures, respectively.
This is a natural framework to enforce fixed magnetic moment magnitudes on the symmetry-equivalent sites as well as to exploit the spin-orbit coupling (SOC)-induced hierarchy of energy scales.
To examine the present scheme, we analyze the MAGNDATA database and find that 77\% of the reported structures are reproducible at the SSG level, among which 82\% are fully reproduced within the oriented SSG scheme, regardless of their spin-only group types or propagation vectors.
When the $k$ index is fixed to its experimental value, the number of magnetic structures generated is on average about nine, also demonstrating the high efficiency of this scheme.
Next, to quantitatively assess computational and predictive performance, we perform spin density functional theory calculations for 283 materials, first carrying out self-consistent calculations for SSA structures without SOC, followed by fixed-charge calculations including SOC for the descendant oriented SSA structures.
The experimental magnetic structures are reproduced as energetically most stable in 83\% of cases at the SSG level without SOC and in 76\% of cases at the oriented SSG level with SOC, showing that the fixed-charge scheme enables accurate evaluation of SOC-induced energy differences at low computational cost.
The characteristic energy scale among oriented SSA structures is only approximately 0.29~meV per magnetic atom, about 300 times smaller than that of distinct SSA structures, suggesting that calculations with SOC confined to the lowest-energy manifold of oriented SSA structures are sufficient for reliable predictions.
These results demonstrate that oriented SSG-based enumeration, combined with the two-step calculations for SSA and oriented SSA structures, provides an efficient and robust route for large-scale magnetic-structure prediction.
\end{abstract}

\maketitle

\section{Introduction}

Spintronics exploits the electron spin degree of freedom in addition to its charge, enabling device concepts with high speed and low power consumption~\cite{Zutic2004,Hirohata2020}.
Most existing spintronic devices rely on ferromagnetic (FM) materials, whereas antiferromagnets (AFMs) have long been considered unsuitable, since their nearly vanishing net magnetization complicates detection and control of their magnetic domains.
Meanwhile, AFMs offer intrinsic advantages, including negligible stray fields, ultrafast spin dynamics, and robustness against external perturbations~\cite{Gomonay2014,Jungwirth2016,Baltz2018,Fukami2020,Xiong2022}.
In fact, recent experimental advances have demonstrated key functionalities such as tunnel magnetoresistance~\cite{Chen2023,Qin2023,Shi2024,Chen2024} and current-induced magnetization or N\'eel-vector switching~\cite{Wadley2016,Grzybowski2017,Bodnar2018,Chen2018,Meinert2018,Baldrati2019,Tsai2020,DuttaGupta2020,Arpaci2021,Hajiri2021,Higo2022,Grzybowski2022,Krishnaswamy2022,Deng2022,Xu2023,Song2024,Zhang2025,Yoon2025}, establishing AFM spintronics as a rapidly expanding research field.

To date, a broad range of AFM materials has been proposed as candidates for spintronic applications.
A prototypical example is the $\mathrm{Mn}_3X$ family, in which noncollinear AFM order gives rise to a small but finite canted magnetic moment~\cite{Nakatsuji2015,Kuroda2017,Nayak2016,Kiyohara2016,Iwaki2020,An2020}.
Despite being antiferromagnetic, these materials possess the same magnetic symmetry as conventional FMs and have been shown in recent studies to be compatible with established FM device architectures~\cite{Chen2023,Qin2023,Shi2024,Tsai2020,Higo2022,Yoon2025,Wu2024,Tsai2026}.
More recently, strong interest has turned to \textit{altermagnets}~\cite{Smejkal2022PRX1,Smejkal2022PRX2,Feng2024,Song2025,Tamang2025,Rathore2025,Jungwirth2025arXiv,Bhowal}, a class of collinear AFMs with zero net magnetization in the absence of spin-orbit coupling (SOC) yet exhibiting even-parity spin splitting~\cite{Smejkal2022PRX2,Wu2007,Hayami2019,Naka2019,Yuan2020,Yuan2021,Ma2021,Yuan2023}.
Beyond these examples, additional symmetry-driven magnetic classes have also been identified, including $p$-wave magnets with odd-parity spin splitting~\cite{Hellenes2024arXiv,Brekke2024,Song2025Nature,Yamada2025} and $Q$ magnets, which feature global spin degeneracy but localized spin splitting, and exhibit nontrivial responses rooted in nonsymmorphic parent space groups~\cite{Matsuda2025,Yu2025-ls}.
These systems are predicted to host unconventional phenomena driven by spin ordering, such as tunnel magnetoresistance effects, without relying on sizable magnetization, thereby substantially broadening the landscape of magnetic materials beyond the conventional FM paradigm.

First-principles calculations have already successfully predicted nontrivial transport phenomena such as the anomalous Hall effect in noncollinear AFMs~\cite{PhysRevLett.112.017205,Kubler_2014}, later confirmed experimentally~\cite{Nakatsuji2015,Nayak2016}.
Large-scale efforts now aim to determine magnetic ground states using high-throughput spin density functional theory (SDFT) calculations, including collinear calculations~\cite{PhysRevMaterials.1.034404,Horton2019,koretsune2020,doi:10.1021/jacs.3c00284,PhysRevMaterials.9.054403,doi:10.1021/acs.jpclett.8b02783} and noncollinear ones~\cite{ZHENG2021107659,Tellez-Mora2024,Sodequist2024}.

The rapid expansion of magnetic classes and the success of first-principles calculations call for systematic frameworks for predicting magnetic structures.
The traditional framework is representation analysis (RA)~\cite{Bertaut:a05871,BERTAUT1981267,IZYUMOV1979239}, which constructs symmetry-adapted magnetic structures from the irreducible representations (IRs) of the nonmagnetic space group.
RA is equivalent and complementary to considering possible magnetic space groups from the nonmagnetic space group~\cite{Opechowski:a07941,Petricek:pz5077,Rodriguez-Carvajal:gar5007}.
A key assumption in RA is that ground-state magnetic structures can be described by a small number of IRs, which is empirically valid in many cases~\cite{Mita2021}.

Although RA has been established as a standard method for magnetic-structure analysis and is offered, for example, in \textsc{isodistort}~\cite{StokesISODISTORT,Campbell:wf5017} and \textsc{maxmagn}~\cite{annurev:/content/journals/10.1146/annurev-matsci-070214-021008}, it has a limitation.
Namely, symmetry-equivalent magnetic atoms in the nonmagnetic structure are not guaranteed to have equal-magnitude magnetic moments in the generated magnetic structures, especially for multidimensional IRs.
For such multidimensional IRs, a linear combination of multiple basis functions is chosen to generate maximal magnetic space groups (MSGs)~\cite{annurev:/content/journals/10.1146/annurev-matsci-070214-021008}, known as the epikernel or isotropy subgroup~\cite{doi:10.1142/0751,E_Ascher_1977}.

Cluster multipole (CMP) theory is an alternative framework in which magnetic structures are systematically generated and classified according to IRs of the parent point group symmetry~\cite{Suzuki2017,Suzuki2019,Yanagi2023}.
This framework will be useful for interpreting the generated magnetic structures and corresponding order parameters.
Based on this framework, systematic analyses of databases combined with first-principles calculations have been employed for high-throughput materials prediction~\cite{Mita2021,Nomoto2024}.
CMP partly resolves the limitation of RA by employing physically motivated vector spherical harmonics as trial basis functions.
However, it still suffers from the fundamental limitation of RA mentioned above.
As a consequence, exhaustive and efficient enumeration of candidate structures is hindered, motivating the development of alternative formulations.

In parallel, the concept of the spin space group (SSG) has attracted renewed attention as a powerful framework for describing magnetic structures, especially in the context of altermagnets~\cite{Liu2022PRX,Watanabe2024SSG,Jiang2024PRX,Xiao2024PRX,Chen2024PRX}.
An SSG is defined as a subgroup of the direct product of a parent space group and the $O(3)$ group, in which spatial and spin operations are treated independently~\cite{doi:10.1063/1.1708514,Brinkman1966,Litvin1973,Litvin1974,Litvin1977,Opechowski1986}.
Physically, SSGs correspond to the symmetry of magnets in the absence of SOC and, therefore, encode higher symmetry information than conventional MSGs.
As a result, SSGs enable a systematic description of spin correlations that are difficult to analyze within the conventional framework based on RA and MSGs~\cite{Rodriguez-Carvajal:gar5007,izyumov1991neutron}.
In addition, as shown in pioneering works on spin-current conductivity~\cite{Zelezny2017,Zhang2018-yj}, comparison between SSGs and MSGs provides a natural way to isolate SOC-induced responses and to assess the symmetry-allowed magnitude of spin-dependent physical quantities, thereby clarifying various physical phenomena that do not rely on SOC.
Recently, the representation theory of SSGs~\cite{Watanabe2024SSG,Jiang2024PRX,Xiao2024PRX,Chen2024PRX,10.21468/SciPostPhys.18.3.109,zhang2025irssgopensourcesoftwarepackage,PhysRevB.111.134407,McClarty2024Landau,Chen2025} and algorithms for deriving symmetry-adapted response tensors~\cite{Etxebarria:cam5007,XIAO2026109872,elcoro2026automaticcalculationsymmetryadaptedtensors} have been developed.

\begin{figure}[t]
\centering
\includegraphics[width=\linewidth]{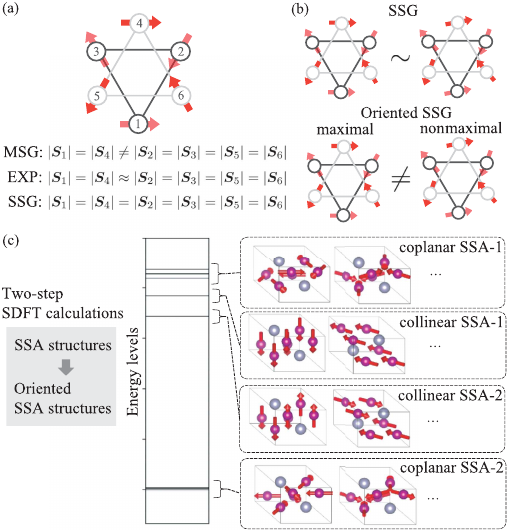}
\caption{(a)~Magnetic structure of the noncollinear antiferromagnet $\mathrm{Mn}_3\mathrm{Sn}$.
The orbital part of the magnetic point group is $mmm$, which constrains the moment sizes to be equal within sites 1--4 and within sites 2--3--5--6 but not between these two sets.
In contrast, the orbital part of the spin point group is $6/mmm$, identical to that of the parent point group, allowing all moment sizes to be constrained to be equal. (b)~Equivalence under spin rotation.
Within the SSG framework, the two spin-rotated structures cannot be distinguished.
Once oriented SSGs are considered, however, the left structure is identified as a maximal MSG, whereas the right one is nonmaximal, allowing the former to be selected. (c)~Work flow of magnetic-structure prediction.
SDFT calculations without SOC are first performed for the SSA structures.
The resulting spin-rotated states are then used as input for SDFT calculations with SOC for the oriented SSA structures.
This two-step work flow exploits the SOC energy hierarchy to reduce computational cost and improve numerical stability through the fixed-charge treatment.}
\label{fig:concept}
\end{figure}

In this work, we develop an SSG-based algorithm for magnetic-structure generation and perform benchmark analyses of magnetic-structure databases together with first-principles calculations.
Figure~\ref{fig:concept} illustrates the concept of the generation scheme and the overall work flow of magnetic-structure prediction.
First, we show that the SSG formulation enables full enumeration of magnetically inequivalent structures up to global $O(3)$ transformations~\cite{Jiang2024PRX,Xiao2024PRX,Chen2024PRX}, which resolves the aforementioned limitation of RA by construction as discussed in Secs.~\ref{sec:ssg_generation} and \ref{sec:ssg_family_choices}.
In the following, we refer to these structures as spin-symmetry-adapted (SSA) structures.
Second, based on the recently introduced concept of oriented SSG~\cite{Liu2025arXiv}, we demonstrate a systematic scheme for enumerating magnetic structures inequivalent up to the parent space group as in conventional MSG classifications, which we term oriented SSA structures.
We then apply this framework to the MAGNDATA database~\cite{MAGNDATA1} and, in combination with high-throughput SDFT calculations, benchmark the predictive performance of AFM structure prediction.
A key principle of the proposed scheme is the hierarchy of energy scales between SSA and oriented SSA structures.
Because energy differences among oriented SSA structures originate from SOC, they are typically much smaller than those between distinct SSA structures.
This hierarchy enables an efficient two-step computational strategy~\cite{Li2025arXiv}: SDFT calculations without SOC are first used to identify the most stable SSA structure, followed by SOC calculations with fixed-charge densities to resolve the oriented SSA structure manifold.
We quantitatively assess the validity of this strategy and demonstrate its effectiveness for large-scale magnetic-structure prediction.

The organization of this paper is as follows.
In Sec.~\ref{sec:ssg}, we present an algorithm for magnetic-structure generation based on the SSG framework.
In Sec.~\ref{sec:msg}, we provide a rigorous formulation of the oriented SSG introduced in Ref.~\cite{Liu2025arXiv} and present a method for generating magnetic structures along this concept.
In Sec.~\ref{sec:examples}, we apply the proposed algorithm to representative AFM systems and discuss the resulting magnetic structures.
In Sec.~\ref{sec:application}, materials listed in the MAGNDATA database are classified according to the oriented SSG scheme, and the range of applicability is discussed.
In Sec.~\ref{sec:sdft}, we present a benchmark of magnetic-structure prediction using high-throughput SDFT calculations.
Finally, Sec.~\ref{sec:conclusion} summarizes our conclusions.

\section{\label{sec:ssg}Enumerating spin-symmetry-adapted structures}

To unambiguously describe our scheme for generating SSA structures, we first briefly review terminology and the group structure of SSGs \cite{Litvin1974,Shinohara2024} in Sec.~\ref{sec:ssg_group}.
Then, we present a procedure to enumerate SSGs from a space group and a spin-only group in Sec.~\ref{sec:ssg_enumeration}; the enumeration is parametrized by the $k$ index that controls the magnetic unit-cell size, yielding a finite and controllable candidate set.
We describe a generation scheme of commensurate SSA structures from a given space group, a spin-only group, and Wyckoff positions of magnetic sites in Sec.~\ref{sec:ssg_generation}; the SSG-based generation guarantees identical moment magnitudes on sites belonging to the same Wyckoff position.
Finally, we discuss practical choices of the space group used in the generation scheme in Sec.~\ref{sec:ssg_family_choices}.
Although such a generation scheme with SSG has already been reported, for the zero-propagation vector case in Ref.~\cite{Jiang2024PRX} and for specific applications in Ref.~\cite{Li2025arXiv}, the present self-contained description covers arbitrary propagation vectors and provides a clear and implementable algorithm.

\subsection{\label{sec:ssg_group}Group structure of spin space group}

Let $\mathcal{X}$ be a subgroup of the direct product of the Euclidean group $E(3)$ and the three-dimensional orthogonal group $O(3)$.
If the following $\mathcal{G}(\mathcal{X})$ is a space group, $\mathcal{X}$ is called a SSG:
\begin{align}
    \mathcal{G}(\mathcal{X}) = \myset{g \in E(3)}{\exists \mathbf{U} \in O(3) \;\text{such that}\; (g, \mathbf{U}) \in \mathcal{X}},
\end{align}
where $g = (\mathbf{R}, \boldsymbol{\tau})$ denotes a spatial operation and $\mathbf{U}$ a spin rotation.
For a spin space group $\mathcal{X}$, we call $\mathcal{G}(\mathcal{X})$ a family space group of $\mathcal{X}$.
A maximal space subgroup of a spin space group $\mathcal{X}$ is defined as
\begin{align}
    \mathcal{H}(\mathcal{X}) = \myset{g \in E(3)}{(g, \mathbf{I}) \in \mathcal{X}},
\end{align}
where $\mathbf{I}$ is the identity rotation in $O(3)$.
The maximal space subgroup $\mathcal{H}(\mathcal{X})$ is a normal subgroup of $\mathcal{G}(\mathcal{X})$.

A family spin point group of spin space group $\mathcal{X}$ is defined as
\begin{align}
    \mathcal{B}(\mathcal{X}) = \myset{\mathbf{U} \in O(3)}{\exists g \in E(3) \;\text{such that}\; (g, \mathbf{U}) \in \mathcal{X}}.
\end{align}
A spin-only group of spin space group $\mathcal{X}$ is defined as
\begin{align}
    \mathcal{B}_{\mathrm{so}}(\mathcal{X}) = \myset{\mathbf{U} \in O(3)}{((\mathbf{I}, \mathbf{0}), \mathbf{U}) \in \mathcal{X}}.
\end{align}
The spin-only group $\mathcal{B}_{\mathrm{so}}(\mathcal{X})$ is a normal subgroup of $\mathcal{B}(\mathcal{X})$.
The direct product group $\mathcal{H}(\mathcal{X}) \times \mathcal{B}_{\mathrm{so}}(\mathcal{X})$ is a normal subgroup of $\mathcal{X}$.

From Goursat's lemma \cite{Goursat1889OrthogonalSubstitutions,Litvin1974}, there is a group isomorphism between the quotient groups
\begin{align}
    \label{eq:goursat}
    \mathcal{X} / [\mathcal{H}(\mathcal{X}) \times \mathcal{B}_{\mathrm{so}}(\mathcal{X})]
        \cong \mathcal{G}(\mathcal{X}) / \mathcal{H}(\mathcal{X})
        \cong \mathcal{B}(\mathcal{X}) / \mathcal{B}_{\mathrm{so}}(\mathcal{X}),
\end{align}
where we let $U: \mathcal{G}(\mathcal{X}) / \mathcal{H}(\mathcal{X}) \rightarrow \mathcal{B}(\mathcal{X}) / \mathcal{B}_{\mathrm{so}}(\mathcal{X})$ be the isomorphism.
Then, spin space group $\mathcal{X}$ is uniquely characterized by the triplet $[\mathcal{H}(\mathcal{X}), \mathcal{B}_{\mathrm{so}}(\mathcal{X}), U]$.

From Hermann's theorem \cite{Hermann+1929+533+555}, there exists a unique group $\mathcal{M}(\mathcal{X})$ satisfying the following two conditions:
(i) $\mathcal{M}(\mathcal{X})$ is a translationengleiche subgroup (t-subgroup) of $\mathcal{G}(\mathcal{X})$, where their translation subgroups are identical;
(ii) $\mathcal{H}(\mathcal{X})$ is a klassengleiche subgroup (k-subgroup) of $\mathcal{M}(\mathcal{X})$, where their point groups are identical.
Here, $\mathcal{H}(\mathcal{X})$ is a normal subgroup of $\mathcal{M}(\mathcal{X})$, because $\mathcal{H}(\mathcal{X}) \trianglelefteq \mathcal{G}(\mathcal{X})$ and $\mathcal{M}(\mathcal{X}) \leq \mathcal{G}(\mathcal{X})$.
Also, $\mathcal{M}(\mathcal{X})$ is a normal subgroup of $\mathcal{G}(\mathcal{X})$, because $\mathcal{M}(\mathcal{X})$ is uniquely defined
as a subgroup of $\mathcal{G}(\mathcal{X})$ with the above two conditions.



\subsection{\label{sec:ssg_enumeration}Enumeration of spin space groups from family space group and spin-only group}

Conversely, the group isomorphism in Eq.~\eqref{eq:goursat} provides a systematic way to enumerate spin space groups $\mathcal{X}$ from a given space group $\mathcal{G}$ and spin-only group $\mathcal{B}_{\mathrm{so}}$.
For every normal space subgroup $\mathcal{H}$ of $\mathcal{G}$, if there exists an injective homomorphism $U: \mathcal{G} / \mathcal{H} \rightarrow \mathcal{N}(\mathcal{B}_{\mathrm{so}}) / \mathcal{B}_{\mathrm{so}}(\mathcal{X})$, then the triplet $(\mathcal{H}, \mathcal{B}_{\mathrm{so}}, U)$ uniquely determines a spin space group, where $\mathcal{N}(\mathcal{B}_{\mathrm{so}})$ is the normalizer of $\mathcal{B}_{\mathrm{so}}$ in $O(3)$.
For later convenience, we denote a translation subgroup of $\mathcal{G}$ as
\begin{align}
    \mathcal{T}(\mathcal{G}) = \myset{(\mathbf{I}, \mathbf{t})}{(\mathbf{I}, \mathbf{t}) \in \mathcal{G}}
\end{align}
and the coset decomposition of $\mathcal{G}$ by $\mathcal{T}(\mathcal{G})$ as
\begin{align}
    \mathcal{G} = \bigcup_{i=1}^{n} g_i \mathcal{T}(\mathcal{G}),
\end{align}
where $n = |\mathcal{G} : \mathcal{T}(\mathcal{G})|$ and $g_i = (\mathbf{R}_i, \boldsymbol{\tau}_i)$ are representatives of cosets.

When a spin space group is a stabilizer of a magnetic structure, its spin-only group $\mathcal{B}_{\mathrm{so}}$ is classified into four types up to transformations \cite{Litvin1974}: nonmagnetic [$\mathcal{B}_{\mathrm{so}} \cong O(3)$], collinear [$\mathcal{B}_{\mathrm{so}} \cong \mathbb{Z}_2 \ltimes SO(2)$], coplanar ($\mathcal{B}_{\mathrm{so}} \cong \mathbb{Z}_2$), and noncoplanar ($\mathcal{B}_{\mathrm{so}} \cong 1$).
For nonmagnetic cases, all enumerated spin space groups are a direct product of a normal space subgroup of $\mathcal{G}$ and $O(3)$.
As this is trivial, we henceforth focus on collinear, coplanar, and noncoplanar spin-only groups.

A spin-structure dimension $d$ is defined as the dimension of the vector space spanned by magnetic moments \cite{OPECHOWSKI197793}; $d=1, 2, 3$ correspond to collinear, coplanar, and noncoplanar spin-only groups, respectively.
The dimension of the natural representation of $\mathcal{B}(\mathcal{X}) / \mathcal{B}_{\mathrm{so}}(\mathcal{X})$ is equal to the spin-structure dimension $d$. In the following, spin space groups are generated separately according to the spin dimension, namely, for collinear, coplanar, and noncoplanar cases. As long as magnetic structures are generated for all spin dimensions, this distinction does not introduce any bias.

Because there are infinitely many space subgroups for a given space group $\mathcal{G}$, we first consider the intermediate space group $\mathcal{M}$ between $\mathcal{G}$ and $\mathcal{H}$ such that
(i) $\mathcal{M}$ is a normal t-subgroup of $\mathcal{G}$
and (ii) $\mathcal{H}$ is a normal k-subgroup of $\mathcal{M}$ with a given index of $N_k = |\mathcal{M} : \mathcal{H}|$.
We call $N_k$ the $k$ index, which corresponds to how many times a magnetic unit cell is enlarged compared to the original unit cell.
There are only a finite number of t-subgroups $\mathcal{M}$ for a given space group $\mathcal{G}$.

Therefore, inputs $(\mathcal{G}, \mathcal{B}_{\mathrm{so}}, N_k)$ give a finite number of spin space groups $\mathcal{X}$ with family space group $\mathcal{G}(\mathcal{X}) = \mathcal{G}$ and spin-only group $\mathcal{B}_{\mathrm{so}}(\mathcal{X}) = \mathcal{B}_{\mathrm{so}}$.
A resulting spin space group $\mathcal{X}$ and associated groups can be expressed as
\begin{align}
    \mathcal{M} &= \bigcup_{m=1}^{N_k} (\mathbf{I}, \mathbf{t}_m) \mathcal{H}, \\
    \mathcal{G} &= \bigcup_{I=1}^{N_t} g_{i_{I}} \mathcal{M}, \\
    \label{eq:ssg_construction}
    \mathcal{X} &= \bigcup_{I=1}^{N_t} \bigcup_{m=1}^{N_k} (g_{i_{I}} (\mathbf{I}, \mathbf{t}_m), \mathbf{U}_{I m}) (\mathcal{H} \times \mathcal{B}_{\mathrm{so}}),
\end{align}
where $N_t = |\mathcal{G} : \mathcal{M}|$, $\{ i_{1}, \cdots, i_{N_{t}} \} \subseteq \{ 1, \cdots, n \}$, and $\{(\mathbf{I}, \mathbf{t}_m)\}$ and $\{ g_{i_{I}} \}$ are coset representatives of $\mathcal{M} / \mathcal{H}$ and $\mathcal{G} / \mathcal{M}$, respectively.
The spin rotation $\mathbf{U}_{I m}$ is arbitrarily chosen from a coset $U \left( g_{i_{I}}(\mathbf{I}, \mathbf{t}_m) \mathcal{H} \right)$.

We consider representative spin space groups up to transformations in $\mathcal{G} \times O(3)$.
Thus, we identify two spin space groups $\mathcal{X}_1$ and $\mathcal{X}_2$ constructed from the same triplet $(\mathcal{H}, \mathcal{B}_{\mathrm{so}}, U)$ if there exists $(g, \mathbf{U}) \in \mathcal{G} \times O(3)$ such that
\begin{align}
    (g, \mathbf{U})^{-1} \mathcal{X}_1 (g, \mathbf{U}) = \mathcal{X}_2.
\end{align}
By this definition, we choose $\mathcal{B}_{\mathrm{so}}$ with a rotation parallel to the $z$ axis for collinear cases, and a reflection perpendicular to the $z$ axis for coplanar cases, without loss of generality.

Note that these conjugacy classes of spin space groups are different from their isomorphism classes.
The latter considers equivalence up to transformations in $\mathcal{N}(\mathcal{G}) \times O(3)$, where $\mathcal{N}(\mathcal{G})$ is the affine normalizer of $\mathcal{G}$ in $E(3)$.
We choose the former equivalence, because it is suitable for exhaustively generating candidate magnetic structures.
For example, each axis of an orthorhombic space group is taken as inequivalent under the former equivalence, and magnetic structures polarized along each axis can be generated separately.

A detailed algorithm to enumerate spin space groups in Eq.~\eqref{eq:ssg_construction} up to transformations in $\mathcal{G} \times O(3)$ is provided in Appendix~\ref{appx:ssg_construction}.

\subsection{\label{sec:ssg_generation}Generation of spin-symmetry-adapted structures from spin space group}

For a given family space group $\mathcal{G}$, a spin-only group $\mathcal{B}_{\mathrm{so}}$, and $k$ index $N_k$, we first enumerate spin space groups $\mathcal{X}$ using the procedure in Sec.~\ref{sec:ssg_enumeration}.
By construction, each Wyckoff position of $\mathcal{G}$ belongs to a single orbit of $\mathcal{X}$, because a symmetry operation in $\mathcal{G}$ has a one-to-one correspondence to that in $\mathcal{X}$ through Eq.~\eqref{eq:ssg_construction}.
Thus, magnetic sites belonging to the same Wyckoff position of $\mathcal{G}$ have identical magnitudes of magnetic moments in a magnetic structure stabilized by $\mathcal{X}$.

We next consider the action of spin symmetry operations in $\mathcal{X}$ on magnetic moments of a magnetic structure.
We denote the $\kappa$th magnetic site in the $\boldsymbol{\ell}$th unit cell as $\mathbf{r}(\boldsymbol{\ell}\kappa)$, where $\kappa \in \{1, \cdots, m \}$ and $\boldsymbol{\ell} \in \mathbb{Z}^3$.
We write the magnetic moment at the site $\mathbf{r}(\boldsymbol{\ell}\kappa)$ as $\mathbf{m}_{\mathbf{r}(\boldsymbol{\ell}\kappa)}$.
For a spin symmetry operation $(g, \mathbf{U}) \in \mathcal{X}$, its action on the magnetic moments $\mathbf{m}$ is defined as
\begin{align}
    [ (g, \mathbf{U}) \mathbf{m} ]_{\mathbf{r}(\boldsymbol{\ell}\kappa)}
        = \mathbf{U} \, \mathbf{m}_{g^{-1} \mathbf{r}(\boldsymbol{\ell}\kappa)}.
\end{align}
This action provides a representation matrix of $\mathcal{X}$ on the magnetic moments as
\begin{align}
    \label{eq:representation_magnetic_moments}
    &[ (g, \mathbf{U}) \mathbf{m} ]_{r_{\mu}(\boldsymbol{\ell}\kappa)} \nonumber \\
        &= \sum_{\boldsymbol{\ell}' \in \mathbb{Z}^3} \sum_{\kappa'=1}^{m} \sum_{\mu'=1}^{3}
            \Gamma_{ \boldsymbol{\ell} \kappa \mu; \boldsymbol{\ell}' \kappa' \mu' } ((g, \mathbf{U})) \, \mathbf{m}_{\mathbf{r}_{\mu'}(\boldsymbol{\ell}'\kappa')},
\end{align}
where $r_{\mu}(\boldsymbol{\ell}\kappa)$ is the $\mu$th component of $\mathbf{r}(\boldsymbol{\ell}\kappa)$ and the representation matrix is given by
\begin{align}
    \Gamma^{\mathrm{site}}_{\boldsymbol{\ell} \kappa; \boldsymbol{\ell}' \kappa'} (g)
        &= \delta_{\mathbf{r}(\boldsymbol{\ell}\kappa), g \mathbf{r}(\boldsymbol{\ell}'\kappa')}, \\
    \Gamma_{ \boldsymbol{\ell} \kappa \mu; \boldsymbol{\ell}' \kappa' \mu' } ((g, \mathbf{U}))
        &= \Gamma^{\mathrm{site}}_{\boldsymbol{\ell} \kappa; \boldsymbol{\ell}' \kappa'} (g) \, [\mathbf{U}]_{\mu \mu'}.
\end{align}
When the spin symmetry operation can be viewed as a magnetic symmetry operation with $(\det \mathbf{R}_g) \mathbf{R}_g = (\det \mathbf{U}) \mathbf{U}$ \cite{Shinohara2024}, the above representation matrices are identical to those for magnetic space groups in Ref.~\cite{refId0}, where we denote the rotation part of $g$ as $\mathbf{R}_g$.

Symmetry-adapted magnetic moments $\mathbf{m}^{(\alpha)}$ of $\mathcal{X}$ are invariant under all spin symmetry operations in $\mathcal{X}$:
\begin{align}
    \label{eq:symmetry_adapted_magnetic_moments}
    (g, \mathbf{U}) \mathbf{m}^{(\alpha)} = \mathbf{m}^{(\alpha)}.
\end{align}
We formally refer to a magnetic structure with $\mathbf{m}^{(\alpha)}$ as a SSA structure of $\mathcal{X}$.
As shown in Appendix~\ref{appx:projection_operator}, such symmetry-adapted magnetic moments can be obtained by applying the projection operator of $\mathcal{X}$.
If multiple $p_{\mathcal{X}} \,(> 1)$ linearly independent symmetry-adapted magnetic moments exist, we label them as $\mathbf{m}^{(\alpha=1)}, \cdots, \mathbf{m}^{(\alpha=p_{\mathcal{X}})}$.

\subsection{\label{sec:ssg_family_choices}Practical choices of family space group}

For a given nonmagnetic structure, the natural first choice of the family space group in Sec.~\ref{sec:ssg_generation} is its space group $\mathcal{G}_{0}$.
However, a combination of $\mathcal{G}_{0}$ and a specified spin-only group $\mathcal{B}_{\mathrm{so}}$ may admit only the trivial solution in Eq.~\eqref{eq:symmetry_adapted_magnetic_moments}, yielding no nonzero symmetry-adapted magnetic moments.
When this occurs, we consider its maximal t-subgroups $\mathcal{G}_{0}'$ that preserve the multiplicities \cite{Wondratschek1993} of the given Wyckoff positions of magnetic sites and use $\mathcal{G}_{0}'$ as the family space group.
This procedure is repeated until we obtain nonzero symmetry-adapted magnetic moments.
Because the given Wyckoff positions are not split in $\mathcal{G}_{0}'$, the resulting symmetry-adapted magnetic moments have identical magnitudes for magnetic sites in the same given Wyckoff position.
We consider representative spin space groups up to transformations in $\mathcal{N}_{\mathcal{G}_{0}}(\mathcal{G}_{0}') \times O(3)$, where $\mathcal{N}_{\mathcal{G}_{0}}(\mathcal{G}_{0}')$ is the normalizer of $\mathcal{G}_{0}'$ in $\mathcal{G}_{0}$, to avoid generating equivalent spin space groups.

\section{\label{sec:msg}Enumerating symmetry-adapted magnetic structures by subduing SSG to MSG}

Until now, we have discussed SSGs up to $O(3)$ transformations in the spin space, and the spin axis of a resulting symmetry-adapted magnetic structure is not fixed, in general.
However, in many cases, the spin axes are aligned along specific crystallographic directions or high-symmetry directions due to SOC.
To describe such situations, we here present a systematic procedure to enumerate SSGs with maximal magnetic space group correspondences from a given SSG.
We call such enumerated SSGs oriented spin space groups (oriented SSGs) hereafter, following Ref.~\cite{Liu2025arXiv}, and use them to describe both a SSG and a MSG of a magnetic structure simultaneously.

We define a family point group of $\mathcal{X}$ as
\begin{align}
    \mathcal{P}(\mathcal{X}) = \myset{\mathbf{R}}{((\mathbf{R}, \boldsymbol{\tau}), \mathbf{U}) \in \mathcal{X}}.
\end{align}
We call $\mathcal{X}$ triclinic if $\mathcal{P}(\mathcal{X}) = 1 \text{ or } \overline{1}$.
When $\mathcal{X}$ is triclinic, there is no restriction on spin axes.
Thus, we assume $\mathcal{X}$ is not triclinic in what follows.

We denote the time-reversal operation as $\theta$ and set
\begin{align}
    \theta_s = \begin{cases}
        1 & (s = 1), \\
        \theta & (s = -1).
    \end{cases}
\end{align}
We write a MSG induced from $\mathcal{X}$ with spin transformation $\mathbf{Q} \in O(3)$ as
\begin{align}
    &\mathcal{M}(\mathcal{X}, \mathbf{Q}) \nonumber \\
        &= \myset{
            (\mathbf{R}, \boldsymbol{\tau})\theta_{\det \mathbf{U}}
        }{
            \begin{array}{l}
                ((\mathbf{R}, \boldsymbol{\tau}), \mathbf{U}) \in \mathcal{X}, \\
                (\det \mathbf{R}) \mathbf{R} = (\det \mathbf{U}) \mathbf{Q}^{-1} \mathbf{U} \mathbf{Q}
            \end{array}
        },
\end{align}
where spin operation $((\mathbf{R}, \boldsymbol{\tau}), \mathbf{U})$ is mapped to a magnetic operation if $(\det \mathbf{R}) \mathbf{R} = (\det \mathbf{U}) \mathbf{U}$ \cite{Shinohara2024}.

In this paper, we define an oriented SSG of $\mathcal{X}$ as a pair $(\mathcal{X}, \mathbf{Q})$ with equivalence relation $(\mathcal{X}, \mathbf{Q}) \sim (\mathcal{X}, \mathbf{Q}')$ if $\mathcal{M}(\mathcal{X}, \mathbf{Q})$ and $\mathcal{M}(\mathcal{X}, \mathbf{Q}')$ are conjugate by certain $g \in \mathcal{G}(\mathcal{X})$.
Furthermore, we consider only oriented SSGs such that $\mathcal{M}(\mathcal{X}, \mathbf{Q})$ is maximal among all MSGs induced from $\mathcal{X}$ by varying $\mathbf{Q} \in O(3)$.

\subsection{\label{sec:msg_noncoplanar}Noncoplanar oriented spin space group}


First, we consider a noncoplanar SSG $\mathcal{X}$ with $\mathcal{B}_{\mathrm{so}}(\mathcal{X}) = 1$.
For a subgroup $\mathcal{Y}$ of $\mathcal{X}$ with $\mathcal{Y} \trianglerighteq \mathcal{T}(\mathcal{H}(\mathcal{X}))$, $\mathcal{Y}$ and $\mathcal{M}(\mathcal{Y}, \mathbf{Q})$ have a one-to-one correspondence for some $\mathbf{Q} \in O(3)$ if $(\det \mathbf{R}) \mathbf{R} = (\det \mathbf{U}) \mathbf{Q}^{-1} \mathbf{U} \mathbf{Q}$ for all $((\mathbf{R}, \boldsymbol{\tau}), \mathbf{U}) \in \mathcal{Y}$.
This condition can be seen as an equivalence of two orthogonal representations of $\mathcal{Y} / \mathcal{T}(\mathcal{H}(\mathcal{X}))$:
\begin{align}
    &\Gamma_{\mathrm{r}} : \mathcal{Y} / \mathcal{T}(\mathcal{H}(\mathcal{X})) \to SO(3); \nonumber \\
    &\quad ((\mathbf{R}, \boldsymbol{\tau}), \mathbf{U}') \mathcal{T}(\mathcal{H}(\mathcal{X})) \mapsto (\det \mathbf{R}) \mathbf{R}, \\
    &\Gamma_{\mathrm{s}} : \mathcal{Y} / \mathcal{T}(\mathcal{H}(\mathcal{X})) \to SO(3); \nonumber \\
    &\quad ((\mathbf{R}, \boldsymbol{\tau}), \mathbf{U}') \mathcal{T}(\mathcal{H}(\mathcal{X})) \mapsto (\det \mathbf{U}') \mathbf{U}'.
\end{align}
If there exists a real intertwiner $\mathbf{Q}_{\ast} \in O(3)$ between $\Gamma_{\mathrm{r}}$ and $\Gamma_{\mathrm{s}}$, SSG $\mathcal{Y}$ can be identified with a MSG:
\begin{align}
    \mathcal{M}(\mathcal{Y}, \mathbf{Q}_{\ast})
        &= \myset{
            (\mathbf{R}, \boldsymbol{\tau}) \theta_{\det \mathbf{U}}
        }{
            ((\mathbf{R}, \boldsymbol{\tau}), \mathbf{U}) \in \mathcal{Y}
        }.
\end{align}
Such a real intertwiner can be numerically constructed by the algorithm described in Appendix~\ref{appx:orthogonal_intertwiner}, if it exists.
Thus, by enumerating subgroups of $\mathcal{X} / \mathcal{T}(\mathcal{H}(\mathcal{X}))$ and finding a real intertwiner between $\Gamma_{\mathrm{r}}$ and $\Gamma_{\mathrm{s}}$ for each subgroup, we can enumerate inequivalent MSGs induced from $\mathcal{X}$.
We further filter the enumerated MSGs to keep only those with maximal symmetry, which correspond to a finite number of oriented SSGs.

\subsection{\label{sec:msg_collinear}Collinear oriented spin space group}

For a collinear SSG $\mathcal{X}$, if its family point group and a normalizer of its spin-only group intersect trivially as $\mathcal{P}(\mathcal{X}) \cap \mathcal{N}(\mathcal{B}_{\mathrm{so}}) = 1$, no nontrivial rotation symmetry operation survives in the induced MSG for any $\mathbf{Q} \in O(3)$, which implies triclinic MSG.
Because there is no restriction on spin axes in triclinic MSGs, we enumerate candidate spin axes for oriented spin-only groups from $\mathbf{R} \in \mathcal{P}(\mathcal{X}) \setminus \{ \mathbf{I}, \mathord{-}\mathbf{I} \}$ in what follows.
We denote a transformation to align with a candidate spin axis as $\mathbf{Q}' \in SO(3)$ and a transformed SSG of $\mathcal{X}$ by $\mathbf{Q}'$ as $\mathcal{X}'$.
For notational simplicity, we assume the spin axis of $\mathcal{B}_{\mathrm{so}}(\mathcal{X}')$ is along the $z$ axis in what follows.

For $((\mathbf{R}, \boldsymbol{\tau}), \mathbf{U}') \in \mathcal{X}'$, we decompose spatial and spin rotation into two-dimensional rotations around the $z$ axis and a sign flip along the $z$ axis, if possible:
\begin{align}
    \label{eq:spatial_rotation_decomposition}
    \mathbf{R}_{xy} \oplus p_z &= \mathbf{R}
        \quad [\mathbf{R}_{xy} \in O_{xy}(2), p_z \in \mathbb{Z}_2], \\
    \label{eq:spin_rotation_decomposition}
    \mathbf{U}'_{xy} \oplus u'_z &= \mathbf{U}'
        \quad [\mathbf{U}'_{xy} \in O_{xy}(2), u'_z \in \mathbb{Z}_2],
\end{align}
where $O_{xy}(2)$ is a two-dimensional orthogonal group acting on the $xy$ plane.
Because $\mathcal{B}_{\mathrm{so}} = O_{xy}(2) \oplus 1$ and $\mathcal{N}(\mathcal{B}_{\mathrm{so}}) / \mathcal{B}_{\mathrm{so}} = 1 \oplus \mathbb{Z}_2$, a transformation $\mathbf{Q} \in \mathcal{N}(\mathcal{B}_{\mathrm{so}})$ acts on $\mathcal{N}(\mathcal{B}_{\mathrm{so}}) / \mathcal{B}_{\mathrm{so}}$ as identity.
Thus, every $\mathbf{Q} \in \mathcal{N}(\mathcal{B}_{\mathrm{so}})$ gives the same MSG $\mathcal{M}(\mathcal{X}', \mathbf{Q})$ and we do not need to enumerate $\mathbf{Q}$ further.
We directly filter the enumerated MSGs $\mathcal{M}(\mathcal{X}', \mathbf{Q}')$ for each candidate spin axis to keep only those with maximal symmetry, which correspond to a finite number of oriented SSGs.

\subsection{\label{sec:msg_coplanar}Coplanar oriented spin space group}

For a coplanar SSG $\mathcal{X}$, it suffices to enumerate candidate spin axes only from $\mathbf{R} \in \mathcal{P}(\mathcal{X}) \setminus \{ \mathbf{I}, \mathord{-}\mathbf{I} \}$ as in the collinear case.
We denote a transformation to align with a candidate spin axis as $\mathbf{Q}' \in SO(3)$ and a transformed SSG of $\mathcal{X}$ by $\mathbf{Q}'$ as $\mathcal{X}'$.
For notational simplicity, we assume the spin axis of $\mathcal{B}_{\mathrm{so}}(\mathcal{X}')$ is along the $z$ axis in what follows.

We decompose spatial and spin rotation parts of spin symmetry operations of $\mathcal{X}'$ as in Eqs.~\eqref{eq:spatial_rotation_decomposition} and \eqref{eq:spin_rotation_decomposition} if possible.
We denote by $\mathcal{Y}'$ a subgroup of $\mathcal{X}'$ such that all spin symmetry operations are decomposed as above:
\begin{align}
    \label{eq:coplanar_fsg_coset_decomposition}
    \mathcal{Y}'
        &= \bigcup_{j}
            ((\mathbf{R}_{j}, \boldsymbol{\tau}_{j}), \mathbf{U}'_{j})
            \{\mathcal{T}(\mathcal{H}(\mathcal{X}')) \times \mathcal{B}_{\mathrm{so}}(\mathcal{X}')\}.
\end{align}

To enumerate MSGs induced from subgroups of $\mathcal{Y}'$, we use the fact that a MSG has a one-to-one correspondence to a pair of its family space group $\mathcal{G}'$ and maximal space subgroup $\mathcal{H}'$ \cite{spglibv2} as follows.
We first enumerate subgroups $\mathcal{G}'$ of $\mathcal{G}(\mathcal{Y}')$ with $\mathcal{G}' \trianglerighteq \mathcal{T}(\mathcal{H}(\mathcal{X}'))$.
For each $\mathcal{G}'$, we further enumerate its normal subgroups $\mathcal{H}' \trianglelefteq \mathcal{G}'$ with $|\mathcal{G}' / \mathcal{H}'| \leq 2$.
For a pair of $(\mathcal{G}', \mathcal{H}')$, we assign a spin rotation for coset representative $(\mathbf{R}_{j}, \boldsymbol{\tau}_{j})$ from $\mathbf{U}'_j \mathcal{B}_{\mathrm{so}}(\mathcal{X}')$ such that $\mathcal{H}'$ corresponds to identity and $\mathcal{G}' \setminus \mathcal{H}'$ corresponds to time-reversal operations:
\begin{align}
    \mathbf{V}_{j}'
        &= \begin{cases}
            \mathbf{U}'_j
                &\quad [ (\mathbf{R}_j, \boldsymbol{\tau}_j) \in \mathcal{H}', \det \mathbf{U}'_j = 1 ], \\
            \mathbf{U}'_j \boldsymbol{\sigma}_z
                &\quad [ (\mathbf{R}_j, \boldsymbol{\tau}_j) \in \mathcal{H}', \det \mathbf{U}'_j = \mathord{-}1 ], \\
            \mathbf{U}'_j \boldsymbol{\sigma}_z
                &\quad [ (\mathbf{R}_j, \boldsymbol{\tau}_j) \in \mathcal{G}' \setminus \mathcal{H}', \det \mathbf{U}'_j = 1 ], \\
            \mathbf{U}'_j
                &\quad [ (\mathbf{R}_j, \boldsymbol{\tau}_j) \in \mathcal{G}' \setminus \mathcal{H}', \det \mathbf{U}'_j = \mathord{-}1 ], \\
        \end{cases}
\end{align}
where $\boldsymbol{\sigma}_z \in \mathcal{B}_{\mathrm{so}}(\mathcal{X}')$ is a mirror operation perpendicular to the $z$ axis and is used to adjust the determinant of spin rotations.
The pair $(\mathcal{G}', \mathcal{H}')$ with assigned spin rotations $\{ \mathbf{V}'_j \}$ has a one-to-one correspondence to a MSG if $(\det \mathbf{R}_j) \mathbf{R}_j = (\det \mathbf{V}'_j) \mathbf{Q}^{-1} \mathbf{V}'_j \mathbf{Q}$ for all $(\mathbf{R}_j, \boldsymbol{\tau}_j) \mathcal{T}(\mathcal{H}(\mathcal{X}')) \in \mathcal{G}' / \mathcal{T}(\mathcal{H}(\mathcal{X}'))$  with a certain $\mathbf{Q} \in O(3)$.

Because $\mathbf{R}_j$ and $\mathbf{V}'_j$ are decomposed into two-dimensional rotations around the $z$ axis and a sign flip along the $z$ axis as
\begin{align}
    \mathbf{R}_{j, xy} \oplus p_{j, z} &= \mathbf{R}_j \quad &[\mathbf{R}_{j, xy} \in O_{xy}(2), p_{j, z} \in \mathbb{Z}_2], \\
    \mathbf{V}'_{j, xy} \oplus v'_{j, z} &= \mathbf{V}'_j \quad &[\mathbf{V}'_{j, xy} \in O_{xy}(2), v'_{j, z} \in \mathbb{Z}_2],
\end{align}
the above condition requires the equivalence of $(z, z)$ components:
\begin{align}
    \label{eq:coplanar_msg_zz_condition}
    (\det \mathbf{R}_j) p_{j, z} = (\det \mathbf{V}'_j) v'_{j, z}
\end{align}
for all $(\mathbf{R}_j, \boldsymbol{\tau}_j) \mathcal{T}(\mathcal{H}(\mathcal{X}')) \in \mathcal{G}' / \mathcal{T}(\mathcal{H}(\mathcal{X}'))$.
In addition, the following two-dimensional representations must be equivalent:
\begin{align}
    &\Gamma_{\mathrm{r}, xy} : \mathcal{G}' / \mathcal{T}(\mathcal{H}(\mathcal{X}')) \to O_{xy}(2); \nonumber \\
    &\quad (\mathbf{R}_j, \boldsymbol{\tau}_j) \mathcal{T}(\mathcal{H}(\mathcal{X}')) \mapsto (\det \mathbf{R}_j) \mathbf{R}_{j, xy}, \\
    &\Gamma_{\mathrm{s}, xy} : \mathcal{G}' / \mathcal{T}(\mathcal{H}(\mathcal{X}')) \to O_{xy}(2); \nonumber \\
    &\quad (\mathbf{R}_j, \boldsymbol{\tau}_j) \mathcal{T}(\mathcal{H}(\mathcal{X}')) \mapsto (\det \mathbf{V}'_j) \mathbf{V}'_{j, xy}.
\end{align}
An intertwiner $\mathbf{Q}_{xy} \in O_{xy}(2)$ between $\Gamma_{\mathrm{r}, xy}$ and $\Gamma_{\mathrm{s}, xy}$ can be constructed as shown in Appendix~\ref{appx:orthogonal_intertwiner}, if it exists.
Thus, by enumerating pairs of $(\mathcal{G}', \mathcal{H}')$ with Eq.~\eqref{eq:coplanar_msg_zz_condition} and finding a real intertwiner between $\Gamma_{\mathrm{r}, xy}$ and $\Gamma_{\mathrm{s}, xy}$, we can enumerate inequivalent MSGs induced from $\mathcal{X}'$.
We further filter the enumerated MSGs to keep only those with maximal symmetry, which correspond to a finite number of oriented SSGs.

\subsection{\label{sec:msg_chirality}Chirality of spin space group}

So far, the transformation $\mathbf{Q}$ between SSGs has been considered in $O(3)$.
However, we should care about the parity of $\mathbf{Q}$ when discussing chirality of coplanar magnetic structures.
Motivated by the definition of spatial chirality \cite{https://doi.org/10.1002/hlca.200390109}, we say that an object is spin chiral if it cannot be superimposed on its time-reversed counterpart by a proper spin rotation.
Conversely, if such a mapping exists, we call the object spin achiral.
For SSGs, this definition is rephrased as follows: A SSG $\mathcal{X}$ is spin chiral if $\mathcal{X}$ and $(1, \overline{1})^{-1} \mathcal{X} (1, \overline{1})$ are not equivalent by any spin transformation $\mathbf{Q} \in SO(3)$.
However, as $(1, \overline{1})^{-1} (g, \mathbf{U}) (1, \overline{1}) = (g, \mathbf{U})$ for all $(g, \mathbf{U}) \in \mathcal{X}$, they are always equivalent by any $\mathbf{Q} \in O(3)$.
Thus, a SSG is always spin achiral.

That being said, there are still discussions on chirality for coplanar magnetic structures with SOC \cite{Simonet2012,Cheong2022,Pradhan2023}.
Vector spin chirality is commonly used to capture such chirality as an ad hoc indicator \cite{J_Villain_1977,doi:10.1143/JPSJ.68.3185}.
Thus, we further consider chirality of SSGs by restricting spin transformations $\mathbf{Q}$ to $SO(d) \oplus \mathbf{I}_{3-d}$ with spin-structure dimension $d=1, 2$.
We refer to such chirality for collinear ($d=1$) and coplanar ($d=2$) objects as spin axichirality and spin planochirality, respectively.
The terms ``axichirality'' and ``planochirality'' are borrowed from the corresponding concept in spatial chirality \cite{Arnaut01011997,doi:https://doi.org/10.1002/9783527625345.ch1}.
No collinear SSG is spin axichiral, because $O(1)$ is Abelian.
Thus, we focus on spin planochirality and provide a procedure to judge spin planochirality of coplanar SSGs in Appendix~\ref{appx:spin_planochirality}.
For a spin-planochiral coplanar SSG $\mathcal{X}$, we generate oriented SSGs from both $\mathcal{X}$ and its ``enantiomorph'' SSG up to orientation-preserving spin transformation in $SO(2) \oplus 1$.

\section{\label{sec:examples}Representative examples}

We apply and demonstrate the present enumeration scheme to three representative magnetic structures: collinear MnTe in Sec.~\ref{sec:examples_collinear}, coplanar Mn$_{3}$Sn in Sec.~\ref{sec:examples_coplanar}, and noncoplanar CoTa$_{3}$S$_{6}$ in Sec.~\ref{sec:examples_noncoplanar}.

\subsection{\label{sec:examples_collinear}Collinear MnTe}

The magnetic structure of MnTe reported in Ref.~\cite{Kunitomi1964} is assigned to MSG $Cmcm$ (BNS No. 63.457) and collected in entry no.~0.800 of \textsc{MAGNDATA} \cite{MAGNDATA1}.
Its family space group is $P6_{3}/mmc$ (No. 194), and the magnetic Mn atoms are located at Wyckoff position $2a$.
Its magnetic structure is collinear within the $ab$ plane.
The magnetic unit cell is the same as the nonmagnetic one.

\begin{figure}[tb]
    \centering
    \includegraphics[width=0.95\columnwidth]{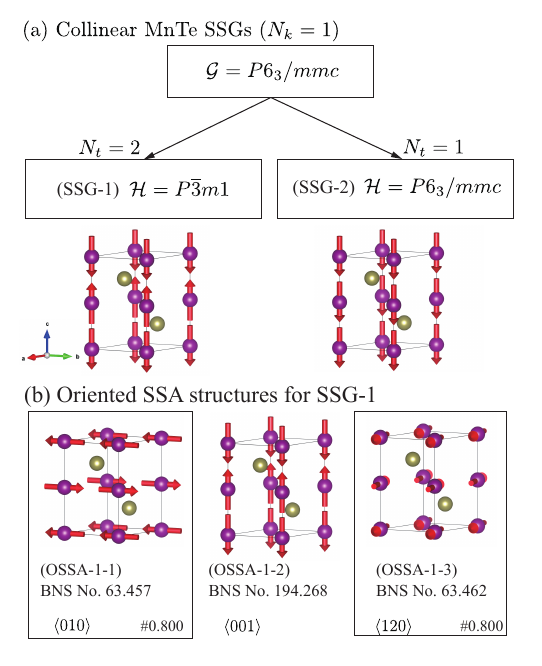}
    \caption{
        Enumeration of collinear magnetic structures of MnTe.
        (a) SSGs compatible with the nonmagnetic family space group $\mathcal{G} = P 6_{3}/m m c$ (No. 194), collinear, and $N_k = 1$.
        Two SSGs are obtained: SSG-1 with normal space subgroup $\mathcal{H} = P \overline{3} m 1$ (No. 164), and SSG-2 with $\mathcal{H} = P 6_{3} / m m c$ (No. 194).
        For each SSG, its SSA structure is shown beneath the corresponding box.
        (b) Collinear oriented SSA structures derived from SSG-1.
        Three distinct oriented SSA structures---OSSA-1-1, OSSA-1-2, and OSSA-1-3---are obtained, corresponding to magnetic moments oriented along the crystallographic directions $\langle 0 1 0 \rangle$, $\langle 0 0 1 \rangle$, and $\langle 1 2 0 \rangle$, respectively.
        Each magnetic space group type of the oriented SSA structures is indicated by the BNS numbers \cite{belov1957neronova}.
        The magnetic moments lying in the $ab$ plane (OSSA-1-1 and OSSA-1-3) are consistent with the experimentally reported one (No.~0.800 in \textsc{MAGNDATA}).
    }
    \label{fig:example_collinear}
\end{figure}

Figure~\ref{fig:example_collinear}(a) illustrates the enumeration procedure for collinear SSGs of MnTe with $N_k = 1$.
Because the magnetic unit cell coincides with the nonmagnetic one, only SSGs with $N_k = 1$ are considered.
Under this constraint, two collinear SSGs compatible with the family space group $\mathcal{G} = P 6_{3}/m m c$ are obtained:
SSG-1, which yields an antiferromagnetic order, and SSG-2, which yields a ferromagnetic order.
Although the illustrative SSA structures are drawn with magnetic moments along the $c$ axis, the SSG alone does not fix the absolute direction of the magnetic moments.

To resolve the magnetic moment directions, we further derive oriented SSA structures from SSG-1 as illustrated in Fig.~\ref{fig:example_collinear}(b).
This generates three inequivalent collinear oriented SSA structures corresponding to distinct crystallographic directions.
Two of them place the magnetic moments in the $ab$ plane, in agreement with the experimentally reported magnetic structure of MnTe.

For comparison, we apply RA to MnTe with $\mathbf{k}=\Gamma$ using \textsc{ISODISTORT} \cite{StokesISODISTORT,Campbell:wf5017}.
Four magnetic irreducible representations contribute at the magnetic Mn $2a$ site: two one-dimensional irreps, $m\Gamma_{2}^{+}$ and $m\Gamma_{4}^{+}$, which yield collinear orders along the $c$ axis [a ferromagnetic structure with MSG $P6_{3}/mm'c'$ (BNS No.~194.270) and an antiferromagnetic structure with MSG $P6_{3}'/m'm'c$ (BNS No.~194.268), respectively]; and two two-dimensional irreps, $m\Gamma_{5}^{+}$ and $m\Gamma_{6}^{+}$, each of which admits two maximal isotropy subgroups corresponding to in-plane moments along $\langle 0 1 0 \rangle$ and along $\langle 1 2 0 \rangle$.
The $m\Gamma_{5}^{+}$ branch produces the antiferromagnetic in-plane structures with MSGs $Cmcm$ (BNS No.~63.457) and $Cm'c'm$ (BNS No.~63.462), while the $m\Gamma_{6}^{+}$ branch produces the ferromagnetic in-plane structures with MSGs $Cm'cm'$ (BNS No.~63.464) and $Cmc'm'$ (BNS No.~63.463).
The resulting six collinear magnetic structures are in one-to-one correspondence with the six oriented SSA structures derived from SSG-1 and SSG-2 above.
We emphasize that this coincidence between RA combined with maximal isotropy subgroups and our oriented SSA enumeration is specific to MnTe and does not hold for collinear structures, in general.
As shown in Appendix~\ref{appx:stat_ssg_equivalent_sites}, even for collinear magnetic structures, the SSG and the MSG can yield different crystallographic orbits of magnetic sites, because spin-space-group operations enforce equal-magnitude constraints across sites that the MSG would otherwise split.
In such cases, an enumeration based on maximal magnetic isotropy subgroups in RA differs from the SSG-based oriented SSA enumeration.

\subsection{\label{sec:examples_coplanar}Coplanar \texorpdfstring{$\mathrm{Mn}_{3}\mathrm{Sn}$}{Mn3Sn}}

Two magnetic structures of Mn$_{3}$Sn are collected in entries No.~0.199 and No.~0.200 of \textsc{MAGNDATA} \cite{MAGNDATA1}, reported in Ref.~\cite{P_J_Brown_1990}.
They are assigned to MSGs $Cmc'm'$ (BNS No. 63.463) and $Cm'cm'$ (BNS No. 63.464), respectively.
Their family space group is $P6_{3}/mmc$ (No. 194), and the magnetic Mn atoms are located at Wyckoff position $6h$.
Their magnetic structures are coplanar within the $ab$ plane.
The magnetic unit cell is the same as the nonmagnetic one, $N_k = 1$.

\begin{figure*}[tb]
    \centering
    \includegraphics[width=0.95\textwidth]{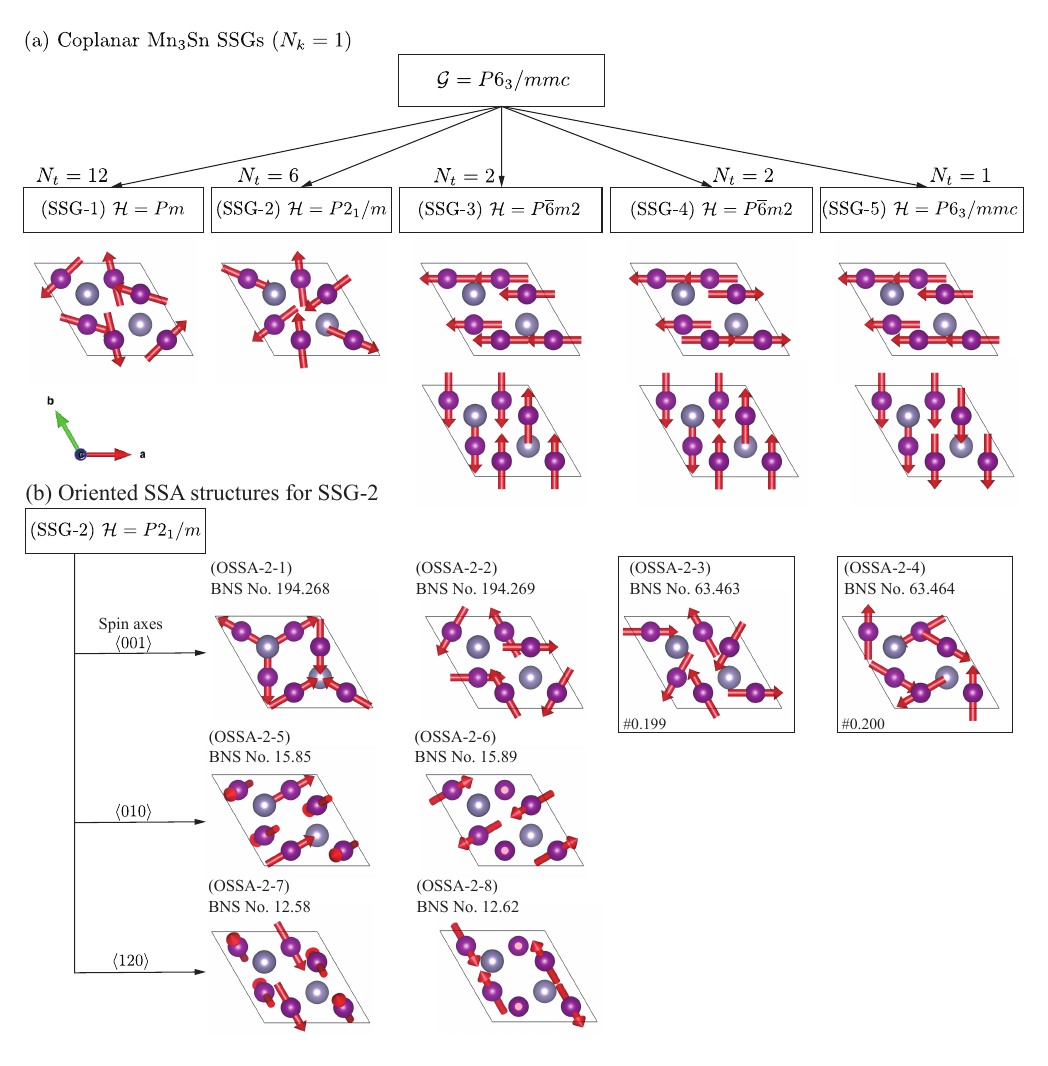}
    \caption{
        Enumeration of coplanar magnetic structures of Mn$_{3}$Sn.
        (a) SSGs compatible with the nonmagnetic family space group $\mathcal{G} = P 6_{3}/m m c$ (No. 194), coplanar, and $N_k = 1$.
        Five SSGs are obtained: SSG-1 with normal space subgroup $\mathcal{H} = P m$ (No. 6), SSG-2 with $\mathcal{H} = P 2_{1}/m$ (No. 11), SSG-3 and SSG-4 with $\mathcal{H} = P \overline{6} m 2$ (No. 187), and SSG-5 with $\mathcal{H} = P 6_{3}/m m c$ (No. 194).
        For each SSG, SSA structures from linearly independent symmetry-adapted magnetic moment bases are shown beneath the corresponding boxes.
        (b) Coplanar oriented SSA structures derived from SSG-2 and its enantiomorphic one.
        For three inequivalent crystallographic orientations of the coplanar spin axis, eight oriented SSA structures (OSSA-2-1 to OSSA-2-8) are obtained in total.
        For $\langle 0 0 1 \rangle$, four oriented SSA structures appear, among which OSSA-2-3 (BNS No. 63.463) and OSSA-2-4 (BNS No. 63.464) correspond to the experimentally reported magnetic structures of Mn$_{3}$Sn in \textsc{MAGNDATA} entries No.~0.199 and No.~0.200, respectively.
        For $\langle 0 1 0 \rangle$ and $\langle 1 2 0 \rangle$, two oriented SSA structures are generated for each direction.
    }
    \label{fig:example_coplanar_SSG}
\end{figure*}

Figure~\ref{fig:example_coplanar_SSG}(a) illustrates the enumeration procedure for coplanar SSGs of Mn$_{3}$Sn with $N_k = 1$.
Because the magnetic unit cell coincides with the nonmagnetic one, only SSGs with $N_k = 1$ are considered.
Under this constraint, five coplanar SSGs compatible with the family space group $\mathcal{G} = P 6_{3}/m m c$ are obtained: SSG-1 to SSG-5.
Although SSG-3 and SSG-4 are enumerated from the same normal space subgroup $\mathcal{H} = P \overline{6} m 2$, their spin rotation parts are inequivalent.
SSG-1 and SSG-2 yield one-dimensional symmetry-adapted magnetic moments, respectively.
SSG-3 to SSG-5 yield two-dimensional symmetry-adapted magnetic moments with collinear bases.

We further derive oriented SSA structures from SSG-2 as illustrated in Fig.~\ref{fig:example_coplanar_SSG}(b).
Because SSG-2 is spin planochiral, we generate oriented SSA structures from both SSG-2 and its enantiomorphic one as discussed in Sec.~\ref{sec:msg_chirality}.
Three distinct crystallographic orientations of the coplanar spin-only group contribute: $\langle 0 0 1 \rangle$, $\langle 0 1 0 \rangle$, and $\langle 1 2 0 \rangle$.
For $\langle 0 0 1 \rangle$, four oriented SSA structures are generated: OSSA-2-1 to OSSA-2-4.
OSSA-2-3 and OSSA-2-4 correspond to the experimentally reported magnetic structures of Mn$_{3}$Sn in entries No.~0.199 and No.~0.200 of \textsc{MAGNDATA}, respectively.
While OSSA-2-3 and OSSA-2-4 are often referred to as inverse triangular, they are equivalent to normal triangular OSSA-2-2 and OSSA-2-1, respectively, by a global spin transformation $C_{6z} C_{2x}$.
Therefore, the distinction by spin planochirality should be used for coplanar Mn$_{3}$Sn when we discuss these magnetic structures.
For $\langle 0 1 0 \rangle$ and $\langle 1 2 0 \rangle$, two oriented SSA structures are generated for each orientation.
Because the enantiomorphic SSG yields isomorphic oriented SSGs in these cases, the spin planochirality does not affect the resulting oriented SSA structures.

For comparison, we apply RA to Mn$_{3}$Sn with $\mathbf{k}=\Gamma$ using \textsc{ISODISTORT} \cite{StokesISODISTORT,Campbell:wf5017}.
The magnetic site representation of Mn at the $6h$ Wyckoff position decomposes into six one-dimensional irreducible representations ($m\Gamma_{1}^{-}$, $m\Gamma_{2}^{\pm}$, $m\Gamma_{3}^{\pm}$, and $m\Gamma_{4}^{+}$) each appearing with multiplicity one, and four two-dimensional irreducible representations ($m\Gamma_{5}^{\pm}$ and $m\Gamma_{6}^{\pm}$), of which $m\Gamma_{5}^{+}$ and $m\Gamma_{6}^{-}$ appear with multiplicity one while $m\Gamma_{5}^{-}$ and $m\Gamma_{6}^{+}$ appear with multiplicity two.
Each one-dimensional irrep generates a unique magnetic structure with a type-I or type-III MSG whose family space group is $P 6_{3}/m m c$ (No.~194), and each two-dimensional irrep admits two maximal isotropy subgroups corresponding to coplanar moments in the $ab$ plane, all of which are type-I or type-III MSGs whose family space group is $C m c m$ (No.~63).

A characteristic feature of Mn$_{3}$Sn is that the magnetic site representation of $6h$ contains $m\Gamma_{5}^{-}$ and $m\Gamma_{6}^{+}$ each with multiplicity two, so RA generates two linearly independent basis vectors per order-parameter direction for these irreps.
Each basis vector individually corresponds to a magnetic configuration with unequal Mn moment magnitudes on the $6h$ orbit, and only a specific linear combination of the two basis vectors recovers the physically expected equal-magnitude configuration.
In the SSG framework, this linear combination is fixed by construction: The spin-space-group operations of the SSG enforce equal magnitudes across the $6h$ orbit, so the symmetry-adapted moment basis of SSG-2 already corresponds to the equal-magnitude subspace.
RA itself does not naturally impose the equal-magnitude condition; incorporating it requires identifying the appropriate linear combination among the multiplicity-induced basis vectors, which is not straightforward, in general.
The inverse-chiral structures OSSA-2-3 and OSSA-2-4, which correspond to the experimental No.~0.199 and No.~0.200 (MSG 63.463 and 63.464 in the $m\Gamma_{6}^{+}$ branch), can also be reached from RA by taking such a linear combination of the two basis vectors.
Conversely, spin planochirality, which we resolve via the enantiomorphic SSG, is invisible to RA at the level of magnetic space groups, so the two enumerations are not in one-to-one correspondence even after the equal-magnitude condition is enforced.

\subsection{\label{sec:examples_noncoplanar}Noncoplanar \texorpdfstring{$\mathrm{Co}\mathrm{Ta}_{3}\mathrm{S}_{6}$}{CoTa3S6}}

The magnetic structure of CoTa$_{3}$S$_{6}$ reported in Ref.~\cite{Takagi2023} is assigned to MSG $P32'1$ (BNS No. 150.27) and collected in entry No.~3.25 of \textsc{MAGNDATA} \cite{MAGNDATA1}.
Its family space group is $P6_{3}22$ (No. 182), and the magnetic Co atoms are located at Wyckoff position $2d$.
Its magnetic structure is noncoplanar, and the magnetic unit cell is 4 times larger than the nonmagnetic one, $N_k = 4$.

\begin{figure*}[tb]
    \centering
    \includegraphics[width=0.95\textwidth]{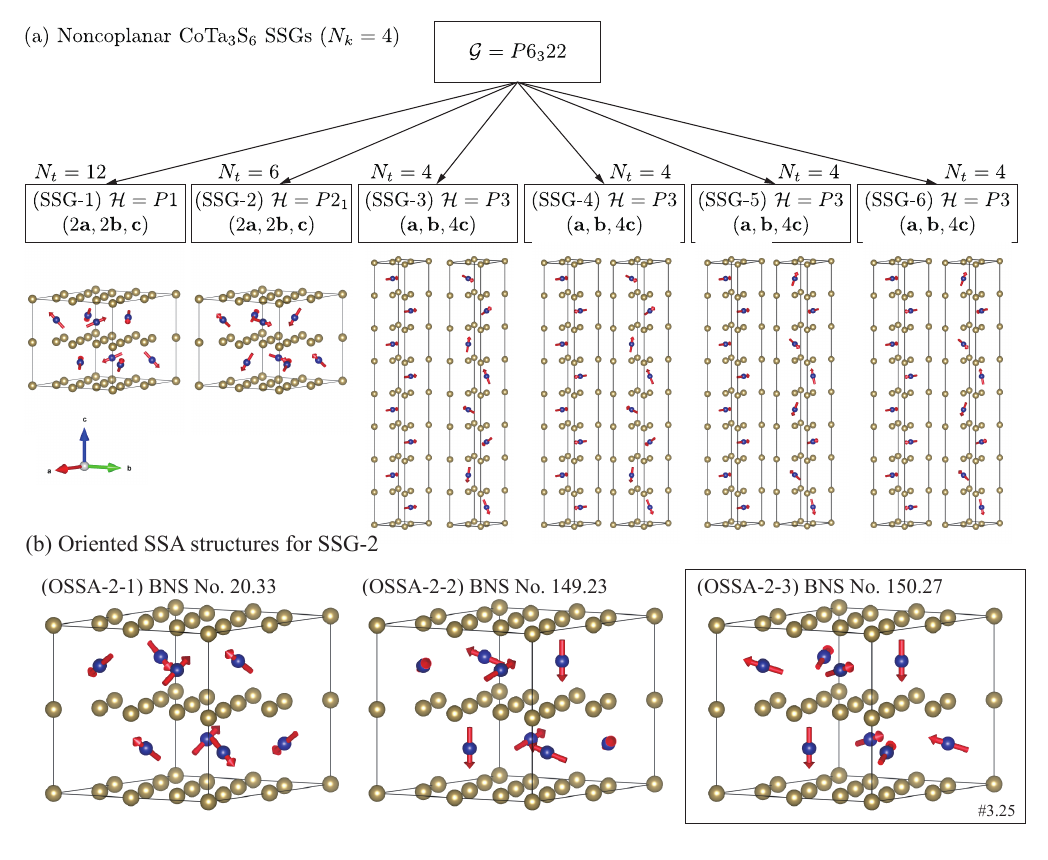}
    \caption{
        Enumeration of noncoplanar magnetic structures of CoTa$_{3}$S$_{6}$.
        (a) SSGs compatible with the nonmagnetic family space group $\mathcal{G} = P 6_{3} 2 2$ (No. 182), noncoplanar, and $N_k = 4$.
        Six SSGs are obtained: SSG-1 with normal space subgroup $\mathcal{H} = P 1$ (No. 1) and $2 \times 2 \times 1$ magnetic unit cell, SSG-2 with $\mathcal{H} = P 2_{1}$ (No. 4) and $2 \times 2 \times 1$ magnetic unit cell, and SSG-3 to SSG-6 with $\mathcal{H} = P 3$ (No. 143) and $1 \times 1 \times 4$ magnetic unit cell.
        For each SSG, SSA structures from linearly independent symmetry-adapted magnetic moment bases are shown beneath the corresponding boxes.
        (b) Noncoplanar oriented SSA structures generated from SSG-2.
        Three distinct oriented SSA structures (OSSA-2-1 to OSSA-2-3) are obtained, among which OSSA-2-3 (BNS No. 150.27) corresponds to the experimentally reported magnetic structure of CoTa$_{3}$S$_{6}$ in \textsc{MAGNDATA} entry No.~3.25.
        The nonmagnetic atoms of sulfur are not shown for clarity.
    }
    \label{fig:example_noncoplanar}
\end{figure*}

Figure~\ref{fig:example_noncoplanar} illustrates the enumeration procedure for noncoplanar SSGs of CoTa$_{3}$S$_{6}$ with $N_k = 4$.
Under this constraint, six noncoplanar SSGs compatible with the family space group $\mathcal{G} = P 6_{3} 2 2$ are obtained: SSG-1 to SSG-6.
Although SSG-3 to SSG-6 are enumerated from the same normal space subgroup $\mathcal{H} = P 3$ with $1 \times 1 \times 4$ magnetic unit cell, their spin rotation parts are inequivalent.
SSG-1 and SSG-2 yield one-dimensional symmetry-adapted magnetic moments, respectively.
SSG-3 to SSG-6 yield two-dimensional symmetry-adapted magnetic moments.

We further derive oriented SSA structures from SSG-2 as illustrated in Fig.~\ref{fig:example_noncoplanar}.
This generates three inequivalent noncoplanar oriented SSA structures (OSSA-2-1 to OSSA-2-3), among which OSSA-2-3 corresponds to the experimentally reported all-in-all-out magnetic structure of CoTa$_{3}$S$_{6}$ in entry No.~3.25 of \textsc{MAGNDATA}.

For comparison, we apply RA to CoTa$_{3}$S$_{6}$ at the M point $\mathbf{k}=(1/2,0,0)$ using \textsc{ISODISTORT} \cite{StokesISODISTORT,Campbell:wf5017}.
Because the oriented SSA scheme and RA take different inputs for the propagation vector, we restrict the comparison to the $2\times 2\times 1$ magnetic unit cell of nonmagnetic CoTa$_{3}$S$_{6}$ by selecting maximal isotropy subgroups in which all three arms of the M star $(1/2,0,0)$, $(0,1/2,0)$, and $(1/2,1/2,0)$ are simultaneously active, since activating only a single arm yields a smaller magnetic cell that cannot reproduce the $2\times 2\times 1$ supercell of \textsc{MAGNDATA} entry No.~3.25.
At the Co $2d$ Wyckoff position, four three-dimensional irreducible representations $m\mathrm{M}_{1}$, $m\mathrm{M}_{2}$, $m\mathrm{M}_{3}$, and $m\mathrm{M}_{4}$ contribute, of which $m\mathrm{M}_{1}$ and $m\mathrm{M}_{4}$ appear with multiplicity one in the magnetic site representation while $m\mathrm{M}_{2}$ and $m\mathrm{M}_{3}$ appear with multiplicity two.
The corresponding maximal isotropy subgroups all share the family space group $P6_{3}22$ (No.~182): $m\mathrm{M}_{1}$ gives the type-I MSG $P6_{3}22$ (BNS No.~182.179), while $m\mathrm{M}_{2}$, $m\mathrm{M}_{3}$, and $m\mathrm{M}_{4}$ give the type-III MSGs $P6_{3}2'2'$ (BNS No.~182.183), $P6_{3}'22'$ (BNS No.~182.182), and $P6_{3}'2'2$ (BNS No.~182.181), respectively.

The multiplicity-one irreps $m\mathrm{M}_{1}$ and $m\mathrm{M}_{4}$ generate coplanar magnetic structures with equal Co moment magnitudes by construction, while the multiplicity-two irreps $m\mathrm{M}_{2}$ and $m\mathrm{M}_{3}$ each yield two linearly independent basis vectors per order-parameter direction whose individual configurations place unequal moments on the $2d$ orbit.
None of these single-irrep maximal isotropy subgroups, however, reproduces the experimentally reported all-in--all-out structure of No.~3.25, whose MSG is $P32'1$ (BNS No.~150.27).
As shown by the prior RA analyses of CoTa$_{3}$S$_{6}$ in Refs.~\cite{Yanagi2023,Takagi2023}, reaching the all-in--all-out structure with equal Co moment magnitudes requires combining basis vectors from different irreps, not merely from the multiplicity-induced bases of a single irrep.
By contrast, the SSG framework naturally accommodates spin symmetry operations and the equal-magnitude condition on crystallographically equivalent sites by construction, without any posthoc multi-irrep mixing.

\section{Application to MAGNDATA database}
\label{sec:application}

We apply the generation scheme to the MAGNDATA database, which contains experimentally reported magnetic structures.
We first outline a classification scheme and present the resulting analysis.
We then discuss materials that are not readily addressed within our framework.

\subsection{Classification scheme}
\label{sec:method_classification}

The MAGNDATA database comprises 2186~magnetic structures in total.
Structures satisfying any of the following conditions fall outside the scope of the present study and are excluded from the analysis: (i)~disordered structures containing sites with partial occupancy, (ii)~non-AFM structures for which the ratio of the total magnetization to the absolute magnetization exceeds 1\%, and (iii)~structures in which magnetic atoms occupy multiple inequivalent Wyckoff positions of the parent space group.
Here, we clarify two points regarding these criteria.
First, condition (ii)~is introduced for benchmarking magnetic-structure generation, since the present method does not generate canted antiferromagnetic and ferrimagnetic states.
Nevertheless, such structures remain within the scope of magnetic-structure prediction through our method, because, when stable, they can be stabilized by SDFT calculations starting from a noncanted state.
Indeed, how well SDFT performs in these cases is an interesting question in itself, and we leave this for future study.
Second, structures classified under condition (iii)~contain magnetic atoms that are not symmetry related; such systems lie beyond the applicability of any symmetry-based approach, and it is, in principle, necessary to impose \textit{ad~hoc} constraints on the magnetic structures.
Thus, we simply excluded them from further consideration.
Quantitatively, condition (ii)~excludes 342 of the 1764 structures that remain after condition (i).
Among them, 207 structures also contain magnetic atoms on multiple inequivalent Wyckoff positions, as is typical for ferrimagnets, and are, therefore, excluded by condition (iii)~irrespective of condition (ii); removing condition (ii)~would, thus, enlarge the dataset only from 1023 to 1158 materials.

After applying these screening criteria, 1023 structures remain, which are classified into the following five classes.
Throughout this study, symmetry identification is performed using \textsc{moyopy} and \textsc{spinspg} packages~\cite{Shinohara2024,moyo} with symprec $=1\times10^{-4}$ and mag\_symprec $=1\times10^{-2}$. 
\begin{itemize}
\item[(i)] lower\_family\_translation\_group---structures in which the translation group of the family space group is smaller than that of the parent space group;
\item[(ii)] lower\_family\_point\_group---structures in which the point group of the family space group is smaller than that of the parent space group;
\item[(iii)] multiple\_basis---cases in which the totally symmetric representation of the SSG admits multiple linearly independent basis functions;
\item[(iv)] ssg\_matched---structures outside the above three classes, which are reproducible as SSA structures up to $O(3)$ transformations;
\item[(v)] oriented\_ssg\_matched---structures that belong to the ssg\_matched class and are reproducible as oriented SSA structures.
\end{itemize}
Note that mcif files in MAGNDATA contain either the crystal structure corresponding to the magnetically ordered phase or that of the disordered phase.
In the former case, symmetry-identification procedures may fail to correctly determine the parent space group and the primitive cell owing to distortions induced by magnetic ordering, which in some cases affect the above symmetry classification.
[For example, the magnetic structure of 3.26\_CoNb3S6 is generated by our method as a triple-$\bm{Q}$ state when the crystal structure of the paramagnetic phase is used.
In contrast, the corresponding MAGNDATA mcif file is based on the lower-symmetry structure of the magnetic phase and, therefore, is excluded from the present analysis according to criterion (iii).] Although such cases could, in principle, be incorporated into the analysis by providing structural information for the disordered phase, we do not apply such a procedure in the present study for simplicity.
We also note that, when the basis is unique, the allowed SSA structures are restricted to either simple ferromagnetic states or antiferromagnetic states with zero net magnetization.

The benchmark analysis using SDFT calculations in Sec.~\ref{sec:sdft} is carried out for materials belonging to ssg\_matched and oriented\_ssg\_matched.
Magnetic structures classified as lower\_family\_translation\_group, lower\_family\_point\_group, and multiple\_basis are discussed in Sec.~\ref{sec:classification_example}.

\subsection{Summary of classification}
\label{sec:results_classification}

The summary of the classification is presented in Table~\ref{tab:category}, and the complete list is provided in Appendix~\ref{appx:full_list}.
Following the screening conditions described in Sec.~\ref{sec:method_classification}, we find that, among the 2186~magnetic structures listed in MAGNDATA, 1023 structures are suitable for the present analysis.
Of these, 623 structures are classified as ssg\_matched, which should be reproducible, at least up to $O(3)$, within the scheme described in Sec.~\ref{sec:ssg}.
Indeed, by explicitly performing the generation, we confirm that the corresponding structures reported in MAGNDATA are successfully reproduced.
Among these 623 ssg\_matched structures, 511 belong to maximal MSGs of the corresponding SSGs and are classified as oriented\_ssg\_matched.
This indicates that the SOC-induced locking of the spin axis is correctly reproduced for 82\% of the materials, which constitutes one of the main results of this work.

\begin{table}[t]
\centering
\caption{Classification of the 1023~magnetic structures listed in MAGNDATA and considered in this study.}
\label{tab:category}
\begin{tabular*}{\linewidth}{@{\extracolsep{\fill}}lc@{}}
\toprule
Lower\_family\_translation\_group & 66 \\
Lower\_family\_point\_group & 166 \\
Multiple\_basis & 168 \\
ssg\_matched & 623 \\
Oriented\_ssg\_matched & 511 \\
\bottomrule
\end{tabular*}
\end{table}

We now comment on the data excluded by condition (iii)~about the multiple Wyckoff positions.
This condition excludes 606 cases, which may appear to be a large fraction of the dataset.
However, this number should not be regarded as a reliable estimate of the materials excluded by the present framework, because MAGNDATA often reports lower-symmetry structures in the magnetic phase.
A more rigorous estimate would require combining MAGNDATA with another crystallographic database.
Such a systematic structure-matching analysis is technically nontrivial and is left for future study.

As shown below, magnetic structures classified as lower\_family\_point\_group are also reproducible by using all $\mathcal{G}'_0$ (see Sec.~\ref{sec:ssg_family_choices} for its definition) as the target family space group while maintaining a finite number of generation trials.
Including these cases, the overall reproducibility at the SSG level reaches $(623+166)/1023\simeq77\%$, which supports the reliability of the present scheme.
In contrast, the cases classified as multiple\_basis require continuous tuning of internal parameters; consequently, it is fundamentally difficult to predict them within a symmetry-based framework using a finite number of trials.
In addition, as discussed below, the structures classified as lower\_family\_translation\_group cannot be addressed by the present method.
Together, these cases account for 234 out of the 1023 structures.

\subsection{Examples not classified as ssg\_matched}
\label{sec:classification_example}

Here, we present representative examples of magnetic structures that are not classified as ssg\_matched.

\subsubsection{\texorpdfstring{$\mathrm{DyB}_4$}{DyB4}}

\begin{figure}[t]
\centering
\includegraphics[width=\linewidth]{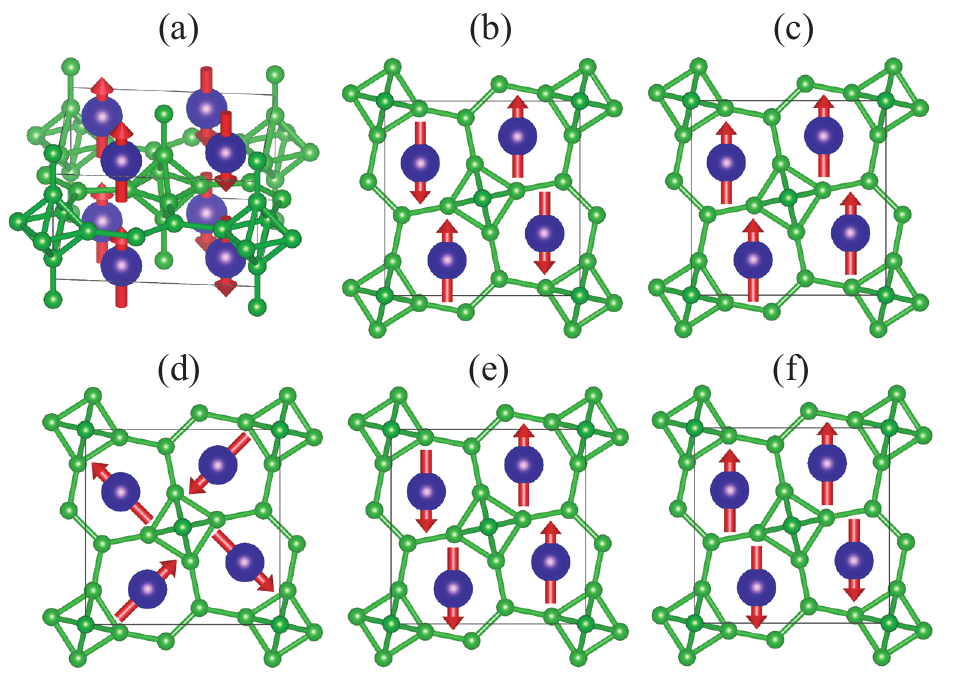}
\caption{Magnetic structures of $\mathrm{DyB}_4$. (a)~Magnetic structures of $\mathrm{DyB}_4$ reported in MAGNDATA entry No.~0.22. (b)--(d)~Three SSA structures with $N_k=1$, whose family space group $\mathcal{G}=P4/mbm$ (No.~127) is identical to the space group. (e),(f) Two SSA structures with the family space group $\mathcal{G}=Pbam$ (No.~55).}
\label{fig:DyB4}
\end{figure}

Figure~\ref{fig:DyB4}(a) shows the magnetic structure of $\mathrm{DyB}_4$ in MAGNDATA entry No.~0.22, and Figs.~\ref{fig:DyB4}(b)--\ref{fig:DyB4}(d) show the SSA structures with $N_k=1$.
The parent point group of $\mathrm{DyB}_4$ is $4/mmm$, and the SSA structures share the same family point group.
Consistent with this symmetry, the structures shown in Figs.~\ref{fig:DyB4}(b)--\ref{fig:DyB4}(d) are invariant under, for example, $[(4^+_{001},\bm{0}),1]$, $[(4^+_{001},\bm{0}),\bar{1}]$, and $[(4^+_{001},\bm{0}),4^-_{001}]$, respectively.
In contrast, the experimental structure shown in Fig.~\ref{fig:DyB4}(a) lacks the spatial fourfold symmetry and, therefore, falls into the lower\_family\_point\_group class.
However, the point group $mmm$ is a subgroup of $4/mmm$ that does not induce Wyckoff-position splitting at the Dy site and is the family point group in Fig.~\ref{fig:DyB4}(a).
Accordingly, this structure can be reproduced while keeping the fixed moment-size condition.
Figures~\ref{fig:DyB4}(e) and \ref{fig:DyB4}(f) show magnetic structures obtained by using a maximal t-subgroup $\mathcal{G}'_0=Pbam$ as the target family space group, with Fig.~\ref{fig:DyB4}(f) being identical to Fig.~\ref{fig:DyB4}(a).
As illustrated by this example, magnetic structures belonging to the lower\_family\_point\_group class are tractable within the present approach.
In the case of $\mathrm{DyB}_4$, relaxing the condition on the target family space group increases the number of SSA structures only from three to five.

For cases with $N_k>1$, there exist combinations of a space group and Wyckoff positions for which no SSA structure has a family space group identical to the parent space group.
Following the treatment described in Sec.~\ref{sec:ssg_family_choices}, we then search for maximal t-subgroups and attempt magnetic-structure generation.
We find that 27 out of 166 cases are successfully reproduced using this procedure.

\subsubsection{\texorpdfstring{$\mathrm{CoSO}_4$}{CoSO4}}

\begin{figure}[t]
\centering
\includegraphics[width=\linewidth]{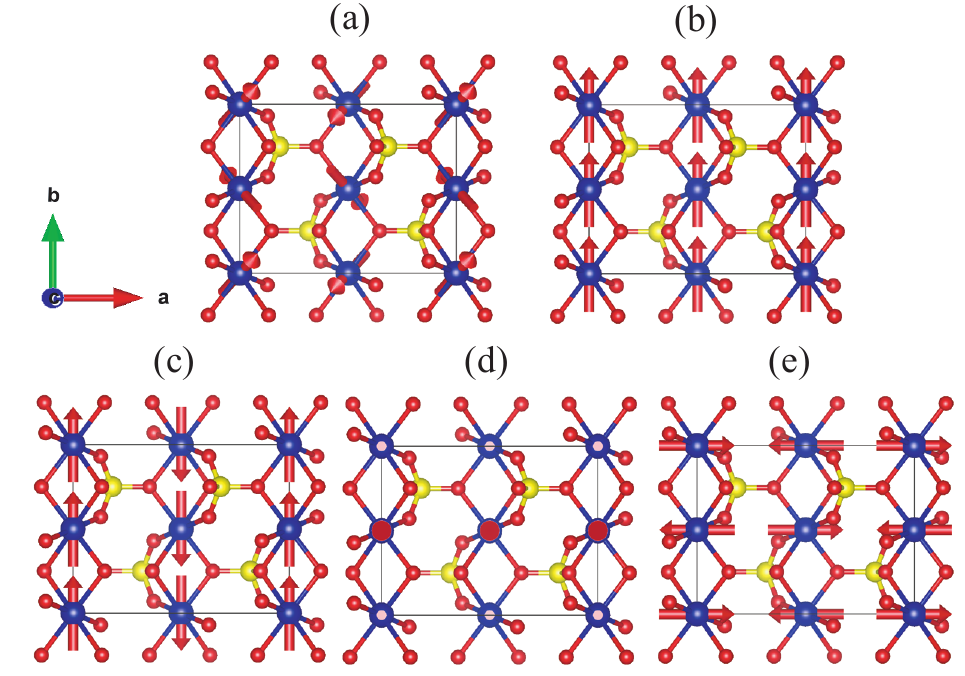}
\caption{Magnetic structures of $\mathrm{CoSO}_4$. (a)~Magnetic structures of $\mathrm{CoSO}_4$ reported in MAGNDATA entry No.~0.96. (b)--(e)~Four SSA structures with $N_k=1$, whose family space group $\mathcal{G}=Pnma$ (No.~62) is identical to the space group. (c)--(e)~correspond to collinear AFMs characterized by modulation directions $[100]$, $[010]$, and $[110]$, respectively.}
\label{fig:CoSO4}
\end{figure}

Figure~\ref{fig:CoSO4}(a) shows the magnetic structure of $\mathrm{CoSO}_4$ reported in MAGNDATA entry No.~0.96, and Figs.~\ref{fig:CoSO4}(b)--\ref{fig:CoSO4}(e) show the SSA structures with $N_k=1$.
It is readily seen that the magnetic structure shown in Fig.~\ref{fig:CoSO4}(a) can be expressed as a linear combination of those in Figs.~\ref{fig:CoSO4}(c)--\ref{fig:CoSO4}(e) and, therefore, belongs to the multiple\_basis class.
In such cases, the coefficients in the linear combination are continuous parameters, and, consequently, an identical magnetic structure cannot be generated by any symmetry-based construction.
Nevertheless, in situations where one magnetic configuration is dominant and the others are only weakly induced as in canted AFMs, the correct magnetic structure may still emerge during self-consistent calculations.
We, therefore, emphasize that not all multiple\_basis cases are entirely intractable within the present framework.

\begin{figure}[t]
\centering
\includegraphics[width=\linewidth]{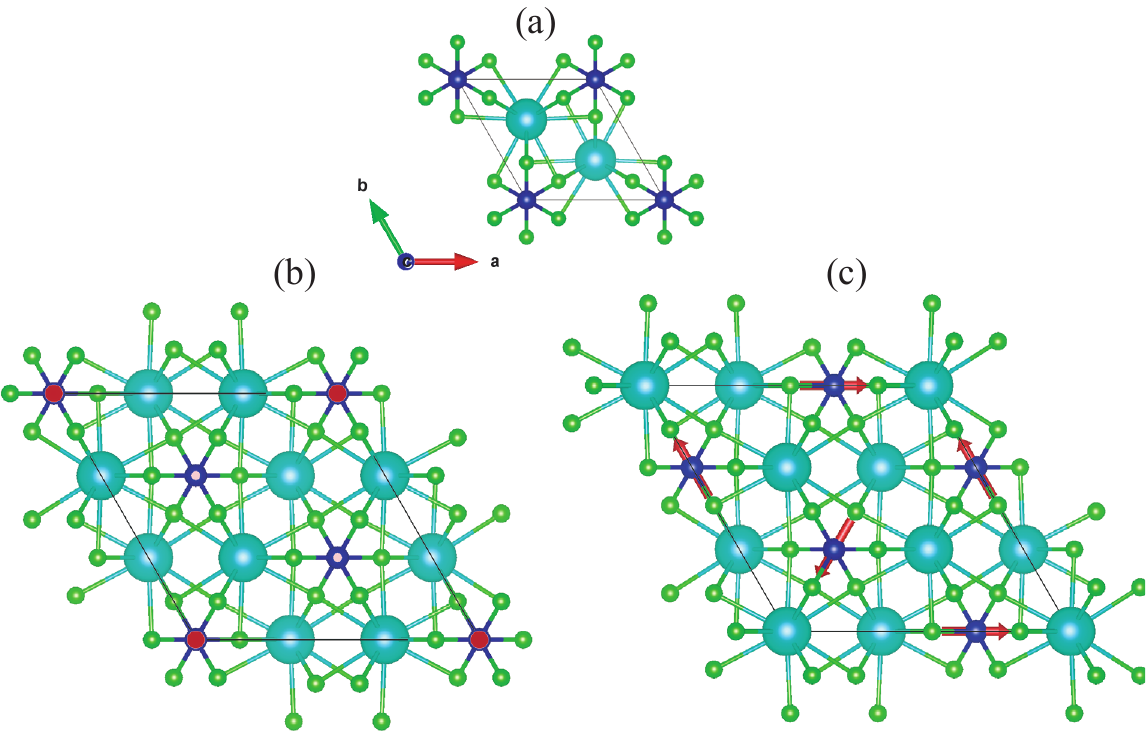}
\caption{Magnetic structure of $\mathrm{CsCoCl}_3$. (a)~Primitive unit cell of $\mathrm{CsCoCl}_3$. (b)~Magnetic structure of $\mathrm{CsCoCl}_3$ reported in MAGNDATA entry No.~1.0.9, whose family space group is $\mathcal{G}=P6_3/mcm$ (No.~193). (c)~An SSA structure with $N_k=3$ and family space group $\mathcal{G}=P6_3/mmc$ (No.~194), which is identical to the space group.}
\label{fig:CsCoCl3}
\end{figure}

\subsubsection{\texorpdfstring{$\mathrm{CsCoCl}_3$}{CsCoCl3}}

Here, we consider the case of $\mathrm{CsCoCl}_3$ (Fig.~\ref{fig:CsCoCl3}).
The primitive unit cell is shown in Fig.~\ref{fig:CsCoCl3}(a), and the magnetic structure in MAGNDATA entry No.~1.0.9 is shown in Fig.~\ref{fig:CsCoCl3}(b).
Figure~\ref{fig:CsCoCl3}(c) presents one of the SSA structures with $N_k=3$.
Both structures shown in Figs.~\ref{fig:CsCoCl3}(b) and \ref{fig:CsCoCl3}(c) have a magnetic unit cell that is 3 times larger than the primitive cell.
However, while the SSA structure preserves the family space group $\mathcal{G}=P6_3/mmc$ (No.~194), which is identical to the space group of the nonmagnetic structure, the experimental magnetic structure has a different family space group $\mathcal{G}=P6_3/mcm$ (No.~193).
The SSA structure is characterized by a helical spin texture associated with a spin translation vector $\bm{k}=(1/3,1/3,0)$, whereas the experimental structure is not because the up-up-down collinear spin alignment cannot be represented by any spin translation.
Consequently, the family translation group of $\mathrm{CsCoCl}_3$ is reduced by a factor of 3, and the structure is assigned to the lower\_family\_translation\_group class.

Within the framework described in Sec.~\ref{sec:ssg}, we consider the space group or its maximal t-subgroups as the family space group of the SSA structures, and, thus, cell expansions not associated with spin translations cannot be treated.
Therefore, the magnetic structures in the lower\_family\_translation\_group class are not reproducible by the present approach.
We finally note that some cases assigned to this class originate from ambiguities in identifying the primitive cell, as discussed at the end of Sec.~\ref{sec:method_classification}; therefore, not all 66 cases in this class are necessarily beyond the reach of the present approach.

\section{High-throughput calculations}
\label{sec:sdft}

In this section, we present benchmarks for the first-principles magnetic-structure prediction method using the proposed scheme.
For materials classified as ssg\_matched, we generate SSA and oriented SSA structures and then perform high-throughput SDFT calculations.
Based on the results, we discuss the computational efficiency and the predictive performance of the proposed scheme.

\subsection{Details of the computation}
\label{sec:method_calc}

We used the \textsc{atomate}2 package~\cite{atomate,atomate2} to manage the calculations, which internally utilizes the \textsc{pymatgen}, \textsc{custodian}, and \textsc{fireworks} libraries~\cite{pymatgen,fireworks}.
Electronic structure calculations were carried out using the Vienna \textit{Ab~Initio} Simulation Package (\textsc{vasp})~\cite{Kresse1996}.
Throughout this study, we employed the Perdew-Burke-Ernzerhof exchange-correlation functional~\cite{Perdew1996} with projector augmented-wave pseudopotentials~\cite{Bloechel1994,Kresse1999}, without applying a Hubbard-$U$ correction.

In practice, self-consistent field (SCF) calculations without SOC were first performed for the SSA structures.
Subsequently, SOC was included, and non-SCF calculations were carried out for the descendant oriented SSA structures, employing fixed-charge densities obtained from the preceding SCF calculations.
We note that oriented SSA structures can be obtained by global $SO(3)$ spin rotations applied to the corresponding SSA structures, for which \textsc{vasp} provides an SCF-to-non-SCF work flow via the \textsc{saxis} flag.
However, because \textsc{saxis} merely defines the spin-quantization axis and does not allow arbitrary $SO(3)$ rotations, we modified the code to accept the Euler angles as input and to apply the corresponding rotations to the magnetic moments prior to the non-SCF calculations.
This extension significantly improved both the computational efficiency and the stability of the oriented SSG-based calculations.

The remaining \textsc{vasp} input parameters for the SCF calculations were set as follows: $\texttt{ENCUT}=1.2$ times the maximum ENMAX value in the POTCAR file, $\texttt{ALGO}=\texttt{ALL}$, $\texttt{ISEARCH}=0$, $\texttt{ISMEAR}=0$, $\texttt{SIGMA}=0.02$, $\texttt{LASPH}=\texttt{True}$, $\texttt{LMIXTAU}=\texttt{True}$, $\texttt{LMAXMIX}=6$, $\texttt{PREC}=\texttt{Accurate}$, $\texttt{LNONCOLLINEAR}=\texttt{True}$, $\texttt{NELM}=200$, $\texttt{KPAR}=6$, $\texttt{NPAR}=2$, and EDIFF = $2\times10^{-5}$.
For the subsequent non-SCF calculations, only the following parameters were modified: $\texttt{ALGO}=\texttt{Normal}$, $\texttt{EDIFF}=1\times10^{-7}$, $\texttt{ICHARG}=11$, and $\texttt{LSORBIT}=\texttt{True}$.
The pseudopotentials were chosen according to the default recommendations of PBE\_64, while open-core pseudopotentials were used for $4f$ and $5f$ elements.
Brillouin-zone integrations were performed using a $\Gamma$-centered $\bm{k}$-point grid generated by \textsc{pymatgen}, with a reciprocal density of $120\,\text{\AA}^{-3}$.
The number of message passing interface processes was fixed to 24 for all calculations.
For unconverged cases, additional calculations were performed with input modifications guided by the default error-handling settings implemented in the \textsc{custodian} code.

\subsection{Summary of the calculations}
\label{sec:sdft_summary}

From the 623 structures classified as ssg\_matched, we further selected a subset of materials according to the following criteria: (i)~the number of atoms in the magnetic unit cell is fewer than 30; (ii)~$4f$ and $5f$ elements are not magnetic; and (iii)~in cases where multiple entries share the same crystal structure, only one representative is retained.
Applying these criteria, we obtained a total of 283 distinct materials; the full list is provided in Appendix~\ref{appx:full_list}.
Among them, 226 materials additionally satisfy the oriented\_ssg\_matched condition.
For the 283 ssg\_matched materials, we generated 876 SSA structures and 2370 descendant oriented SSA structures.
The computational time and memory requirements for this generation step were negligible compared with those for the subsequent calculations.
Then, 876 SCF calculations without SOC and 2364 non-SCF calculations with SOC were performed in this study; the remaining six oriented SSA structures were not calculated, because the parent SCF calculation did not converge.
For simplicity, the structure generation was restricted to magnetic structures with the same $k$ index $N_k$ as reported in MAGNDATA.

\begin{figure}[t]
\centering
\includegraphics[width=\linewidth]{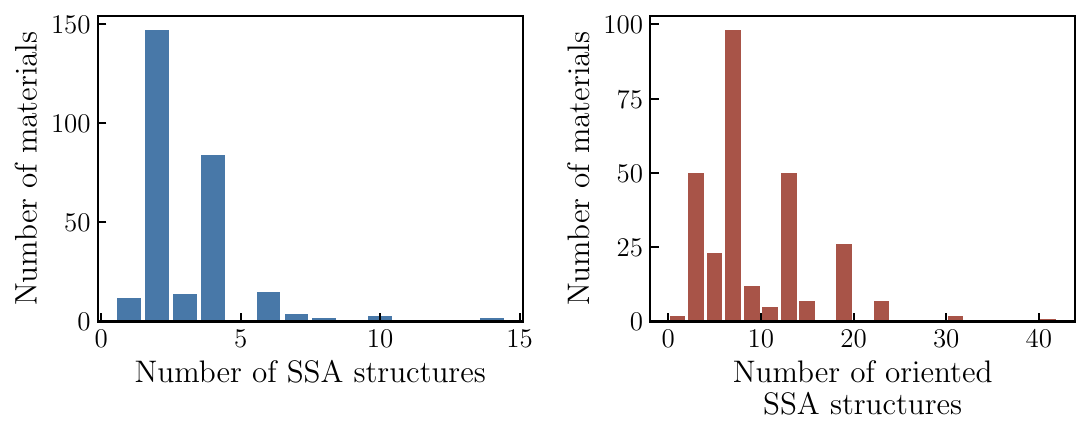}
\caption{Distribution of the number of generated candidate structures per material for the 283 benchmark materials: (left) SSA structures and (right) descendant oriented SSA structures.}
\label{fig:candidate_distribution}
\end{figure}

\begin{figure*}[t]
\centering
\includegraphics[width=\linewidth]{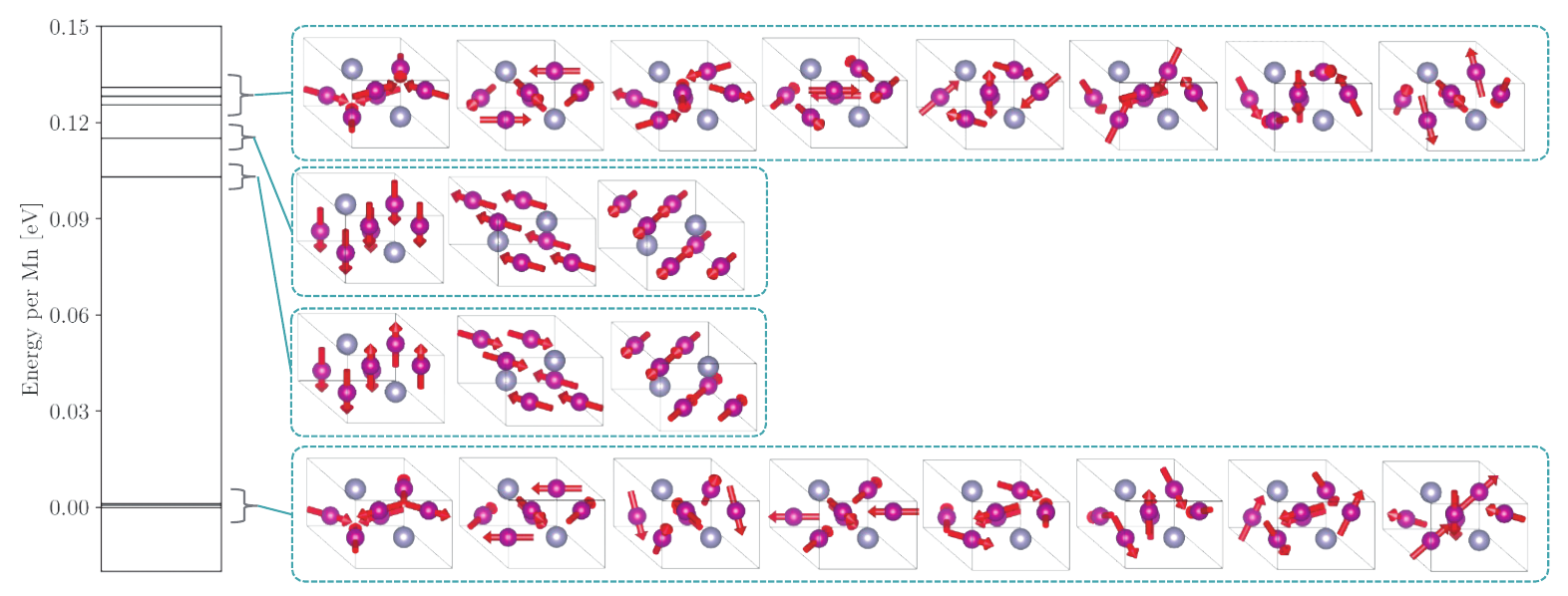}
\caption{Oriented SSA structures with $N_k=1$ and their energy levels of $\mathrm{Mn}_3\mathrm{Sn}$.
The structures enclosed by dashed lines are generated from the same SSG; in particular, the lowest-energy manifold corresponds to the SSG-2 structure shown in Fig.~\ref{fig:example_coplanar_SSG}.}
\label{fig:Mn3Sn}
\end{figure*}

Figure~\ref{fig:candidate_distribution} shows the distribution of the number of generated structures per material.
The number of SSA structures ranges from 1 to 14, and that of oriented SSA structures ranges from 1 to 42.
These demonstrate that the symmetry-based truncation adopted in this work keeps the candidate space per material very small.
We note that 12 materials possess only a single SSA candidate.
For these materials, the SSA-level assessment is trivial, and we exclude these from the SSA-level benchmark, leaving 271 materials; they are retained in the oriented-SSA benchmark and the benchmark relaxing the $N_k$ condition in Sec.~\ref{sec:sdft_kindex}, in which every material has at least two candidates.
As a concrete example, for $\mathrm{Mn}_3\mathrm{Sn}$ (entry No.~0.199), the generation scheme with $N_k=1$ yields exactly four SSA structures (two collinear and two coplanar, the latter shown in Fig.~\ref{fig:example_coplanar_SSG}) and 22 descendant oriented SSA structures (three, three, eight, and eight for the respective SSA structures).
All of the four SSA and 22 oriented SSA structures were calculated, and the complete set of the 22 oriented SSA structures with their energies is displayed in Fig.~\ref{fig:Mn3Sn}.
Here, we note that the number of generated magnetic structures depends on practical choices made in the construction procedure.
These include, for example, how far reductions of the family space group are allowed when generating SSA structures and how low-symmetry an MSG is permitted when generating oriented SSA structures.
The choice of such truncation criteria is ultimately practical, reflecting a balance between completeness and computational cost.
In the present work, we adopt the simplest setting, considering only maximal family space groups without Wyckoff-position splitting and maximal MSGs.
The predictive performance reported below should, therefore, be understood as the result obtained under this specific practical setting rather than as a universal accuracy bound of the framework itself.

Table~\ref{tab:sdft_summary} summarizes the computational performance of SDFT calculations.
Under the conditions described in Sec.~\ref{sec:method_calc}, only one calculation failed to converge (entry No.~1.461).
In contrast, all non-SCF calculations converged successfully.
Even among the converged cases, the magnetic structure sometimes changes during the calculation.
In the present case, the converged magnetic structure deviates by more than 10\% from the input structures in 57 out of 876 SCF calculations and in 189 out of 2364 non-SCF calculations.
We note that the latter includes cases in which the magnetic structure had already changed during the SCF stage, and, therefore, the fraction of changes attributable solely to the non-SCF calculations is much smaller.
Here, the deviation is measured by the normalized overlap between the two spin structures and collapse to an essentially nonmagnetic state ($|m|<0.1\,\mu_{\mathrm B}$ for all atoms) is also counted as deviating.

\begin{table}[t]
\centering
\caption{Summary of computational performance of SDFT calculations. SSG and oriented SSG correspond to SCF without SOC and non-SCF with SOC calculations, respectively.}
\label{tab:sdft_summary}
\begin{tabular*}{\linewidth}{@{\extracolsep{\fill}}lcc@{}}
\toprule
 & SSG & Oriented SSG \\
\midrule
Number of calculations & 876 & 2364 \\
Unconverged calculations & 1 & 0 \\
Magnetic-structure change $>10\%$ & 57 & 189 \\
\bottomrule
\end{tabular*}
\end{table}

Regarding the computational cost, the 876 SCF calculations required a total of approximately 650~hours (wall-clock time) on 24 cores, whereas the non-SCF calculations consumed about 400~hours.
Considering that the number of non-SCF calculations is roughly 3 times larger than that of SCF and that the convergence criteria (EDIFF in the input) are 200 times tighter, this result clearly demonstrates the high efficiency of the fixed-charge approach adopted in this study.
We also note that, without fixing the charge density, electronic-structure calculations with SOC may relax into local minima that differ from those obtained in the calculations without SOC.
From this perspective, even though it can lead to an underestimation of SOC effects, the fixed-charge approach is essential for achieving stable total-energy evaluations, as recommended for magnetic anisotropy calculations in the \textsc{vasp} documentation.

\subsection{Predictive performance}
\label{sec:predictive}

Here, we present results for the predictive performance of the SDFT calculations.
Related analyses based on the cluster multipole method were previously reported by some of the present authors~\cite{Mita2021}, and, more recently, RA-based analyses were reported in Ref.~\cite{Zhou2026arXiv}; both studies focused exclusively on structures with $N_k=1$.
In contrast, the calculations performed here include 175 cases with $N_k=1$ and 108 cases with $N_k>1$.

The results are summarized in Table~\ref{tab:predictive_performance}, and the full list is presented in Appendix~\ref{appx:sdft_list}.
In this study, we filter out calculations that fail to preserve the target configuration before the energy comparison, and a material none of whose calculations preserved the initial structure is counted as a failure.
Also, we do not explicitly include the nonmagnetic state as a candidate configuration.

\begin{table}[t]
\centering
\caption{Summary of the predictive performance of SDFT calculations. $\bar{\Delta}$ is the average energy difference in units of meV per magnetic atom, estimated from the standard deviation.}
\label{tab:predictive_performance}
\begin{tabular*}{\linewidth}{@{\extracolsep{\fill}}lcc@{}}
\toprule
 & SSG & Oriented SSG \\
\midrule
Number of materials & 271 & 226 \\
Correctly identified materials & 225 & 172 \\
$\bar{\Delta}$ & 103 & 0.29 \\
\bottomrule
\end{tabular*}
\end{table}

Among the 271 materials classified as ssg\_matched, 225 materials (approximately 83\%) were correctly identified as having one of their experimentally observed magnetic structures as energetically most stable.
Among the 226 materials classified as oriented\_ssg\_matched, 172 materials (approximately 76\%) were found to have oriented SSA structures identical to the reported one, lying within the set of energetically most stable configurations.
These prediction accuracies are comparable to the 78\% reported in Ref.~\cite{Zhou2026arXiv}.
We note that energy comparisons among oriented SSA structures require resolving small energy differences induced by SOC.
Given the intrinsic accuracy of SDFT, it is necessary to introduce an appropriate energy tolerance to judge the agreement with the reported structures.
In the present analysis, we adopt a threshold of $30\,\mu\mathrm{eV}$ per magnetic atom, chosen as one-tenth of the typical SOC energy scale (see the following discussion).
From these results, the reliability of SDFT for assessing the stability of magnetic structures is on the order of 80\%, in terms of both SSG and oriented SSG descriptions.
It should be noted, however, that this figure is not absolute, as it may vary depending on the choice of the exchange-correlation functional.

Note that, for 0.607\_RuO2, 1.5\_YBa2Cu3O6+d, and 1.420\_YBa2Cu3O6, all SSA calculations collapsed to an essentially nonmagnetic state ($|m|\lesssim0.05\,\mu_{\mathrm B}$ per magnetic atom), consistent with the well-known failure of generalized gradient approximation to reproduce their magnetically ordered states~\cite{Smolyanyuk2024,Brahimi2024arXiv}.
Although we do not explicitly include the nonmagnetic state in the calculations, these materials are correctly counted as failures through the filtering.

Next, we discuss the energy differences among SSA structures and those among oriented SSA structures, which are required to validate the two-step prediction: SCF calculations without SOC are first used to identify the most stable SSA structure, followed by non-SCF calculations with SOC to resolve the oriented SSA structure manifold.
As discussed above, energy differences among oriented SSA structures are expected to scale with SOC and are, therefore, much smaller than those among SSA structures.
As an example, Fig.~\ref{fig:Mn3Sn} shows the magnetic structures of $\mathrm{Mn}_3\mathrm{Sn}$ together with the corresponding energy differences.
As can be seen, $\mathrm{Mn}_3\mathrm{Sn}$ exhibits a clear separation of energy scales: The inclusion of SOC does not lead to a reordering of the energy levels among different SSA structures.
More systematic results are shown in Table~\ref{tab:predictive_performance}.
Here, we focus only on cases where the converged magnetic structures agree with the initial ones.
The average energy difference $\bar{\Delta}$ among SSA structures, estimated from the standard deviation, is approximately 103~meV per magnetic atom.
In contrast, $\bar{\Delta}$ among oriented SSA structures is about $290\,\mu\mathrm{eV}$ per magnetic atom, indicating a separation of energy scales by nearly a factor of 300.
In fact, among the 226 materials classified as oriented\_ssg\_matched, only two cases (entries No.~0.2 and No.~1.382) were found in which the energetically most stable SSA structure was altered by the inclusion of SOC.
This clear hierarchy demonstrates that, at least for the $d$-electron magnetic materials considered in this study, the proposed scheme is well justified.

\subsection{\texorpdfstring{Benchmark relaxing the $N_k$ condition}{Benchmark relaxing the Nk condition}}
\label{sec:sdft_kindex}

In the benchmarks presented so far, $N_k$ was fixed to that of the experimental structure.
Here, we examine the predictive performance when this restriction is relaxed.
To this end, we performed calculations at the SSA level under the following settings: For materials whose experimental structure has $N_k=1$, the search space was extended up to $N_k=2$; for materials whose experimental structure has $N_k\geq 2$, all $N_k$ values up to the experimental one were considered.
All other computational conditions were the same as in Sec.~\ref{sec:method_calc}.
For reference, the experimental $N_k$ values of the 283 materials are distributed as $N_k=1$: 175, $N_k=2$: 102, $N_k=4$: 4, and $N_k=6$: 2.
With this extension, the total number of SSA structures increases by approximately a factor of 3, from 876 to 2408.
The accuracy at the SSA level becomes $208/283$ (approximately 73\%).
We note that the reciprocal density of the $k$ grid, $120\,\text{\AA}^{-3}$, was optimized for the original dataset; the performance for the present dataset, which involves a wider range of supercell sizes, may, therefore, be improved by employing a higher reciprocal density.

Note that the protocol above still refers to the experimental $N_k$ through the material-dependent upper bound of the search space; a fully blind predictive benchmark should evaluate all candidates within a fixed choice of $N_k$, independently of any experimental information.
Mimicking such a fully blind search by fixing the truncation to $N_k\leq 2$ for every material, where the six materials whose experimental structures have $N_k>2$ are counted as failures, the accuracy becomes $203/283$ (approximately 72\%).
This result indicates that our method remains effective even for practical magnetic-structure prediction in which $N_k$ is not known \textit{a~priori}, whereas magnetic structures with $N_k$ beyond the adopted truncation are outside the reach of the search.

\subsection{Remarks on practical high-throughput magnetic-structure prediction}
\label{sec:remarks_ht}

Before closing this section, we make several remarks on the practical use of the present framework in high-throughput magnetic-structure searches.

\subsubsection{Finite-temperature effect}

Stable magnetic structures may depend on temperature, and some compounds exhibit multiple competing magnetic phases.
In general, finite-temperature effects are expected to modify the magnetic state only within an energy scale comparable to the temperature.
Therefore, metastable SSA and oriented-SSA states within several tens of meV of the lowest-energy state may reasonably be regarded as candidate magnetic structures relevant at finite temperatures.
An alternative approach is to use the stable and metastable structures obtained within the present framework to construct spin models, thereby enabling a more direct analysis of finite-temperature magnetism.
The development of spin model construction based on SSA and oriented-SSA structures is an interesting subject for future study.

\subsubsection{\texorpdfstring{Effect of the Hubbard $U$ correction}{Effect of the Hubbard U correction}}

Since magnetic materials are often correlated systems, $\mathrm{SDFT}+U$ is frequently used in practical calculations.
To examine the sensitivity of the present benchmark to this effect, we also considered $\mathrm{SDFT}+U$ calculations for the same dataset; the details are presented in Appendix~\ref{appx:sdft_u}.
At the SSA level, the correctly predicted magnetic structures from SDFT and $\mathrm{SDFT}+U$ overlap by more than 90\%, indicating that the energetic ordering is not strongly affected by $U$ for most compounds in the present benchmark.
Thus, calculations without $U$ are a reasonable default in high-throughput applications.
If one wishes to reduce the risk of missing possible magnetic structures, supplementary $\mathrm{SDFT}+U$ calculations may nevertheless be useful for selected classes of compounds.

\subsubsection{Supercell size}

In the present framework, once the supercell size is specified by $N_k$, magnetic structures with all compatible cells, or, equivalently, all allowed propagation vectors, are generated automatically.
The method can generate magnetic structures with large $N_k$; however, accurate SDFT calculations and reliable energy comparisons for large cells are generally not realistic, and such systems, therefore, fall outside the practical search range of the present approach.
An exception is provided by simple spiral structures, for which calculations without SOC can still be carried out using the frozen-magnon method~\cite{Sandratskii1998}.

\subsubsection{\texorpdfstring{Cases of $4d$ and $5d$ magnets}{Cases of 4d and 5d magnets}}

Although many of the 283 materials used in the benchmark calculations are $3d$ transition-metal magnets, the dataset also includes 20 systems with relatively strong SOC, namely, 11 containing $4d$ elements and nine containing $5d$ elements.
At the SSA level, the prediction fails for seven of these 20 materials, indicating lower accuracy than for the $3d$ compounds.
However, this deterioration does not appear to originate primarily from SOC-induced energy-level splitting.
In fact, in the $\mathrm{SDFT}+U$ calculations discussed in Appendix~\ref{appx:sdft_u}, only three of the 20 materials are mispredicted, which is comparable to the overall prediction accuracy of approximately 80\%.
Since the main role of $+U$ is to correct the underestimated magnetic moment size, these results suggest that our two-step method remains applicable to $4d$ and $5d$ systems with accuracy comparable to that for $3d$ systems.
We note that our dataset also includes compounds with $4p$, $5p$, and $6p$ elements from groups 13--17: 59 compounds with $4p$, 29 with $5p$, and 18 with $6p$ elements.
This also supports the applicability of our method to a broad range of materials, including systems with relatively strong SOC.

\subsubsection{Self-consistency of SOC in SDFT}

The present work flow employs a fixed-charge scheme, following the recommendation in the \textsc{vasp} documentation, although this treatment is known to underestimate the effect of SOC.
For noncollinear magnetic structures with SOC, fully self-consistent calculations are often numerically challenging because of the complex energy landscape containing many local metastable electronic structures.
As summarized in Appendix~\ref{appx:sdft_soc}, our direct calculations show that fully self-consistent treatment of SOC substantially worsens the convergence behavior and also reduces the predictive performance.
These results suggest that, for the purpose of magnetic-structure prediction, the present fixed-charge treatment offers a more practical and robust description.

\subsubsection{Wyckoff-position splittings}

In the present work, the family space groups used to construct SSGs are restricted to those without Wyckoff-position splitting of the magnetic atoms, thereby imposing the equal-moment-size condition.
In principle, this restriction could be relaxed to generate magnetic structures with different moment sizes and use them as additional candidates in the search.
However, we do not recommend adopting this extension in the present high-throughput work flow for the following reasons.

First, when this condition is relaxed, the relative angles and magnitudes of the magnetic moments on the split sites are not fixed by symmetry.
As a result, such structures cannot be generated uniquely on symmetry grounds alone, and additional \textit{ad~hoc} choices would be required.
Second, a large difference in moment size between magnetic atoms on symmetry-equivalent sites is generally energetically unfavorable.
This is because such a large difference typically requires a difference in valence and, consequently, a strong charge-density modulation; if stable, such a modulation would appear above the magnetic transition.
In practice, including such unstable configurations among the candidate magnetic structures can severely degrade the convergence and stability of the calculations, which are among the most critical factors in high-throughput applications.
Indeed, our analysis of MAGNDATA identified only 52 out of 1764 structures in which symmetry-equivalent sites have different moment sizes, as discussed in Appendix~\ref{appx:stat_ssg_equivalent_sites}.
Although this number should not be regarded as a fully exact estimate, since MAGNDATA sometimes contains crystal structures whose symmetry has already been lowered by magnetic ordering, the small number indicates that such cases are uncommon.
Finally, we note that small differences in moment size may occur even on symmetry-equivalent sites.
Such cases, however, can still be reached by SDFT calculations starting from an equal-moment initial structure and can, therefore, be explored while retaining the nonsplitting constraint on Wyckoff positions.

\section{Conclusion}
\label{sec:conclusion}

We have presented a framework for the systematic generation of magnetic structures inspired by the concept of oriented SSGs.
A key advantage of the SSG-based formulation is that it can naturally enforce the fixed-moment condition on symmetry-equivalent sites, thereby enabling an exhaustive enumeration of magnetically inequivalent structures up to global $O(3)$ transformations.
Building on the oriented SSG concept, we further introduced a practical procedure for constructing magnetic structures that are inequivalent under the parent space group and are, thus, suitable for resolving SOC-induced energy splittings.

Applying the method to the MAGNDATA database, we classified 1023 reported magnetic structures into five categories and showed typical examples for which our scheme has limited applicability.
For a computationally tractable set of 283 materials, we performed SDFT calculations using a two-step work flow~\cite{Li2025arXiv}: SCF calculations without SOC for the SSA structures followed by fixed-charge non-SCF calculations with SOC for the descendant oriented SSA structures obtained through global $SO(3)$ rotations of the parent SSA structures.
This work flow provides high numerical stability and favorable computational efficiency, as summarized in Sec.~\ref{sec:sdft_summary}.

The benchmark results quantify the practical reliability of first-principles magnetic-structure prediction in a broad dataset.
At the SSG level, the experimentally reported structures are identified as energetically stable in 83\% of the examined materials.
At the oriented SSG level, the corresponding success rate is 76\%.
Importantly, we found a clear separation of energy scales: Typical energy differences among distinct SSA structures are around 100~meV per magnetic atom, whereas those among oriented SSA structures are around 0.3~meV per magnetic atom, indicating an SOC-induced hierarchy by roughly a factor of 300.
Consistent with this hierarchy, only a very small fraction of materials exhibit a reordering of the lowest-energy SSA structure upon inclusion of SOC.

These findings establish oriented SSG-based enumeration, combined with first-principles calculations within a two-step strategy, as a robust and efficient framework for large-scale magnetic-structure prediction.
The proposed approach is applicable to systematic searches for AFMs with targeted symmetry properties and provides a quantitative baseline for assessing the accuracy limits of SDFT-based magnetic-structure predictions.
Future extensions to magnetic structures classified in lower\_family\_translation\_group and multiple\_basis classes will further broaden the applicability and strengthen symmetry-guided high-throughput approaches.

An implementation of the present algorithm is distributed under a permissive free software license in \textsc{spinforge}~\cite{Spinforge}.
Figures of magnetic structures are created using \textsc{vesta}~\cite{Momma:db5098}.

\begin{acknowledgments}
This work is supported by JSPS-KAKENHI (No.~JP21H04990, No.~JP23K13058, No.~JP24K00581, No.~JP25K21684, No.~JP25H00420, No.~JP25H01246, No.~JP25H01252, and No.~JP25H02115), JST-ERATO (No.~JPMJER2503), JST-MIRAI (No.~JPMJMI20A1), JST-CREST (No.~JPMJCR23O4), JST-ASPIRE (No.~JPMJAP2317), and the RIKEN TRIP initiative (RIKEN Quantum, Advanced General Intelligence for Science Program, Many-body Electron Systems).
The authors acknowledge the use of the AI-based writing tool ChatGPT (OpenAI, version 5.2) and Claude (Anthropic, version 4.6) for language polishing.
The authors take full responsibility for the content of this manuscript.
\end{acknowledgments}

\section*{Data availability}
The data that support the findings of this article are not publicly available upon publication because it is not technically feasible and/or the cost of preparing, depositing, and hosting the data would be prohibitive within the terms of this research project.
The data are available from the authors upon reasonable request.

\appendix

\section{\label{appx:ssg_construction}Detailed algorithm for enumerating spin space groups}

We continue the discussion in Sec.~\ref{sec:ssg_enumeration} to enumerate SSGs from a given family space group $\mathcal{G}$, a spin-only group $\mathcal{B}_{\mathrm{so}}$, and a $k$ index $N_k$.
Although the following algorithm is essentially the same as that in Refs.~\cite{Xiao2024PRX,Chen2024PRX,Jiang2024PRX}, we here provide a self-contained and more implementation-friendly description usable for on-the-fly SSG enumeration.
This simplification partly arises from restricting equivalence to conjugacy under $\mathcal{G} \times O(3)$ rather than group isomorphism under $\mathcal{N}(\mathcal{G}) \times O(3)$.

The overall procedure to enumerate SSGs is as follows.
We introduce notation for space group $\mathcal{G}$ in Sec.~\ref{appx:ssg_construction_G}.
We enumerate normal t-subgroups $\mathcal{M}$ of $\mathcal{G}$ in Sec.~\ref{appx:ssg_construction_M}.
We enumerate normal k-subgroups $\mathcal{H}$ of $\mathcal{M}$ with $\mathcal{H} \trianglelefteq \mathcal{G}$ in Sec.~\ref{appx:ssg_construction_H}.
Finally, we enumerate injective homomorphisms $U: \mathcal{G} / \mathcal{H} \rightarrow \mathcal{N}(\mathcal{B}_{\mathrm{so}}) / \mathcal{B}_{\mathrm{so}}$ by reducing the problem to enumeration of faithful orthogonal representations of $\mathcal{G} / \mathcal{H}$ in Sec.~\ref{appx:ssg_construction_U}.

\subsection{\label{appx:ssg_construction_G}Group structure of family space group}

We prepare notation for the family space group $\mathcal{G}$.
We write the point group of $\mathcal{G}$ as
\begin{align}
    \mathcal{P}(\mathcal{G}) = \myset{\mathbf{R}_i}{(\mathbf{R}_i, \boldsymbol{\tau}_i) \mathcal{T}(\mathcal{G}) \in \mathcal{G} / \mathcal{T}(\mathcal{G})}.
\end{align}
We denote a lattice formed by translation subgroup $\mathcal{T}(\mathcal{G})$ as
\begin{align}
    \mathcal{L}_{\mathcal{G}} = \myset{\mathbf{t}}{(\mathbf{I}, \mathbf{t}) \in \mathcal{T}(\mathcal{G})}.
\end{align}
Let $\boldsymbol{\tau}_{\bullet}: \mathcal{P}(\mathcal{G}) \rightarrow \mathbb{R}^3$ be a translation part of $\mathcal{G}$ with $\boldsymbol{\tau}_{\mathbf{R}_i} = \boldsymbol{\tau}_i$.
For coset representatives $g_i$ and $g_j$, we choose an index $ij \in \{ 1, \ldots, n \}$ such that $\mathbf{R}_{ij} = \mathbf{R}_i \mathbf{R}_j \, [\in \mathcal{P}(\mathcal{G})]$.
Then, we define a factor system for a group extension of $\mathcal{P}(\mathcal{G})$ by $\mathcal{T}(\mathcal{G})$ as
\begin{align}
    \label{eq:coset_multiplication_G}
    g_i g_j &= (\mathbf{I}, \boldsymbol{\sigma}(i, j)) g_{ij}
\end{align}
with
\begin{align}
    \label{eq:factor_system_G}
    \boldsymbol{\sigma}(i, j) = \boldsymbol{\tau}_{i} + \mathbf{R}_i \boldsymbol{\tau}_{j} - \boldsymbol{\tau}_{ij}.
\end{align}

\subsection{\label{appx:ssg_construction_M}Normal \texorpdfstring{$t$}{t}-subgroup of \texorpdfstring{$\mathcal{G}$}{G}}

We enumerate normal subgroups $\mathcal{P}'$ of $\mathcal{P}(\mathcal{G})$ by brute-force search with memoization.
Since the number of subgroups of an order-$n$ group is bounded by $\exp\left( O[(\log n)^2] \right)$ \cite{dfa04ab5-fd84-3659-a4f7-00d497f426ff} and $|\mathcal{P}(\mathcal{G})| \leq 48$ for three-dimensional crystallographic point groups, this step is computationally tractable.

For each normal subgroup $\mathcal{P}'$, we denote the coset decomposition of $\mathcal{P}(\mathcal{G})$ by $\mathcal{P}'$ as
\begin{align}
    \mathcal{P}(\mathcal{G}) = \bigcup_{I=1}^{N_t} \mathbf{R}_{i_{I}} \mathcal{P}',
\end{align}
where $N_t = |\mathcal{P}(\mathcal{G}) : \mathcal{P}'|$ and $\{ \mathbf{R}_{i_{I}} \}_{I=1}^{N_{t}}$ are representatives of the cosets.
Then, we construct a t-subgroup $\mathcal{M}$ of $\mathcal{G}$ as
\begin{align}
    \mathcal{M} = \bigcup_{\mathbf{R}_h \in \mathcal{P}'} g_h \mathcal{T}(\mathcal{G}).
\end{align}
While $\mathcal{M}$ is a t-subgroup of $\mathcal{G}$ by construction, we need to check whether $\mathcal{M}$ is normal in $\mathcal{G}$.
That is, for all pairs of $g_{i_{I}} \mathcal{M} \in \mathcal{G} / \mathcal{M}$ and $g_h \mathcal{T}(\mathcal{G}) \in \mathcal{M} / \mathcal{T}(\mathcal{G})$, we require $g_{i_{I}}^{-1} g_h g_{i_{I}} \in \mathcal{M}$.
This condition is equivalent to requiring that
\begin{align}
    \label{eq:normal_M_G}
    \boldsymbol{\tau}_h + \mathbf{R}_h \boldsymbol{\tau}_{i_{I}} - \boldsymbol{\tau}_{i_{I}}
        = \mathbf{R}_{i_{I}} \boldsymbol{\tau}_{\mathbf{R}_{i_{I}}^{-1} \mathbf{R}_h \mathbf{R}_{i_{I}}}
        \, \mod \mathcal{L}_{\mathcal{G}},
\end{align}
where we use $\mathbf{R}_{i_{I}}^{-1} \mathbf{R}_h \mathbf{R}_{i_{I}} \in \mathcal{P}'$ because $\mathcal{P}'$ is normal in $\mathcal{P}(\mathcal{G})$.
In the following, we denote $\mathcal{P}'$ satisfying Eq.~\eqref{eq:normal_M_G} as $\mathcal{P}_{\mathcal{M}}$.

Note that, for coset representatives $\mathbf{R}_{i_I}$ and $\mathbf{R}_{i_J}$, their multiplication may not be a coset representative of $\mathcal{P}(\mathcal{G}) / \mathcal{P}(\mathcal{M})$.
Here $IJ$ labels the product coset in $\mathcal{P}(\mathcal{G}) / \mathcal{P}(\mathcal{M})$, and $i_{IJ}$ is the index of its chosen representative.
We denote the index $h_{IJ} \in \{ 1, \ldots, n\}$ such that
\begin{align}
    \label{eq:point_group_factor_system_G_M}
    \mathbf{R}_{i_{IJ}} = \mathbf{R}_{i_I} \mathbf{R}_{i_J} \mathbf{R}_{h_{IJ}}.
\end{align}

\subsection{\label{appx:ssg_construction_H}Normal space subgroup of \texorpdfstring{$\mathcal{G}$}{G}}

We enumerate normal sublattices $\mathcal{L}'$ of $\mathcal{L}_{\mathcal{G}}$ under $\mathcal{P}(\mathcal{G})$ with a given $k$ index $N_k = |\mathcal{L}_{\mathcal{G}} / \mathcal{L}'|$.
Inequivalent sublattices of $\mathcal{L}_{\mathcal{G}}$ are enumerated by Hermite normal forms of $3 \times 3$ integer matrices with determinant $N_k$ \cite{Hart2008}.
The normality condition is expressed as
\begin{align}
    \mathbf{R}_i \mathcal{L}' = \mathcal{L}' \quad [\forall \mathbf{R}_i \in \mathcal{P}(\mathcal{G})],
\end{align}
and it can be checked by verifying that the basis vectors of $\mathbf{R} \mathcal{L}'$ and $\mathcal{L}'$ are related by a unimodular matrix.
We denote the translation subgroup formed by $\mathcal{L}'$ as $\mathcal{T}'$ and coset representatives of $\mathcal{T}(\mathcal{G}) / \mathcal{T}'$ as
\begin{align}
    \mathcal{T}(\mathcal{G}) = \bigcup_{m=1}^{N_k} (\mathbf{I}, \mathbf{t}_m) \mathcal{T}'.
\end{align}

Next, we consider a normal $k$-subgroup $\mathcal{H}$ of $\mathcal{M}$ with translation subgroups $\mathcal{T}'$.
Although point groups of $\mathcal{H}$ and $\mathcal{M}$ are the same, their translation parts may differ up to $\mathcal{L}_{\mathcal{G}}$ \cite{Stokes:vk5013}.
Thus, we enumerate possible additional translation parts $\mathbf{t}_{m_{h}} \, (m_{h} \in \{ 1, \dots, N_{k} \})$ for each coset representative such that
\begin{align}
    \mathcal{H} &= \bigcup_{\mathbf{R}_h \in \mathcal{P}(\mathcal{M})} \tilde{g}_h \mathcal{T}',
\end{align}
where $\tilde{g}_h = (\mathbf{R}_h, \boldsymbol{\tau}_h + \mathbf{t}_{m_h})$.
We denote the translation part of $\tilde{g}_h$ as
\begin{align}
    \tilde{\boldsymbol{\tau}}_{h} = \boldsymbol{\tau}_h + \mathbf{t}_{m_h}.
\end{align}
For each pair of $\tilde{g}_{h_1}$ and $\tilde{g}_{h_2}$, their multiplication
\begin{align*}
    &\tilde{g}_{h_1} \tilde{g}_{h_2} \\
        &= (\mathbf{I}, \mathbf{t}_{m_{h_1}}) g_{h_1} (\mathbf{I}, \mathbf{t}_{m_{h_2}}) g_{h_2} \\
        &= (\mathbf{I}, \mathbf{t}_{m_{h_1}} + \mathbf{R}_{h_1} \mathbf{t}_{m_{h_2}}) g_{h_1} g_{h_2} \\
        &= (\mathbf{I}, \mathbf{t}_{m_{h_1}} + \mathbf{R}_{h_1} \mathbf{t}_{m_{h_2}} + \boldsymbol{\sigma}(h_1, h_2)) g_{h_1 h_2}
            \quad [\because \mbox{Eq.~\eqref{eq:coset_multiplication_G}}]
\end{align*}
implies
\begin{align}
    \label{eq:closure_H}
    \mathbf{t}_{m_{h_1h_2}}
        &= \mathbf{t}_{m_{h_1}} + \mathbf{R}_{h_1} \mathbf{t}_{m_{h_2}} + \boldsymbol{\sigma}(h_1, h_2) \pmod{\mathcal{L}'}
\end{align}
Thus, we obtain a group $\mathcal{H}$ by exhaustively assigning $\mathbf{t}_{m}$ for each generator of $\mathcal{P}(\mathcal{M})$ and checking the condition of Eq.~\eqref{eq:closure_H} for all pairs of elements in $\mathcal{P}(\mathcal{M})$.

Before checking the associativity condition, we first check the normality conditions of $\mathcal{H}$ in $\mathcal{M}$ and $\mathcal{G}$.
If $\mathcal{H}$ is normal in $\mathcal{M}$, then for all pairs of $\mathbf{R}_h \in \mathcal{P}(\mathcal{M})$ and $m \in \{ 1, \dots, N_{k} \}$, we require $(\mathbf{I}, \mathbf{t}_m)^{-1} \tilde{g}_h (\mathbf{I}, \mathbf{t}_m) \in \mathcal{H}$.
The multiplication leads to
\begin{align*}
    (\mathbf{I}, \mathbf{t}_m)^{-1} \tilde{g}_h (\mathbf{I}, \mathbf{t}_m)
        &= (\mathbf{R}_h, \tilde{\boldsymbol{\tau}}_h + \mathbf{R}_h \mathbf{t}_m - \mathbf{t}_m) \\
        &= (\mathbf{I}, \mathbf{R}_h \mathbf{t}_m - \mathbf{t}_m) \tilde{g}_h.
\end{align*}
Thus, the normality condition of $\mathcal{H}$ in $\mathcal{M}$ is expressed as
\begin{align}
    \label{eq:normality_H_M}
    \mathbf{R}_h \mathbf{t}_m
        = \mathbf{t}_m \, \mod \mathcal{L}_{\mathcal{H}}.
\end{align}
Similarly, if $\mathcal{H}$ is normal in $\mathcal{G}$, then, for all pairs of $g_{i_{I}} \mathcal{M} \in \mathcal{G} / \mathcal{M}$ and $\tilde{g}_h \mathcal{T}_{\mathcal{H}} \in \mathcal{H} / \mathcal{T}_{\mathcal{H}}$, we require $(\mathbf{I}, \mathbf{t}_m)^{-1} g_{i_{I}}^{-1} \tilde{g}_h g_{i_{I}} (\mathbf{I}, \mathbf{t}_m) \in \mathcal{H}$.
The multiplication leads to
\begin{align*}
    &(\mathbf{I}, \mathbf{t}_m)^{-1} g_{i_{I}}^{-1} \tilde{g}_h g_{i_{I}} (\mathbf{I}, \mathbf{t}_m) \\
        &= g_{i_{I}}^{-1} (\mathbf{I}, \mathbf{R}_{i_{I}} \mathbf{t}_m)^{-1} \tilde{g}_h (\mathbf{I}, \mathbf{R}_{i_{I}} \mathbf{t}_m) g_{i_{I}} \\
        &= g_{i_{I}}^{-1} \tilde{g}_h g_{i_{I}} \, \mod \mathcal{L}_{\mathcal{H}}
            \quad [\because \mbox{Eq.~\eqref{eq:normality_H_M}}] \\
        &= (\mathbf{R}_{i_{I}}^{-1} \mathbf{R}_h \mathbf{R}_{i_{I}},
            \mathbf{R}_{i_{I}}^{-1} (
                \tilde{\boldsymbol{\tau}}_h
                + \mathbf{R}_h \boldsymbol{\tau}_{i_{I}}
                - \boldsymbol{\tau}_{i_{I}}
            )
        ) \, \mod \mathcal{L}_{\mathcal{H}}.
\end{align*}
Thus, the normality condition of $\mathcal{H}$ in $\mathcal{G}$ is expressed as
\begin{align}
    \label{eq:normality_H_G}
    \tilde{\boldsymbol{\tau}}_h
        + \mathbf{R}_h \boldsymbol{\tau}_{i_{I}}
        - \boldsymbol{\tau}_{i_{I}}
        = \mathbf{R}_{i_{I}} \tilde{\boldsymbol{\tau}}_{\mathbf{R}_{i_{I}}^{-1} \mathbf{R}_h \mathbf{R}_{i_{I}}}
        \, \mod \mathcal{L}_{\mathcal{H}},
\end{align}
where we define $\tilde{\boldsymbol{\tau}}_{\bullet}: \mathcal{P}(\mathcal{M}) \rightarrow \mathbb{R}^3$ as the translation part of $\mathcal{H}$ with $\tilde{\boldsymbol{\tau}}_{\mathbf{R}_h} = \tilde{\boldsymbol{\tau}}_h$.

We consider a group multiplication of cosets of $\mathcal{G} / \mathcal{H}$ by
\begin{widetext}
\begin{align*}
    g_{i_I} g_{i_J} \mathcal{H}
        &= (\mathbf{I}, \boldsymbol{\sigma}(i_I, i_J)) g_{i_I i_J} \mathcal{H}
            \quad [\because \mbox{Eq.~\eqref{eq:coset_multiplication_G}}] \\
        &= (\mathbf{I}, \boldsymbol{\sigma}(i_I, i_J)) g_{i_I i_J} g_{h_{IJ}} g_{h_{IJ}}^{-1} \mathcal{H}
            \quad [\because \mbox{Eq.~\eqref{eq:point_group_factor_system_G_M}}] \\
        &= (\mathbf{I}, \boldsymbol{\sigma}(i_I, i_J) + \boldsymbol{\sigma}( i_I i_J, h_{IJ})) g_{i_{IJ}} g_{h_{IJ}}^{-1} \mathcal{H} \\
        &= (\mathbf{I}, \boldsymbol{\sigma}(i_I, i_J) + \boldsymbol{\sigma}( i_I i_J, h_{IJ})) g_{i_{IJ}} [(\mathbf{I}, -\mathbf{t}_{m_{h_{IJ}}}) \tilde{g}_{h_{IJ}}]^{-1} \mathcal{H} \\
        &= (\mathbf{I}, \boldsymbol{\sigma}(i_I, i_J) + \boldsymbol{\sigma}( i_I i_J, h_{IJ}) + \mathbf{R}_{i_{IJ}} \mathbf{t}_{m_{h_{IJ}}}) g_{i_{IJ}} \mathcal{H}
            \quad [\because \mbox{Eq.~\eqref{eq:normality_H_M}}].
\end{align*}
\end{widetext}
Thus, the group multiplication of $\mathcal{G} / \mathcal{H}$ is expressed as
\begin{align}
    \label{eq:group_multiplication_M_H}
    g_{i_I} g_{i_J} \mathcal{H}
        &= (\mathbf{I}, \mathbf{s}(i_I, i_J)) g_{i_{IJ}} \mathcal{H}
\end{align}
with
\begin{align}
    \mathbf{s}(i_I, i_J)
        &= \boldsymbol{\sigma}(i_I, i_J) + \boldsymbol{\sigma}( i_I i_J, h_{IJ}) + \mathbf{R}_{i_{IJ}} \mathbf{t}_{m_{h_{IJ}}}.
\end{align}
Similarly, the group multiplication of $\mathcal{G} / \mathcal{H}$ is expressed as
\begin{align*}
    &g_{i_I} (\mathbf{I}, \mathbf{t}_m) g_{i_J} (\mathbf{I}, \mathbf{t}_n) \mathcal{H} \\
        &= g_{i_I} g_{i_J} (\mathbf{I}, \mathbf{R}_{i_J}^{-1} \mathbf{t}_m + \mathbf{t}_n) \mathcal{H} \\
        &= (\mathbf{I}, \mathbf{s}(i_I, i_J)) g_{i_{IJ}} (\mathbf{I}, \mathbf{R}_{i_J}^{-1} \mathbf{t}_m + \mathbf{t}_n) \mathcal{H}
            \quad [\because \mbox{Eq.~\eqref{eq:group_multiplication_M_H}}] \\
        &= g_{i_{IJ}} (\mathbf{I}, \mathbf{R}_{i_J}^{-1} \mathbf{t}_m + \mathbf{t}_n + \mathbf{R}_{i_{IJ}}^{-1} \mathbf{s}(i_I, i_J)) \mathcal{H}.
\end{align*}

Finally, we check the associativity condition of $\mathcal{G} / \mathcal{H}$.
For all triplets of coset representatives $g_{i_I}$, $g_{i_J}$, and $g_{i_K}$, we require $(g_{i_I} g_{i_J} \mathcal{H}) g_{i_K} \mathcal{H} = g_{i_I} \mathcal{H} (g_{i_J} g_{i_K} \mathcal{H})$, which leads to the 2-cocycle condition for $\mathbf{s}$:
\begin{align}
    \label{eq:2-cocycle_condition_H}
    &\mathbf{s}(i_{IJ}, i_K) + \mathbf{s}(i_I, i_J) \nonumber \\
        &= \mathbf{s}(i_I, i_{JK}) + \mathbf{R}_{i_I} \mathbf{s}(i_J, i_K)
        \, \mod \mathcal{L}_{\mathcal{H}}.
\end{align}

Thus, we obtain a normal $k$-subgroup $\mathcal{H}$ of $\mathcal{M}$ with $\mathcal{H} \trianglelefteq \mathcal{G}$ by checking Eqs.~\eqref{eq:closure_H}, \eqref{eq:normality_H_M}, \eqref{eq:normality_H_G}, and \eqref{eq:2-cocycle_condition_H}.
In the following, we denote $\mathcal{T}'$ and $\mathcal{L}'$ as $\mathcal{T}_{\mathcal{H}}$ and $\mathcal{L}_{\mathcal{H}}$ for $k$-subgroup $\mathcal{H}$, respectively.

\subsection{\label{appx:ssg_construction_U}Injective homomorphism from quotient group to spin rotation quotient group}

For each normal subgroup $\mathcal{H}$ of $\mathcal{G}$, we consider injective homomorphisms $U: \mathcal{G} / \mathcal{H} \rightarrow \mathcal{N}(\mathcal{B}_{\mathrm{so}}) / \mathcal{B}_{\mathrm{so}}$.
Since the dimension of the natural representation of $\mathcal{B}(\mathcal{X}) / \mathcal{B}_{\mathrm{so}}(\mathcal{X})$ is equal to a spin-structure dimension $d$ and every representation in $\mathbb{R}$ can be transformed to an orthogonal representation, $U$ can be regarded as a $d$-dimensional orthogonal representation of $\mathcal{G} / \mathcal{H}$.
Because two orthogonal representations are equivalent if and only if they are transformed by an orthogonal matrix, the equivalence classes of $U$ as orthogonal representations are identical to those as spin space groups as adopted in Sec.~\ref{sec:ssg_enumeration}.
Consequently, it suffices to enumerate the inequivalent injective homomorphisms $U$ by constructing inequivalent $d$-dimensional faithful orthogonal representations $\tilde{U}: \mathcal{G} / \mathcal{H} \rightarrow O(d)$.

Because orthogonal representations can be uniquely decomposed into irreducible representations in $\mathbb{R}$ [also referred to as physically irreducible representations (PIRs) \cite{PhysRevB.43.11010}], all $\tilde{U}$ can be constructed by combining PIRs of $\mathcal{G} / \mathcal{H}$.
We numerically enumerate all inequivalent PIRs using \textsc{spgrep} \cite{spgrep}.

After constructing $\tilde{U}$, we embed its $d$-dimensional orthogonal matrices into $O(3)$ by direct sum with $(3-d)$-dimensional identity matrices:
\begin{align}
    U(g_{i_I}(\mathbf{I}, \mathbf{t}_m)\mathcal{H})
        = \left(
            \tilde{U}(g_{i_I}(\mathbf{I}, \mathbf{t}_m)\mathcal{H}) \oplus \mathbf{I}_{3-d}
        \right)
        \mathcal{B}_{\mathrm{so}},
\end{align}
where $\mathbf{I}_{3-d}$ is the $(3-d)$-dimensional identity matrix.


\section{\label{appx:projection_operator}Symmetry-adapted magnetic moments from projection operator of spin space group}

We provide details of the projection operator method to obtain symmetry-adapted magnetic moments $\mathbf{m}^{(\alpha)}$ in Eq.~\eqref{eq:symmetry_adapted_magnetic_moments} with spin space group $\mathcal{X}$ in Eq.~\eqref{eq:ssg_construction} and representation matrices $\boldsymbol{\Gamma}$ in Eq.~\eqref{eq:representation_magnetic_moments}.
Because $\mathcal{X}$ is infinite, we construct the projection operator by decomposing it into those of its subgroups, the spin-only group $\mathcal{B}_{\mathrm{so}}$ and the normal space subgroup $\mathcal{H}$.

We first construct the projection operator for $\mathcal{B}_{\mathrm{so}}$:
\begin{align}
    \mathbf{P}^{\mathrm{so}}
        &= \int_{ \mathbf{U} \in \mathcal{B}_{\mathrm{so}} } \mathbf{U} \, d \mu(\mathbf{U}),
\end{align}
where $d \mu$ is the normalized Haar measure.
For simplicity, we choose the $z$ axis as the spin axis for collinear and coplanar spin-only groups.
Then, the projection operator for a collinear spin-only group is given by
\begin{align*}
    \mathbf{P}^{\mathrm{so}}
        &= \frac{1}{2} \left(
                \mathbf{I} + \begin{pmatrix} -1 & 0 & 0 \\ 0 & 1 & 0 \\ 0 & 0 & 1 \end{pmatrix}
            \right)
            \cdot \frac{1}{2 \pi} \int_{0}^{2 \pi} \mathbf{U}^{z}_{\theta} d\theta \\
        &= \begin{pmatrix}
                0 & 0 & 0 \\
                0 & 0 & 0 \\
                0 & 0 & 1
            \end{pmatrix},
\end{align*}
where $\mathbf{U}^{z}_{\theta}$ is a rotation matrix around the $z$ axis by an angle of $\theta$.
For a coplanar spin-only group, the projection operator is given by
\begin{align*}
    \mathbf{P}^{\mathrm{so}}
        &= \frac{1}{2} \left(
                \mathbf{I} + \begin{pmatrix} 1 & 0 & 0 \\ 0 & 1 & 0 \\ 0 & 0 & -1 \end{pmatrix}
            \right) \\
        &= \begin{pmatrix}
                1 & 0 & 0 \\
                0 & 1 & 0 \\
                0 & 0 & 0
            \end{pmatrix}.
\end{align*}
For a noncoplanar spin-only group, the projection operator is simply given by $\mathbf{P}^{\mathrm{so}} = \mathbf{I}$.

Next, we construct the projection operator for $\mathcal{H}$.
We denote representatives of inequivalent unit cells under $\mathcal{T}(\mathcal{H})$ as $\ell = \{ \boldsymbol{\ell}_{1}, \ldots, \boldsymbol{\ell}_{N_k} \}$.
Then, from the periodicity of magnetic moments under $\mathcal{T}(\mathcal{H})$, it suffices to consider unit cells only in $\ell$.
Thus, we consider only coset representatives of $\mathcal{H} / \mathcal{T}(\mathcal{H})$ in constructing the projection operator:
\begin{align}
    P^{\mathcal{H}}_{ \boldsymbol{\ell} \kappa, \boldsymbol{\ell}' \kappa' }
        &= \frac{1}{|\mathcal{P}(\mathcal{H})|} \sum_{\mathbf{R}_h \in \mathcal{P}(\mathcal{H}) } \Gamma^{\mathrm{site}}_{ \boldsymbol{\ell} \kappa, \boldsymbol{\ell}' \kappa' } (g_h).
\end{align}

Finally, we combine the projection operators of $\mathcal{B}_{\mathrm{so}}$ and $\mathcal{H}$ to construct that of $\mathcal{X}$:
\begin{widetext}
\begin{align}
    P^{\mathcal{X}}_{ \boldsymbol{\ell} \kappa \mu, \boldsymbol{\ell}' \kappa' \mu' }
        &= \frac{1}{N_t N_k} \sum_{I=1}^{N_t} \sum_{q=1}^{N_k} \sum_{\kappa''=1}^{m} \sum_{\mu''=1}^{3} \sum_{\boldsymbol{\ell}''\in\ell}
            \Gamma_{\boldsymbol{\ell} \kappa \mu, \boldsymbol{\ell}'' \kappa'' \mu''}( (g_{i_I} (\mathbf{I}, \mathbf{t}_q), \mathbf{U}_{I q}) )
            P^{\mathcal{H}}_{ \boldsymbol{\ell}'' \kappa'', \boldsymbol{\ell}' \kappa' }
            P^{\mathrm{so}}_{ \mu'' \mu' }.
\end{align}
\end{widetext}
Eigenvectors of $\mathbf{P}^{\mathcal{X}}$ with eigenvalue 1 give symmetry-adapted magnetic moments $\mathbf{m}^{(\alpha)}$ in Eq.~\eqref{eq:symmetry_adapted_magnetic_moments}.

\section{\label{appx:orthogonal_intertwiner}Orthogonal intertwiner between two orthogonal representations}

Let $\Gamma_1$ and $\Gamma_2$ be orthogonal $k$-dimensional representations of a finite group $G$.
An intertwiner of $\Gamma_1$ and $\Gamma_2$ is a nonzero real matrix $\mathbf{Q}' \in \mathbb{R}^{k \times k}$ satisfying
\begin{align}
    \boldsymbol{\Gamma}_1 (g) \mathbf{Q}' = \mathbf{Q}' \boldsymbol{\Gamma}_2 (g) \quad (\forall g \in G).
\end{align}
An invertible intertwiner exists if and only if $\Gamma_1$ and $\Gamma_2$ are equivalent.

For any real matrix $\mathbf{B} \in \mathbb{R}^{k \times k}$, the following matrix is an intertwiner for these representations:
\begin{align}
    \mathbf{Q}'
        &\coloneqq \sum_{g \in G} \boldsymbol{\Gamma}_1 (g) \mathbf{B} \boldsymbol{\Gamma}_2 (g)^{-1},
\end{align}
if $\mathbf{Q}'$ is nonzero.
Thus, we can construct an intertwiner by generating a random matrix $\mathbf{B}$ and calculating $\mathbf{Q}'$.

If an intertwiner $\mathbf{Q}'$ exists and is well conditioned, it can be converted to an orthogonal matrix by
\begin{align}
    \mathbf{Q} &= \mathbf{Q}' ({\mathbf{Q}'}^{\top} \mathbf{Q}')^{-1/2},
\end{align}
where $({\mathbf{Q}'}^{\top} \mathbf{Q}')^{-1/2}$ is the inverse square root of the positive definite matrix ${\mathbf{Q}'}^{\top} \mathbf{Q}'$.

\section{\label{appx:spin_planochirality}Spin-planochiral coplanar spin space group}

For simplicity, we consider coplanar SSGs $\mathcal{X}$ with $\mathcal{B}_{\mathrm{so}}(\mathcal{X}) = \{1, \sigma_z\}$ without loss of generality.
We say that $\mathcal{X}$ is spin planochiral if $\mathcal{X}$ and $\mathcal{X}_{\mathrm{e}} = (1, \boldsymbol{\sigma}_{x})^{-1} \mathcal{X} (1, \boldsymbol{\sigma}_{x})$ are not equivalent by any spin transformation $\mathbf{Q} \in SO(2) \oplus 1$, where $\boldsymbol{\sigma}_{x}$ is a mirror operation perpendicular to the $x$ axis and is chosen as one of improper spin rotations out of the coplanar plane.
We denote the triplets of $\mathcal{X}$ and $\mathcal{X}_{\mathrm{e}}$, defined in Sec.~\ref{sec:ssg_enumeration} as $(\mathcal{H}, \mathcal{B}_{\mathrm{so}}, U)$ and $(\mathcal{H}, \mathcal{B}_{\mathrm{so}}, U_{\mathrm{e}})$, respectively.
We write the corresponding two-dimensional faithful orthogonal representations for $U$ and $U_{\mathrm{e}}$, defined in Appendix~\ref{appx:ssg_construction_U}, as $\tilde{U}$ and $\tilde{U}_{\mathrm{e}}$, respectively.
Then, $\mathcal{X}$ is spin planochiral if there is no orientation-preserving intertwiner $\tilde{\mathbf{Q}} \in SO(2)$ between $\tilde{U}$ and $\tilde{U}_{\mathrm{e}}$.

This condition requires both (i) $\tilde{U}$ and $\tilde{U}_{\mathrm{e}}$ are $O(2)$-equivalent and (ii) a centralizer of $\tilde{U}$ has no improper rotation.
The first condition can be checked by the procedure in Appendix~\ref{appx:orthogonal_intertwiner}.
The second condition is satisfied if and only if $\tilde{U}$ is a two-dimensional PIR as follows.
Otherwise, $\tilde{U}$ is constructed from two one-dimensional PIRs, $u_1$ and $u_2$, as $\tilde{U} = u_1 \oplus u_2$ after some transformation.
We can construct improper $\tilde{\mathbf{Q}} = \mathrm{diag}(\mathop{-}1, 1)$ as a centralizer of $\tilde{U}$ by flipping the axis of $u_1$.
Thus, $\tilde{U}$ must be a two-dimensional PIR if $\mathcal{X}$ is spin planochiral.


\section{Complete list of classifications}
\label{appx:full_list}
We present the complete list of classifications in Table~\ref{tab:full_list}. 
Details of the classification scheme and remarks are provided in Sec.~\ref{sec:method_classification}. 
The entries used in the SDFT-based analysis in Sec.~\ref{sec:sdft} are also indicated.

\begingroup
\setlength{\LTcapwidth}{0.94\textwidth}
\begin{longtable*}{ll|ll|ll}
\caption{Complete list of classifications in Sec.~\ref{sec:results_classification}. The column ``no.'' indicates the entries in the \textsc{MAGNDATA} database. 
In the ``class'' column, \texttt{Lt}, \texttt{Lp}, \texttt{M}, \texttt{S}, and \texttt{oS} correspond to the 
\texttt{lower\_family\_translation\_group}, \texttt{lower\_family\_point\_group}, 
\texttt{multiple\_basis}, \texttt{ssg\_matched}, and \texttt{oriented\_ssg\_matched} classes, respectively. 
Entries marked with an asterisk are used in the SDFT-based analysis shown in Sec.~\ref{sec:sdft}.}
\label{tab:full_list}\\
        \toprule
        No. & Class\hspace{0.3cm} & No. & Class \hspace{0.3cm} & No. & Class \\
        \midrule        
        \endfirsthead
        
        \multicolumn{6}{c}{\tablename\ \thetable\ (\textit{cont.})} \\\\
        No. & Class & No. & Class & No. & Class \\
        \midrule
        \endhead

        \endfoot
        \endlastfoot

        \texttt{0.1\_LaMnO3} & \texttt{oS}* & \texttt{0.898\_Mn3IrSi} & \texttt{M} & \texttt{1.416\_Tb2O2S} & \texttt{Lp} \\
        \texttt{0.2\_Cd2Os2O7} & \texttt{oS}* & \texttt{0.899\_Mn3IrGe} & \texttt{M} & \texttt{1.417\_Tb2O2Se} & \texttt{S} \\
        \texttt{0.3\_Ca3LiOsO6} & \texttt{oS}* & \texttt{0.900\_Mn3CoGe} & \texttt{M} & \texttt{1.418\_Cu4O3} & \texttt{Lp} \\
        \texttt{0.6\_YMnO3} & \texttt{oS} & \texttt{0.903\_Pr2PdGe6} & \texttt{M} & \texttt{1.419\_GdIn3} & \texttt{Lp} \\
        \texttt{0.7\_ScMnO3} & \texttt{oS} & \texttt{0.904\_Nd2PdGe6} & \texttt{oS} & \texttt{1.420\_YBa2Cu3O6} & \texttt{oS}* \\
        \texttt{0.8\_ScMnO3} & \texttt{S} & \texttt{0.905\_Tb2PdGe6} & \texttt{M} & \texttt{1.421\_NdRh2Si2} & \texttt{oS} \\
        \texttt{0.9\_GdB4} & \texttt{oS} & \texttt{0.906\_Dy2PdGe6} & \texttt{M} & \texttt{1.422\_ErRh2Si2} & \texttt{oS} \\
        \texttt{0.12\_U3Ru4Al12} & \texttt{M} & \texttt{0.907\_Ho2PdGe6} & \texttt{M} & \texttt{1.423\_UPb3} & \texttt{Lp} \\
        \texttt{0.15\_MnF2} & \texttt{oS}* & \texttt{0.908\_Tb2PtGe6} & \texttt{M} & \texttt{1.424\_UCu5} & \texttt{Lp} \\
        \texttt{0.16\_EuTiO3} & \texttt{oS} & \texttt{0.909\_Er2PtGe6} & \texttt{oS} & \texttt{1.425\_UGeTe} & \texttt{oS} \\
        \texttt{0.17\_FePO4} & \texttt{M} & \texttt{0.910\_TbNiSi2} & \texttt{oS} & \texttt{1.426\_UGeS} & \texttt{oS} \\
        \texttt{0.18\_BaMn2As2} & \texttt{oS}* & \texttt{0.916\_Cd2Os2O7} & \texttt{oS} & \texttt{1.427\_HoCo2Ge2} & \texttt{oS} \\
        \texttt{0.19\_MnTiO3} & \texttt{oS}* & \texttt{0.917\_Sr2ScOsO6} & \texttt{S}* & \texttt{1.428\_UN} & \texttt{Lp} \\
        \texttt{0.20\_MnTe2} & \texttt{oS}* & \texttt{0.918\_Ag2RuO4} & \texttt{M} & \texttt{1.432\_Ba2LuRuO6} & \texttt{Lp} \\
        \texttt{0.21\_PbNiO3} & \texttt{oS}* & \texttt{0.919\_EuMnBi2} & \texttt{oS} & \texttt{1.433\_Ba2YRuO6} & \texttt{Lp} \\
        \texttt{0.22\_DyB4} & \texttt{Lp} & \texttt{0.920\_ThMnPN} & \texttt{oS}* & \texttt{1.438\_BaCoF4} & \texttt{Lp} \\
        \texttt{0.23\_Ca3Mn2O7} & \texttt{oS}* & \texttt{0.921\_ThMnPN} & \texttt{oS} & \texttt{1.439\_BaCoF4} & \texttt{Lt} \\
        \texttt{0.24\_LiMnPO4} & \texttt{oS}* & \texttt{0.922\_ThMnAsN} & \texttt{oS}* & \texttt{1.441\_NaFe3(SO4)2(OH)6} & \texttt{oS} \\
        \texttt{0.25\_NaOsO3} & \texttt{oS}* & \texttt{0.923\_ThMnAsN} & \texttt{oS} & \texttt{1.442\_URu2Si2} & \texttt{oS} \\
        \texttt{0.28\_LiFeSi2O6} & \texttt{M} & \texttt{0.924\_RbRuO4} & \texttt{M} & \texttt{1.445\_Y2BaCuO5} & \texttt{Lp} \\
        \texttt{0.29\_Er2Ti2O7} & \texttt{M} & \texttt{0.926\_Pr2PdGe6} & \texttt{M} & \texttt{1.446\_CeCoAl4} & \texttt{oS} \\
        \texttt{0.30\_YbMnO3} & \texttt{oS} & \texttt{0.927\_Nd2PdGe6} & \texttt{oS} & \texttt{1.448\_HoSi} & \texttt{Lt} \\
        \texttt{0.31\_HoMnO3} & \texttt{oS} & \texttt{0.928\_Dy2PdGe6} & \texttt{M} & \texttt{1.449\_Li2CuW2O8} & \texttt{S}* \\
        \texttt{0.32\_HoMnO3} & \texttt{oS} & \texttt{0.929\_Tb2PdGe6} & \texttt{M} & \texttt{1.452\_FeSn} & \texttt{oS}* \\
        \texttt{0.44\_YMnO3} & \texttt{S} & \texttt{0.930\_Ho2PdGe6} & \texttt{M} & \texttt{1.453\_EuMn2Si2} & \texttt{oS}* \\
        \texttt{0.45\_La2NiO4} & \texttt{Lp} & \texttt{0.931\_Tb2PtGe6} & \texttt{M} & \texttt{1.454\_Mn6Ni16Si7} & \texttt{Lp} \\
        \texttt{0.47\_Gd2Sn2O7} & \texttt{Lp} & \texttt{0.932\_Er2PtGe6} & \texttt{oS} & \texttt{1.455\_Mn6Ni16Si7} & \texttt{M} \\
        \texttt{0.50\_MnTiO3} & \texttt{oS} & \texttt{0.934\_Sr2NiTeO6} & \texttt{S}* & \texttt{1.456\_Sr2CuO2Cu2S2} & \texttt{oS} \\
        \texttt{0.56\_Ba2CoGe2O7} & \texttt{oS}* & \texttt{0.936\_Sr2MnTeO6} & \texttt{S}* & \texttt{1.457\_NdNiMg15} & \texttt{Lp} \\
        \texttt{0.58\_CoAl2O4} & \texttt{oS}* & \texttt{0.937\_Sr2CoTeO6} & \texttt{S} & \texttt{1.458\_CsCo2Se2} & \texttt{oS}* \\
        \texttt{0.59\_Cr2O3} & \texttt{oS}* & \texttt{0.942\_Er2Ge2O7} & \texttt{M} & \texttt{1.460\_PrCuSi} & \texttt{Lp} \\
        \texttt{0.62\_SrMn2V2O8} & \texttt{oS} & \texttt{0.945\_Yb2Ir2O7} & \texttt{oS}* & \texttt{1.461\_Sr2Cr3As2O2} & \texttt{oS}* \\
        \texttt{0.65\_Fe2O3-alpha} & \texttt{oS}* & \texttt{0.947\_YCrO3} & \texttt{oS} & \texttt{1.462\_La2CoPtO6} & \texttt{S} \\
        \texttt{0.66\_Fe2O3-alpha} & \texttt{S} & \texttt{0.950\_LaErO3} & \texttt{M} & \texttt{1.463\_Sr2Fe3Se2O3} & \texttt{M} \\
        \texttt{0.70\_Na3Co(CO3)2Cl} & \texttt{oS} & \texttt{0.955\_Na2Mn(H2C3O4)2(H2O)2} & \texttt{oS} & \texttt{1.464\_U2N2P} & \texttt{oS} \\
        \texttt{0.71\_Li2Ni(SO4)2} & \texttt{oS} & \texttt{0.959\_Cr2TeO6} & \texttt{oS} & \texttt{1.465\_U2N2As} & \texttt{oS} \\
        \texttt{0.72\_CaMnBi2} & \texttt{oS}* & \texttt{0.960\_Fe2TeO6} & \texttt{oS} & \texttt{1.468\_TbMn2Si2} & \texttt{oS}* \\
        \texttt{0.73\_SrMnBi2} & \texttt{oS}* & \texttt{0.961\_LiCrGe2O6} & \texttt{S} & \texttt{1.469\_YMn2Si2} & \texttt{oS}* \\
        \texttt{0.74\_Mn3Cu0.5Ge0.5N} & \texttt{oS}* & \texttt{0.962\_LiCrGe2O6} & \texttt{S} & \texttt{1.470\_UCr2Si2} & \texttt{S} \\
        \texttt{0.75\_Cr2WO6} & \texttt{oS}* & \texttt{0.963\_LiCrGe2O6} & \texttt{S} & \texttt{1.471\_EuCd2As2} & \texttt{oS} \\
        \texttt{0.76\_Cr2TeO6} & \texttt{oS}* & \texttt{0.964\_LiCrGe2O6} & \texttt{S} & \texttt{1.472\_CaOFeS} & \texttt{Lp} \\
        \texttt{0.79\_CaIrO3} & \texttt{oS}* & \texttt{0.966\_V2WO6} & \texttt{oS}* & \texttt{1.473\_CuBr(C4H4N2)2(BF4)} & \texttt{oS} \\
        \texttt{0.80\_U2Pd2In} & \texttt{oS} & \texttt{0.967\_BaMn2V2O8} & \texttt{oS} & \texttt{1.474\_CuCl(C4H4N2)2(BF4)} & \texttt{oS} \\
        \texttt{0.81\_U2Pd2Sn} & \texttt{oS} & \texttt{0.979\_TmVO3} & \texttt{oS}* & \texttt{1.475\_DyNiAl4} & \texttt{oS} \\
        \texttt{0.83\_LiFeP2O7} & \texttt{S}* & \texttt{0.980\_TmVO3} & \texttt{oS} & \texttt{1.476\_Ba2CoO4} & \texttt{M} \\
        \texttt{0.87\_NaFePO4} & \texttt{oS}* & \texttt{0.984\_LuVO3} & \texttt{oS}* & \texttt{1.477\_Ba2CoO4} & \texttt{M} \\
        \texttt{0.88\_LiNiPO4} & \texttt{M} & \texttt{0.987\_BaFe2S2O} & \texttt{M} & \texttt{1.478\_CoTi2O5} & \texttt{Lp} \\
        \texttt{0.89\_BaMn2Bi2} & \texttt{oS}* & \texttt{0.988\_BaFe2Se2O} & \texttt{M} & \texttt{1.479\_U2Ni2Sn} & \texttt{oS} \\
        \texttt{0.92\_CaMn2Sb2} & \texttt{oS}* & \texttt{0.991\_HoFeO3} & \texttt{oS}* & \texttt{1.482\_Er2Fe2Si2C} & \texttt{Lp} \\
        \texttt{0.95\_LiFePO4} & \texttt{oS}* & \texttt{0.1004\_CsO2} & \texttt{M} & \texttt{1.484\_Li2MnGeO4} & \texttt{M} \\
        \texttt{0.96\_CoSO4} & \texttt{M} & \texttt{0.1005\_Mn3RhGe} & \texttt{M} & \texttt{1.486\_CeRhAl4Si2} & \texttt{oS} \\
        \texttt{0.97\_FeSb2O4} & \texttt{M} & \texttt{0.1006\_Mn3IrGe} & \texttt{M} & \texttt{1.487\_CeIrAl4Si2} & \texttt{oS} \\
        \texttt{0.104\_ErVO3} & \texttt{M} & \texttt{0.1010\_C10H6MnN4O4} & \texttt{M} & \texttt{1.488\_CeMn2Si2} & \texttt{oS}* \\
        \texttt{0.107\_Ho2Ge2O7} & \texttt{M} & \texttt{0.1015\_CoTe6O13} & \texttt{oS} & \texttt{1.489\_CeMn2Si2} & \texttt{oS} \\
        \texttt{0.108\_Mn3Ir} & \texttt{oS}* & \texttt{0.1017\_CePdAl3} & \texttt{oS} & \texttt{1.490\_CeMn2Si2} & \texttt{oS} \\
        \texttt{0.109\_Mn3Pt} & \texttt{oS}* & \texttt{0.1018\_SrMnO3} & \texttt{oS}* & \texttt{1.491\_PrMn2Si2} & \texttt{oS}* \\
        \texttt{0.110\_Cr2O3} & \texttt{oS} & \texttt{0.1019\_SrMnO3} & \texttt{oS} & \texttt{1.492\_PrMn2Si2} & \texttt{oS} \\
        \texttt{0.112\_FeBO3} & \texttt{oS}* & \texttt{0.1020\_EuCu2Sb2} & \texttt{oS} & \texttt{1.493\_NdMn2Si2} & \texttt{oS}* \\
        \texttt{0.113\_NiCO3} & \texttt{S}* & \texttt{0.1021\_TmFeO3} & \texttt{oS}* & \texttt{1.494\_NdMn2Si2} & \texttt{oS} \\
        \texttt{0.114\_CoCO3} & \texttt{S}* & \texttt{0.1022\_TmFeO3} & \texttt{oS} & \texttt{1.495\_YMn2Si2} & \texttt{oS} \\
        \texttt{0.115\_MnCO3} & \texttt{S}* & \texttt{0.1023\_Cr2MoO6} & \texttt{oS}* & \texttt{1.496\_YMn2Ge2} & \texttt{oS}* \\
        \texttt{0.116\_FeCO3} & \texttt{oS}* & \texttt{0.1036\_CoTa4Se8} & \texttt{oS}* & \texttt{1.497\_EuMg2Bi2} & \texttt{oS} \\
        \texttt{0.117\_LuFeO3} & \texttt{oS} & \texttt{0.1059\_PrBaMn2O6} & \texttt{oS}* & \texttt{1.498\_Cu6(SiO3)6(H2O)6} & \texttt{M} \\
        \texttt{0.119\_CoSe2O5} & \texttt{M} & \texttt{0.1064\_RuP3SiO11} & \texttt{oS} & \texttt{1.499\_CsFe(MoO4)2} & \texttt{S} \\
        \texttt{0.122\_Li2Mn(SO4)2} & \texttt{M} & \texttt{0.1065\_MnGeN2} & \texttt{oS}* & \texttt{1.504\_GdCuSn} & \texttt{Lp} \\
        \texttt{0.125\_MnGeO3} & \texttt{oS}* & \texttt{0.1066\_MnSiN2} & \texttt{oS}* & \texttt{1.505\_GdAgSn} & \texttt{Lp} \\
        \texttt{0.126\_NpCo2} & \texttt{oS} & \texttt{0.1076\_Li2Ni2W2O9} & \texttt{M} & \texttt{1.506\_GdAuSn} & \texttt{Lp} \\
        \texttt{0.127\_Dy3Al5O12} & \texttt{oS} & \texttt{0.1078\_[Na(OH2)3]Mn(NCS)3} & \texttt{oS} & \texttt{1.507\_NdPd5Al2} & \texttt{Lt} \\
        \texttt{0.128\_FeSO4F} & \texttt{oS}* & \texttt{0.1081\_Mn2Au} & \texttt{oS} & \texttt{1.508\_Mn2AlB2} & \texttt{oS}* \\
        \texttt{0.131\_Mn(N(CN2))2} & \texttt{oS}* & \texttt{0.1082\_LuCrWO6} & \texttt{oS} & \texttt{1.510\_TbNi2Ge2} & \texttt{Lt} \\
        \texttt{0.137\_Cu2V2O7} & \texttt{oS}* & \texttt{0.1083\_RuP3SiO11} & \texttt{oS} & \texttt{1.511\_TbNi2Si2} & \texttt{Lp} \\
        \texttt{0.142\_Fe2TeO6} & \texttt{oS}* & \texttt{0.1085\_PrFeO3} & \texttt{oS}* & \texttt{1.512\_TbCo2Si2} & \texttt{oS} \\
        \texttt{0.143\_Cr2TeO6} & \texttt{oS} & \texttt{0.1090\_NH4CrF3} & \texttt{oS} & \texttt{1.513\_HoCo2Si2} & \texttt{oS} \\
        \texttt{0.144\_Cr2WO6} & \texttt{oS} & \texttt{0.1091\_La2O3Mn2Se2} & \texttt{oS}* & \texttt{1.514\_HoCo2Si2} & \texttt{oS} \\
        \texttt{0.146\_EuZrO3} & \texttt{oS} & \texttt{0.1092\_La2O3Mn2Se2} & \texttt{oS} & \texttt{1.515\_ErCo2Si2} & \texttt{S} \\
        \texttt{0.147\_EuZrO3} & \texttt{oS} & \texttt{0.1093\_La2O3Mn2Se2} & \texttt{oS} & \texttt{1.516\_ErCo2Si2} & \texttt{oS} \\
        \texttt{0.148\_La2LiRuO6} & \texttt{S}* & \texttt{0.1103\_EuSc2Te4} & \texttt{oS} & \texttt{1.519\_CoSO4} & \texttt{M} \\
        \texttt{0.150\_NiS2} & \texttt{oS}* & \texttt{0.1104\_GdCrO3} & \texttt{oS}* & \texttt{1.520\_NiSO4} & \texttt{oS}* \\
        \texttt{0.152\_LiFePO4} & \texttt{M} & \texttt{0.1111\_MnSb2O4} & \texttt{oS}* & \texttt{1.521\_FeSO4} & \texttt{oS}* \\
        \texttt{0.155\_CaMnGe2O6} & \texttt{S} & \texttt{0.1112\_CoNb4Se8} & \texttt{oS}* & \texttt{1.522\_CrVO4} & \texttt{S}* \\
        \texttt{0.156\_CaMnGe2O6} & \texttt{S}* & \texttt{1.0.1\_Ag2CrO2} & \texttt{Lt} & \texttt{1.523\_VPO4} & \texttt{Lt} \\
        \texttt{0.159\_DyCoO3} & \texttt{M} & \texttt{1.0.8\_Ba3MnNb2O9} & \texttt{oS} & \texttt{1.524\_InMnO3} & \texttt{M} \\
        \texttt{0.160\_TbCoO3} & \texttt{M} & \texttt{1.0.9\_CsCoCl3} & \texttt{Lt} & \texttt{1.525\_InMnO3} & \texttt{M} \\
        \texttt{0.161\_CoSe2O5} & \texttt{oS} & \texttt{1.0.14\_CsFeCl3} & \texttt{oS} & \texttt{1.526\_LiCoF4} & \texttt{S}* \\
        \texttt{0.163\_MnPS3} & \texttt{S}* & \texttt{1.0.26\_RbCoBr3} & \texttt{Lt} & \texttt{1.527\_CsNiF3} & \texttt{Lp} \\
        \texttt{0.171\_DyScO3} & \texttt{M} & \texttt{1.0.27\_Li2MnTeO6} & \texttt{Lt} & \texttt{1.530\_CeC2} & \texttt{oS} \\
        \texttt{0.177\_Mn3GaN} & \texttt{oS}* & \texttt{1.0.32\_EuIn2As2} & \texttt{oS} & \texttt{1.531\_PrC2} & \texttt{oS} \\
        \texttt{0.178\_CoF2} & \texttt{oS}* & \texttt{1.0.33\_FeF3} & \texttt{M} & \texttt{1.532\_NdC2} & \texttt{oS} \\
        \texttt{0.180\_MnPSe3} & \texttt{S}* & \texttt{1.0.36\_CsMnI3} & \texttt{Lt} & \texttt{1.533\_TbC2} & \texttt{Lt} \\
        \texttt{0.186\_CeMnAsO} & \texttt{oS}* & \texttt{1.0.37\_CsMnI3} & \texttt{Lt} & \texttt{1.534\_HoC2} & \texttt{Lt} \\
        \texttt{0.189\_CeMn2Ge4O12} & \texttt{oS} & \texttt{1.0.38\_CsCoCl3} & \texttt{Lt} & \texttt{1.535\_UPd2Ge2} & \texttt{Lt} \\
        \texttt{0.193\_LiCoPO4} & \texttt{oS}* & \texttt{1.0.42\_CsNiCl3} & \texttt{oS} & \texttt{1.536\_UPd2Si2} & \texttt{oS} \\
        \texttt{0.194\_UPt2Si2} & \texttt{oS} & \texttt{1.0.45\_Ba3CoSb2O9} & \texttt{oS} & \texttt{1.537\_URh2Si2} & \texttt{oS} \\
        \texttt{0.198\_GdVO4} & \texttt{oS} & \texttt{1.0.58\_Li2MnTeO6} & \texttt{oS} & \texttt{1.538\_Ba2MnTeO6} & \texttt{S}* \\
        \texttt{0.199\_Mn3Sn} & \texttt{oS}* & \texttt{1.1\_Mn3O4} & \texttt{oS} & \texttt{1.539\_KMnP} & \texttt{oS}* \\
        \texttt{0.200\_Mn3Sn} & \texttt{oS} & \texttt{1.2\_CuSe2O5} & \texttt{M} & \texttt{1.540\_KMnP} & \texttt{oS} \\
        \texttt{0.207\_TlFe1.6Se2} & \texttt{M} & \texttt{1.3\_Sr2IrO4} & \texttt{Lp} & \texttt{1.541\_RbMnP} & \texttt{oS}* \\
        \texttt{0.208\_TlFe1.6Se2} & \texttt{S} & \texttt{1.4\_YBa2Cu3O6+d} & \texttt{oS} & \texttt{1.542\_RbMnP} & \texttt{oS} \\
        \texttt{0.209\_TlFe1.6Se2} & \texttt{oS}* & \texttt{1.5\_YBa2Cu3O6+d} & \texttt{oS}* & \texttt{1.543\_RbMnAs} & \texttt{oS}* \\
        \texttt{0.211\_Ca2MnO4} & \texttt{oS}* & \texttt{1.6\_NiO} & \texttt{Lp} & \texttt{1.544\_RbMnAs} & \texttt{oS} \\
        \texttt{0.212\_Sr2Mn3As2O2} & \texttt{oS}* & \texttt{1.8\_CeRu2Al10} & \texttt{oS} & \texttt{1.545\_RbMnBi} & \texttt{oS}* \\
        \texttt{0.215\_BaNi2P2O8} & \texttt{S}* & \texttt{1.9\_Li2VOSiO4} & \texttt{Lp} & \texttt{1.546\_CsMnBi} & \texttt{oS}* \\
        \texttt{0.216\_SrEr2O4} & \texttt{oS} & \texttt{1.13\_Ba3Nb2NiO9} & \texttt{oS} & \texttt{1.547\_CsMnP} & \texttt{oS}* \\
        \texttt{0.217\_LiCrGe2O6} & \texttt{S} & \texttt{1.16\_BaFe2As2} & \texttt{oS}* & \texttt{1.548\_CsMnP} & \texttt{oS} \\
        \texttt{0.222\_CuMnAs} & \texttt{oS}* & \texttt{1.17\_CoV2O6-alpha} & \texttt{S}* & \texttt{1.549\_U2Ni2In} & \texttt{oS} \\
        \texttt{0.229\_Ba2MnSi2O7} & \texttt{oS}* & \texttt{1.18\_MnS2} & \texttt{oS}* & \texttt{1.550\_LiMnAs} & \texttt{oS}* \\
        \texttt{0.230\_K2CoP2O7} & \texttt{oS} & \texttt{1.20\_HoMnO3} & \texttt{Lp} & \texttt{1.551\_LiMnAs} & \texttt{oS} \\
        \texttt{0.236\_CaFe4Al8} & \texttt{M} & \texttt{1.21\_DyCo2Si2} & \texttt{oS} & \texttt{1.552\_LiMnAs} & \texttt{oS} \\
        \texttt{0.237\_Er2Sn2O7} & \texttt{Lp} & \texttt{1.22\_DyCu2Si2} & \texttt{Lp} & \texttt{1.553\_KMnAs} & \texttt{oS}* \\
        \texttt{0.238\_Er2Pt2O7} & \texttt{Lp} & \texttt{1.23\_La2CuO4} & \texttt{oS}* & \texttt{1.554\_KMnAs} & \texttt{oS} \\
        \texttt{0.239\_Ca3LiRuO6} & \texttt{oS}* & \texttt{1.24\_ZnV2O4} & \texttt{Lp} & \texttt{1.556\_FeSn2} & \texttt{oS}* \\
        \texttt{0.243\_Li2Fe(SO4)2} & \texttt{S} & \texttt{1.25\_KFe3(OH)6(SO4)2} & \texttt{oS} & \texttt{1.557\_FeGe2} & \texttt{oS}* \\
        \texttt{0.244\_Li2Co(SO4)2} & \texttt{oS} & \texttt{1.26\_CsFe2Se3} & \texttt{Lp} & \texttt{1.558\_MnSn2} & \texttt{Lp} \\
        \texttt{0.246\_LiFe(SO4)2} & \texttt{oS} & \texttt{1.28\_CrN} & \texttt{Lt} & \texttt{1.559\_MnSn2} & \texttt{Lt} \\
        \texttt{0.252\_Cs2FeCl5.D2O} & \texttt{oS} & \texttt{1.30\_BaCo2V2O8} & \texttt{Lp} & \texttt{1.560\_GeNi2O4} & \texttt{Lp} \\
        \texttt{0.254\_[C(ND2)3]Cu(DCOO)3} & \texttt{oS} & \texttt{1.31\_MnO} & \texttt{Lp} & \texttt{1.561\_GeNi2O4} & \texttt{M} \\
        \texttt{0.255\_[C(ND2)3]Cu(DCOO)3} & \texttt{oS} & \texttt{1.33\_ErAuGe} & \texttt{Lp} & \texttt{1.562\_GeNi2O4} & \texttt{M} \\
        \texttt{0.264\_Fe3(PO4)2} & \texttt{S}* & \texttt{1.34\_HoAuGe} & \texttt{M} & \texttt{1.563\_GeNi2O4} & \texttt{M} \\
        \texttt{0.267\_YbMnBi2} & \texttt{oS}* & \texttt{1.35\_LiErF4} & \texttt{Lp} & \texttt{1.564\_GeCo2O4} & \texttt{M} \\
        \texttt{0.273\_Mn3ZnN} & \texttt{oS}* & \texttt{1.37\_VOCl} & \texttt{S}* & \texttt{1.568\_GdCu2Si2} & \texttt{Lp} \\
        \texttt{0.277\_MgMnO3} & \texttt{oS}* & \texttt{1.39\_LiFeGe2O6} & \texttt{M} & \texttt{1.569\_SrRu2O6} & \texttt{oS} \\
        \texttt{0.279\_Mn3As} & \texttt{oS}* & \texttt{1.42\_La2NiO4} & \texttt{oS}* & \texttt{1.570\_La3OsO7} & \texttt{Lp} \\
        \texttt{0.280\_Mn3As} & \texttt{oS} & \texttt{1.43\_PrNiO3} & \texttt{Lp} & \texttt{1.571\_La3OsO7} & \texttt{Lp} \\
        \texttt{0.284\_KOsO4} & \texttt{oS}* & \texttt{1.45\_NdNiO3} & \texttt{Lp} & \texttt{1.573\_FeSO4} & \texttt{M} \\
        \texttt{0.285\_KRuO4} & \texttt{oS}* & \texttt{1.49\_Ag2NiO2} & \texttt{Lt} & \texttt{1.574\_NdBiPt} & \texttt{Lp} \\
        \texttt{0.290\_CeCu2} & \texttt{oS} & \texttt{1.50\_AgNiO2} & \texttt{Lp} & \texttt{1.575\_ErRh} & \texttt{Lp} \\
        \texttt{0.292\_NiTe2O5} & \texttt{M} & \texttt{1.52\_CaFe2As2} & \texttt{oS}* & \texttt{1.576\_Yb2O2S} & \texttt{oS} \\
        \texttt{0.301\_Sr2CoTeO6} & \texttt{S}* & \texttt{1.56\_Gd2Ti2O7} & \texttt{oS} & \texttt{1.578\_KErSe2} & \texttt{Lp} \\
        \texttt{0.303\_BaCrF5} & \texttt{oS}* & \texttt{1.57\_CuMnO2} & \texttt{S}* & \texttt{1.580\_NiTiO3} & \texttt{S}* \\
        \texttt{0.307\_ScCrO3} & \texttt{oS}* & \texttt{1.58\_La2O2Fe2OSe2} & \texttt{M} & \texttt{1.585\_PrFeAsO} & \texttt{oS}* \\
        \texttt{0.308\_InCrO3} & \texttt{oS}* & \texttt{1.59\_KTb3F12} & \texttt{oS} & \texttt{1.588\_NdFeAsO} & \texttt{oS}* \\
        \texttt{0.309\_TlCrO3} & \texttt{oS}* & \texttt{1.60\_Ca3Co2O6} & \texttt{Lp} & \texttt{1.593\_BaCoSO} & \texttt{Lt} \\
        \texttt{0.315\_ZrMn2Ge4O12} & \texttt{oS} & \texttt{1.61\_MnWO4} & \texttt{Lt} & \texttt{1.594\_BaCoSO} & \texttt{Lt} \\
        \texttt{0.320\_U2Pd2In} & \texttt{oS} & \texttt{1.62\_CuO} & \texttt{Lt} & \texttt{1.595\_CaCoSO} & \texttt{Lt} \\
        \texttt{0.321\_U2Pd2Sn} & \texttt{oS} & \texttt{1.63\_MnPb4Sb6S14} & \texttt{oS} & \texttt{1.596\_TbCuSb2} & \texttt{Lp} \\
        \texttt{0.323\_LaCrO3} & \texttt{oS} & \texttt{1.64\_BaNiF4} & \texttt{Lp} & \texttt{1.597\_TbCuSb2} & \texttt{Lt} \\
        \texttt{0.324\_CdYb2S4} & \texttt{Lp} & \texttt{1.65\_SrFeO2} & \texttt{oS}* & \texttt{1.617\_LiFe(MoO4)2} & \texttt{S} \\
        \texttt{0.325\_CdYb2Se4} & \texttt{Lp} & \texttt{1.66\_Fe(ND3)2PO4} & \texttt{M} & \texttt{1.618\_CoO} & \texttt{Lp} \\
        \texttt{0.326\_Nd2Sn2O7} & \texttt{oS} & \texttt{1.69\_CoO} & \texttt{S}* & \texttt{1.619\_MnS} & \texttt{Lp} \\
        \texttt{0.330\_ErGe3} & \texttt{S} & \texttt{1.70\_CoV2O6} & \texttt{S}* & \texttt{1.620\_NdCu2} & \texttt{Lt} \\
        \texttt{0.334\_CoF3} & \texttt{oS}* & \texttt{1.71\_SrCo2V2O8} & \texttt{M} & \texttt{1.623\_EuMg2Bi2} & \texttt{oS} \\
        \texttt{0.335\_FeF3} & \texttt{oS}* & \texttt{1.77\_Sr2IrO4} & \texttt{M} & \texttt{1.624\_EuSn2P2} & \texttt{oS} \\
        \texttt{0.336\_NdFeO3} & \texttt{oS}* & \texttt{1.79\_Li2CoSiO4} & \texttt{M} & \texttt{1.625\_Sr2Fe3S2O3} & \texttt{oS} \\
        \texttt{0.339\_Nd2Hf2O7} & \texttt{oS} & \texttt{1.80\_Dy2CoGa8} & \texttt{oS} & \texttt{1.627\_KCeS2} & \texttt{Lp} \\
        \texttt{0.340\_Nd2Zr2O7} & \texttt{oS} & \texttt{1.81\_GdIn3} & \texttt{Lp} & \texttt{1.629\_FeGe} & \texttt{oS}* \\
        \texttt{0.345\_Tb2C3} & \texttt{Lp} & \texttt{1.82\_Nd2RhIn8} & \texttt{oS} & \texttt{1.630\_LuMn6Sn6} & \texttt{oS}* \\
        \texttt{0.348\_Bi2CuO4} & \texttt{oS}* & \texttt{1.87\_Tb2CoGa8} & \texttt{oS} & \texttt{1.631\_YMn6Ge6} & \texttt{oS}* \\
        \texttt{0.350\_TbAlO3} & \texttt{M} & \texttt{1.88\_Mn5Si3} & \texttt{M} & \texttt{1.635\_ErFe2Si2} & \texttt{Lt} \\
        \texttt{0.351\_TbFeO3} & \texttt{oS}* & \texttt{1.94\_Ba3LaRu2O9} & \texttt{Lp} & \texttt{1.636\_ErMn2Si2} & \texttt{oS}* \\
        \texttt{0.354\_TbCrO3} & \texttt{oS}* & \texttt{1.95\_BaNd2O4} & \texttt{M} & \texttt{1.637\_ErMn2Si2} & \texttt{oS} \\
        \texttt{0.361\_Sr3LiRuO6} & \texttt{oS}* & \texttt{1.96\_BaNd2O4} & \texttt{M} & \texttt{1.638\_ErMn2Ge2} & \texttt{oS}* \\
        \texttt{0.362\_RbFeCl5(D2O)} & \texttt{oS} & \texttt{1.97\_Li2MnO3} & \texttt{S}* & \texttt{1.639\_ErMn2Ge2} & \texttt{oS} \\
        \texttt{0.363\_KFeCl5(D2O)} & \texttt{oS} & \texttt{1.99\_CsCoCl3(D2O)2} & \texttt{M} & \texttt{1.640\_ErMn2Ge2} & \texttt{oS} \\
        \texttt{0.364\_SrCr2As2} & \texttt{oS}* & \texttt{1.100\_Cu2MnSnS4} & \texttt{Lp} & \texttt{1.641\_Ba2FeSi2O7} & \texttt{oS} \\
        \texttt{0.365\_BaCr2As2} & \texttt{oS}* & \texttt{1.101\_LuMnO3} & \texttt{Lp} & \texttt{1.642\_TlFeS2} & \texttt{S}* \\
        \texttt{0.377\_Mn3Ge} & \texttt{oS}* & \texttt{1.103\_U2Rh2Sn} & \texttt{oS} & \texttt{1.643\_DyOCl} & \texttt{oS} \\
        \texttt{0.378\_UBi2} & \texttt{oS} & \texttt{1.104\_Gd2CuO4} & \texttt{Lp} & \texttt{1.644\_EuSn2As2} & \texttt{oS} \\
        \texttt{0.379\_SmFeO3} & \texttt{oS}* & \texttt{1.107\_Sm2CuO4} & \texttt{Lp} & \texttt{1.648\_Nd2O3} & \texttt{Lp} \\
        \texttt{0.380\_SmFeO3} & \texttt{oS} & \texttt{1.110\_ScMn6Ge6} & \texttt{oS}* & \texttt{1.649\_Sr3ZnIrO6} & \texttt{M} \\
        \texttt{0.382\_LiMnPO4} & \texttt{oS} & \texttt{1.111\_GdBiPt} & \texttt{Lp} & \texttt{1.653\_FeWO4} & \texttt{S}* \\
        \texttt{0.383\_LiCoPO4} & \texttt{oS} & \texttt{1.112\_NiTa2O6} & \texttt{Lt} & \texttt{1.654\_NiNb2O6} & \texttt{M} \\
        \texttt{0.384\_LiCoPO4} & \texttt{S} & \texttt{1.113\_NiSb2O6} & \texttt{Lp} & \texttt{1.655\_FeNb2O6} & \texttt{M} \\
        \texttt{0.385\_LiCoPO4} & \texttt{M} & \texttt{1.114\_Ca4IrO6} & \texttt{M} & \texttt{1.656\_CoNb2O6} & \texttt{Lp} \\
        \texttt{0.388\_Co3Al2Si3O12} & \texttt{M} & \texttt{1.115\_Dy3Ru4Al12} & \texttt{M} & \texttt{1.659\_MnCl2(CO(NH2)2)2} & \texttt{oS} \\
        \texttt{0.398\_Ca2RuO4} & \texttt{M} & \texttt{1.116\_AgMnVO4} & \texttt{M} & \texttt{1.660\_FePb4Sb6S14} & \texttt{M} \\
        \texttt{0.399\_FeOOH} & \texttt{oS}* & \texttt{1.117\_NaFePO4} & \texttt{M} & \texttt{1.663\_Tb2Ni2In} & \texttt{Lp} \\
        \texttt{0.401\_Sr4Fe4O11} & \texttt{oS}* & \texttt{1.118\_GdPO4} & \texttt{M} & \texttt{1.664\_DyVO4} & \texttt{Lp} \\
        \texttt{0.402\_Sr4Fe4O11} & \texttt{oS} & \texttt{1.120\_BaFe2Se3} & \texttt{M} & \texttt{1.665\_Ba3CoNb2O9} & \texttt{oS} \\
        \texttt{0.404\_Sr3NaRuO6} & \texttt{oS}* & \texttt{1.121\_NaFeSO4F} & \texttt{S} & \texttt{1.666\_TbCoGa5} & \texttt{Lp} \\
        \texttt{0.406\_GdNiSi3} & \texttt{oS} & \texttt{1.125\_LaFeAsO} & \texttt{oS}* & \texttt{1.667\_UPtGa5} & \texttt{Lp} \\
        \texttt{0.410\_GdAlO3} & \texttt{oS} & \texttt{1.126\_NaCoSO4F} & \texttt{M} & \texttt{1.668\_HoCoGa5} & \texttt{Lp} \\
        \texttt{0.413\_UGeSe} & \texttt{oS} & \texttt{1.127\_BiNiO(PO4)} & \texttt{M} & \texttt{1.669\_KFe(PO3F)2} & \texttt{M} \\
        \texttt{0.416\_LaCrO3} & \texttt{oS}* & \texttt{1.128\_BiCoO(PO4)} & \texttt{M} & \texttt{1.671\_NpCoGa5} & \texttt{oS} \\
        \texttt{0.417\_LaCrO3} & \texttt{oS} & \texttt{1.129\_AgFe3(SO4)2(OD)6} & \texttt{oS} & \texttt{1.672\_EuZn2As2} & \texttt{oS} \\
        \texttt{0.419\_ErGe2O7} & \texttt{M} & \texttt{1.133\_CuSb2O6} & \texttt{M} & \texttt{1.673\_EuCd2Sb2} & \texttt{oS} \\
        \texttt{0.420\_Sr2LuRuO6} & \texttt{S}* & \texttt{1.134\_Co2C10O8H2} & \texttt{S} & \texttt{1.678\_CrN} & \texttt{oS}* \\
        \texttt{0.421\_EuMnSb2} & \texttt{oS}* & \texttt{1.136\_AgCrS2} & \texttt{Lt} & \texttt{1.681\_PrFe2Al8} & \texttt{M} \\
        \texttt{0.423\_EuMnSb2} & \texttt{oS} & \texttt{1.138\_MgV2O4} & \texttt{M} & \texttt{1.683\_UPdGa5} & \texttt{oS} \\
        \texttt{0.426\_EuMnBi2} & \texttt{oS}* & \texttt{1.139\_Ho2RhIn8} & \texttt{Lp} & \texttt{1.685\_NiCr2O4} & \texttt{Lp} \\
        \texttt{0.427\_Sm2Ti2O7} & \texttt{oS} & \texttt{1.140\_PrMgPb} & \texttt{S} & \texttt{1.689\_LuMn2Ge2} & \texttt{oS}* \\
        \texttt{0.430\_Yb3Pt4} & \texttt{M} & \texttt{1.141\_NdMgPb} & \texttt{S} & \texttt{1.690\_TmMn2Ge2} & \texttt{oS}* \\
        \texttt{0.432\_KMnF3} & \texttt{oS} & \texttt{1.142\_CeMgPb} & \texttt{Lp} & \texttt{1.691\_YMn2Ge2} & \texttt{oS} \\
        \texttt{0.433\_KMnF3} & \texttt{oS}* & \texttt{1.143\_Mn3Pt} & \texttt{Lp} & \texttt{1.692\_YMn2Ge2} & \texttt{oS} \\
        \texttt{0.434\_K2ReI6} & \texttt{S}* & \texttt{1.144\_NH4FeCl2(HCOO)} & \texttt{S} & \texttt{1.694\_TbMn2Ge2} & \texttt{oS}* \\
        \texttt{0.440\_SrCuTe2O6} & \texttt{oS} & \texttt{1.145\_Mn3Ni20P6} & \texttt{Lp} & \texttt{1.695\_Mn3Ni2Si} & \texttt{M} \\
        \texttt{0.444\_YbCl3} & \texttt{S} & \texttt{1.146\_LaCrAsO} & \texttt{oS}* & \texttt{1.696\_HoNiSi2} & \texttt{oS} \\
        \texttt{0.448\_Ce4Ge3} & \texttt{Lp} & \texttt{1.147\_Li2Fe(SO4)2} & \texttt{M} & \texttt{1.699\_GdInCu4} & \texttt{Lt} \\
        \texttt{0.451\_DyRuAsO} & \texttt{oS} & \texttt{1.150\_PrAg} & \texttt{Lp} & \texttt{1.700\_HoInCu4} & \texttt{oS} \\
        \texttt{0.452\_TbRuAsO} & \texttt{oS} & \texttt{1.153\_Mn3GaC} & \texttt{Lp} & \texttt{1.701\_HoCdCu4} & \texttt{Lp} \\
        \texttt{0.453\_DyCoSi2} & \texttt{oS} & \texttt{1.154\_NaFeSi2O6} & \texttt{S} & \texttt{1.702\_YBaCo2O5} & \texttt{oS}* \\
        \texttt{0.454\_PrScSb} & \texttt{oS} & \texttt{1.158\_YMn3Al4O12} & \texttt{oS} & \texttt{1.706\_Ba2MnTeO6} & \texttt{Lp} \\
        \texttt{0.461\_CoRh2O4} & \texttt{oS}* & \texttt{1.159\_Li2Ni(WO4)2} & \texttt{S}* & \texttt{1.707\_Ba2MnWO6} & \texttt{Lp} \\
        \texttt{0.462\_MnAl2O4} & \texttt{oS}* & \texttt{1.160\_UP} & \texttt{Lp} & \texttt{1.709\_CsCrF4} & \texttt{M} \\
        \texttt{0.463\_Co3O4} & \texttt{oS}* & \texttt{1.162\_NdMg} & \texttt{Lp} & \texttt{1.714\_CeAuBi2} & \texttt{oS} \\
        \texttt{0.464\_BaMn2P2} & \texttt{oS}* & \texttt{1.163\_TmPdIn} & \texttt{Lt} & \texttt{1.715\_Sr2CoWO6} & \texttt{S} \\
        \texttt{0.465\_HoCr2Si2} & \texttt{oS}* & \texttt{1.166\_La2LiOsO6} & \texttt{M} & \texttt{1.716\_Sr2MnMoO6} & \texttt{M} \\
        \texttt{0.466\_ThCr2Si2} & \texttt{oS}* & \texttt{1.167\_NiS2} & \texttt{M} & \texttt{1.717\_Sr2MnWO6} & \texttt{M} \\
        \texttt{0.467\_TbPO4} & \texttt{oS} & \texttt{1.168\_Sr2CuTeO6} & \texttt{Lp} & \texttt{1.718\_Ca2MnWO6} & \texttt{M} \\
        \texttt{0.468\_ErB4} & \texttt{Lp} & \texttt{1.169\_CaCoGe2O6} & \texttt{S} & \texttt{1.719\_Ca2MnWO6} & \texttt{M} \\
        \texttt{0.469\_TbB4} & \texttt{Lp} & \texttt{1.171\_Tb2Fe2Si2C} & \texttt{oS} & \texttt{1.721\_UCu5} & \texttt{Lp} \\
        \texttt{0.470\_BaMn2Sb2} & \texttt{oS}* & \texttt{1.172\_NiTa2O6} & \texttt{Lt} & \texttt{1.722\_Ba3LaRu2O9} & \texttt{Lp} \\
        \texttt{0.471\_Ba2Mn3Sb2O2} & \texttt{oS}* & \texttt{1.176\_YbCo2Si2} & \texttt{Lt} & \texttt{1.723\_NaMn2O4} & \texttt{M} \\
        \texttt{0.472\_LaMn2Si2} & \texttt{oS}* & \texttt{1.177\_Sr2CuWO6} & \texttt{Lp} & \texttt{1.724\_Ba2NiTeO6} & \texttt{Lp} \\
        \texttt{0.474\_EuMn2Ge2} & \texttt{oS}* & \texttt{1.180\_Na3Co2SbO6} & \texttt{Lp} & \texttt{1.725\_Ba3NiTa2O9} & \texttt{oS} \\
        \texttt{0.475\_Sr2TbIrO6} & \texttt{S} & \texttt{1.183\_FePS3} & \texttt{S}* & \texttt{1.727\_Tm3Cu4Ge4} & \texttt{Lt} \\
        \texttt{0.476\_Cs2[FeCl5(H2O)]} & \texttt{oS} & \texttt{1.185\_GeCu2O4} & \texttt{Lp} & \texttt{1.728\_Tm3Cu4Sn4} & \texttt{oS} \\
        \texttt{0.482\_SrMn2As2} & \texttt{oS}* & \texttt{1.186\_SrRu2O6} & \texttt{oS}* & \texttt{1.729\_Gd2Fe2Si2C} & \texttt{oS} \\
        \texttt{0.483\_YbMn2Sb2} & \texttt{S}* & \texttt{1.187\_TbRh2Si2} & \texttt{oS} & \texttt{1.730\_Cu2MnSiS4} & \texttt{M} \\
        \texttt{0.484\_U2N2S} & \texttt{oS} & \texttt{1.188\_CeRh2Si2} & \texttt{Lp} & \texttt{1.731\_Cu2FeSiS4} & \texttt{M} \\
        \texttt{0.485\_U2N2Se} & \texttt{oS} & \texttt{1.189\_TbMg3} & \texttt{Lp} & \texttt{1.732\_Cu2MnSnS4} & \texttt{Lp} \\
        \texttt{0.486\_ErCr2Si2} & \texttt{oS}* & \texttt{1.190\_YCr(BO3)2} & \texttt{S}* & \texttt{1.733\_Cu2MnGeS4} & \texttt{M} \\
        \texttt{0.491\_NdB4} & \texttt{S} & \texttt{1.191\_HoCr(BO3)2} & \texttt{S}* & \texttt{1.734\_Cu2FeGeS4} & \texttt{Lp} \\
        \texttt{0.492\_NdB4} & \texttt{M} & \texttt{1.194\_NiWO4} & \texttt{S}* & \texttt{1.735\_Li2FeGeS4} & \texttt{oS} \\
        \texttt{0.498\_LaMn2Si2} & \texttt{oS} & \texttt{1.195\_Er2Ni2In} & \texttt{Lt} & \texttt{1.736\_Mn(N2H5)2(SO4)2} & \texttt{S} \\
        \texttt{0.499\_UCr2Si2C} & \texttt{oS}* & \texttt{1.196\_MnV2O6} & \texttt{M} & \texttt{1.738\_TbNiAl} & \texttt{M} \\
        \texttt{0.501\_LiFe2F6} & \texttt{oS}* & \texttt{1.197\_Fe4Si2Sn7O16} & \texttt{M} & \texttt{1.740\_CeAuSb2} & \texttt{Lt} \\
        \texttt{0.504\_NaCrSi2O6} & \texttt{S}* & \texttt{1.199\_Sc2NiMnO6} & \texttt{S} & \texttt{1.741\_KNiAsO4} & \texttt{S}* \\
        \texttt{0.505\_Pb2VO(PO4)2} & \texttt{oS} & \texttt{1.200\_U2Ni2Sn} & \texttt{oS} & \texttt{1.742\_KNiAsO4} & \texttt{Lp} \\
        \texttt{0.513\_YRuO3} & \texttt{oS}* & \texttt{1.205\_Dy2Fe2Si2C} & \texttt{oS} & \texttt{1.743\_CeRhGe3} & \texttt{Lt} \\
        \texttt{0.518\_TbCr2Si2} & \texttt{oS}* & \texttt{1.206\_Dy2Fe2Si2C} & \texttt{S} & \texttt{1.744\_PrPdSn} & \texttt{Lp} \\
        \texttt{0.519\_HoCr2Si2} & \texttt{oS} & \texttt{1.208\_UAs} & \texttt{Lp} & \texttt{1.746\_YMn2} & \texttt{Lp} \\
        \texttt{0.520\_TbCoO3} & \texttt{M} & \texttt{1.210\_FePSe3} & \texttt{Lp} & \texttt{1.747\_ErAuIn} & \texttt{oS} \\
        \texttt{0.521\_DyCoO3} & \texttt{M} & \texttt{1.211\_Dy2O2S} & \texttt{Lp} & \texttt{1.749\_HoSbTe} & \texttt{Lp} \\
        \texttt{0.523\_CaMn2Sb2} & \texttt{S} & \texttt{1.212\_Dy2O2Se} & \texttt{Lp} & \texttt{1.750\_HoSbTe} & \texttt{Lt} \\
        \texttt{0.524\_MnPSe3} & \texttt{S} & \texttt{1.213\_Ho2O2Se} & \texttt{Lp} & \texttt{1.751\_CaCo3Ti4O12} & \texttt{Lt} \\
        \texttt{0.527\_Er2Si2O7} & \texttt{S} & \texttt{1.214\_Yb2O2Se} & \texttt{S} & \texttt{1.753\_HoBi} & \texttt{Lp} \\
        \texttt{0.528\_CrSb} & \texttt{oS}* & \texttt{1.215\_UP2} & \texttt{oS} & \texttt{1.755\_KErSe2} & \texttt{Lp} \\
        \texttt{0.530\_SrCuTe2O6} & \texttt{oS} & \texttt{1.219\_CuF2} & \texttt{S}* & \texttt{1.757\_Pr2PdAl7Ge4} & \texttt{M} \\
        \texttt{0.552\_Pb2MnO4} & \texttt{M} & \texttt{1.222\_Er2CoGa8} & \texttt{Lp} & \texttt{1.758\_CaMn3V4O12} & \texttt{M} \\
        \texttt{0.553\_K2ReI6} & \texttt{S} & \texttt{1.223\_Tm2CoGa8} & \texttt{Lp} & \texttt{1.759\_ZnFe2O4} & \texttt{Lt} \\
        \texttt{0.562\_Ce2Ni3Ge5} & \texttt{oS} & \texttt{1.224\_CoNb2O6} & \texttt{M} & \texttt{1.760\_ZnFe2O4} & \texttt{Lp} \\
        \texttt{0.563\_Ce2Ni3Ge5} & \texttt{oS} & \texttt{1.225\_ScMn6Ge6} & \texttt{oS} & \texttt{1.761\_ZnFe2O4} & \texttt{Lt} \\
        \texttt{0.564\_U2Rh3Si5} & \texttt{M} & \texttt{1.226\_CeCo2Ge4O12} & \texttt{M} & \texttt{1.763\_BaNiTe2O7} & \texttt{M} \\
        \texttt{0.565\_Ce2Ni3Ge5} & \texttt{M} & \texttt{1.228\_RuCl3} & \texttt{S} & \texttt{1.764\_NdSbTe} & \texttt{Lt} \\
        \texttt{0.566\_TbNiGe2} & \texttt{oS} & \texttt{1.229\_BaMoP2O8} & \texttt{Lp} & \texttt{1.765\_DySbTe} & \texttt{oS} \\
        \texttt{0.571\_CoSO4} & \texttt{M} & \texttt{1.230\_NiPS3} & \texttt{S}* & \texttt{1.767\_Li2CoCl4} & \texttt{oS}* \\
        \texttt{0.575\_ZnFeF5(H2O)2} & \texttt{oS}* & \texttt{1.232\_CuMnSb} & \texttt{Lp} & \texttt{1.769\_Ni2Te3O8} & \texttt{M} \\
        \texttt{0.581\_FeF3} & \texttt{oS} & \texttt{1.233\_CuMnSb} & \texttt{Lp} & \texttt{1.770\_Tb2Ni2In} & \texttt{Lp} \\
        \texttt{0.582\_Fe3F8(H2O)2} & \texttt{oS}* & \texttt{1.235\_Ba(TiO)Cu4(PO4)4} & \texttt{M} & \texttt{1.772\_Pr2PdAl7Ge4} & \texttt{M} \\
        \texttt{0.585\_YbCl3} & \texttt{S} & \texttt{1.237\_VCl2} & \texttt{oS}* & \texttt{1.773\_PrIr3B2} & \texttt{Lp} \\
        \texttt{0.586\_YCrO3} & \texttt{oS}* & \texttt{1.238\_VBr2} & \texttt{oS}* & \texttt{1.774\_BaNd2PtO5} & \texttt{M} \\
        \texttt{0.587\_TmCrO3} & \texttt{oS}* & \texttt{1.239\_MnBr2} & \texttt{Lt} & \texttt{1.775\_CaCu3Ti4O12} & \texttt{M} \\
        \texttt{0.591\_ErCrO3} & \texttt{oS}* & \texttt{1.240\_FeI2} & \texttt{Lt} & \texttt{1.777\_EuAl2Si2} & \texttt{oS} \\
        \texttt{0.592\_DyCrO3} & \texttt{oS}* & \texttt{1.241\_FeCl2} & \texttt{oS}* & \texttt{1.778\_ThCr2Si2C} & \texttt{oS}* \\
        \texttt{0.598\_AlCr2} & \texttt{S}* & \texttt{1.242\_FeBr2} & \texttt{oS}* & \texttt{1.782\_FeBr3} & \texttt{oS}* \\
        \texttt{0.599\_CaMnSi} & \texttt{oS}* & \texttt{1.244\_CrCl3} & \texttt{S}* & \texttt{1.784\_Li2CoCl4} & \texttt{oS} \\
        \texttt{0.600\_CaMnSi} & \texttt{oS} & \texttt{1.245\_CoBr2} & \texttt{S}* & \texttt{1.785\_K2ReCl6} & \texttt{M} \\
        \texttt{0.601\_CaMnGe} & \texttt{S}* & \texttt{1.246\_CoCl2} & \texttt{S}* & \texttt{1.786\_RuBr3} & \texttt{Lp} \\
        \texttt{0.602\_CaMnGe} & \texttt{S} & \texttt{1.247\_NiCl2} & \texttt{S}* & \texttt{1.787\_RuCl3} & \texttt{Lp} \\
        \texttt{0.603\_CaMn2Ge2} & \texttt{oS}* & \texttt{1.248\_NiBr2} & \texttt{S}* & \texttt{1.793\_Ca3Ru2O7} & \texttt{oS} \\
        \texttt{0.604\_CaMn2Ge2} & \texttt{oS} & \texttt{1.249\_K2NiF4} & \texttt{Lp} & \texttt{1.794\_Ca3Ru2O7} & \texttt{oS} \\
        \texttt{0.605\_BaMn2Ge2} & \texttt{oS}* & \texttt{1.250\_KNiF3} & \texttt{oS}* & \texttt{1.795\_BiMn3Cr4O12} & \texttt{oS} \\
        \texttt{0.606\_BaMn2Ge2} & \texttt{oS} & \texttt{1.252\_CaCo2P2} & \texttt{oS}* & \texttt{1.796\_BiMn3Cr4O12} & \texttt{oS} \\
        \texttt{0.607\_RuO2} & \texttt{oS}* & \texttt{1.253\_CeCo2P2} & \texttt{oS}* & \texttt{1.799\_MnO} & \texttt{S}* \\
        \texttt{0.608\_PrMnO3} & \texttt{oS}* & \texttt{1.254\_UNiGa5} & \texttt{oS} & \texttt{1.800\_NiO} & \texttt{Lp} \\
        \texttt{0.609\_NdMnO3} & \texttt{S}* & \texttt{1.255\_UPtGa5} & \texttt{oS} & \texttt{1.802\_CrSBr} & \texttt{oS}* \\
        \texttt{0.611\_BaMnSb2} & \texttt{oS}* & \texttt{1.256\_BaNi2V2O8} & \texttt{S}* & \texttt{1.803\_LiCrTe2} & \texttt{oS}* \\
        \texttt{0.617\_KMnSb} & \texttt{oS}* & \texttt{1.257\_BaNi2As2O8} & \texttt{Lp} & \texttt{1.805\_FeVMoO7} & \texttt{S} \\
        \texttt{0.618\_KMnBi} & \texttt{oS}* & \texttt{1.260\_NaMnGe2O6} & \texttt{Lp} & \texttt{1.806\_CrVMoO7} & \texttt{S} \\
        \texttt{0.619\_LaMnAsO} & \texttt{oS}* & \texttt{1.261\_NpRhGa5} & \texttt{oS} & \texttt{1.807\_FeWO4} & \texttt{S} \\
        \texttt{0.620\_NdMnAsO} & \texttt{oS}* & \texttt{1.262\_NpRhGa5} & \texttt{oS} & \texttt{1.809\_BiCoO3} & \texttt{oS}* \\
        \texttt{0.623\_NdMnAsO} & \texttt{oS} & \texttt{1.263\_Ca3Ru2O7} & \texttt{oS} & \texttt{1.810\_BiCoO3} & \texttt{oS} \\
        \texttt{0.624\_LaMnAsO} & \texttt{oS} & \texttt{1.264\_CoPS3} & \texttt{S}* & \texttt{1.811\_BiCoO3} & \texttt{oS} \\
        \texttt{0.625\_U2Pd2In} & \texttt{oS} & \texttt{1.265\_CuMnSb} & \texttt{Lp} & \texttt{1.821\_Yb3Ga5O12} & \texttt{M} \\
        \texttt{0.626\_NaMnP} & \texttt{oS}* & \texttt{1.267\_Dy2Co3Al9} & \texttt{M} & \texttt{1.822\_Sr2MnSi2O7} & \texttt{M} \\
        \texttt{0.627\_NaMnP} & \texttt{oS} & \texttt{1.271\_CeSbTe} & \texttt{oS} & \texttt{1.823\_Sr2MnSi2O7} & \texttt{M} \\
        \texttt{0.628\_NaMnP} & \texttt{oS} & \texttt{1.272\_CeNiAsO} & \texttt{M} & \texttt{1.824\_PbCo2V2O8} & \texttt{M} \\
        \texttt{0.629\_NaMnAs} & \texttt{oS}* & \texttt{1.278\_Cu(NCS)2} & \texttt{S}* & \texttt{1.826\_NdZnPO} & \texttt{Lp} \\
        \texttt{0.630\_NaMnAs} & \texttt{oS} & \texttt{1.281\_YBaCuFeO5} & \texttt{oS}* & \texttt{1.828\_ZnFe2O4} & \texttt{Lt} \\
        \texttt{0.631\_NaMnSb} & \texttt{oS}* & \texttt{1.286\_Fe2(C2O4)3.4H2O} & \texttt{S} & \texttt{1.830\_FeBi4S7} & \texttt{oS}* \\
        \texttt{0.632\_NaMnSb} & \texttt{oS} & \texttt{1.287\_V2O3} & \texttt{S}* & \texttt{1.833\_Pb(OF)Cu3(SeO3)2(NO3)} & \texttt{M} \\
        \texttt{0.633\_KFeS2} & \texttt{S}* & \texttt{1.288\_CePd2Si2} & \texttt{Lp} & \texttt{1.834\_Na3Co(CO3)2Cl} & \texttt{oS} \\
        \texttt{0.634\_NaMnBi} & \texttt{oS}* & \texttt{1.289\_CePd2Ge2} & \texttt{Lp} & \texttt{1.835\_La2Co2O3Se2} & \texttt{M} \\
        \texttt{0.635\_NaMnBi} & \texttt{oS} & \texttt{1.290\_CeRh2Si2} & \texttt{Lp} & \texttt{1.838\_LaMn2Au4} & \texttt{oS}* \\
        \texttt{0.636\_RbFeS2} & \texttt{S}* & \texttt{1.291\_CeAu2Si2} & \texttt{oS} & \texttt{1.840\_Sr2Mn2O5} & \texttt{M} \\
        \texttt{0.637\_KFeSe2} & \texttt{oS}* & \texttt{1.292\_HoNi2B2C} & \texttt{oS} & \texttt{1.841\_Sr3Fe2O5Cl2} & \texttt{Lp} \\
        \texttt{0.638\_RbFeSe2} & \texttt{oS}* & \texttt{1.293\_NdNi2B2C} & \texttt{Lp} & \texttt{1.842\_Ca2FeO3Cl} & \texttt{oS} \\
        \texttt{0.639\_Mn2Au} & \texttt{oS}* & \texttt{1.294\_HoNi2B2C} & \texttt{oS} & \texttt{1.843\_Ca2FeO3Cl} & \texttt{oS} \\
        \texttt{0.640\_Mn2Au} & \texttt{oS} & \texttt{1.295\_DyNi2B2C} & \texttt{oS} & \texttt{1.844\_Sr2FeO3Cl} & \texttt{oS} \\
        \texttt{0.642\_LaMnO3} & \texttt{oS} & \texttt{1.296\_PrNi2B2C} & \texttt{oS} & \texttt{1.845\_Sr2FeO3Cl} & \texttt{oS} \\
        \texttt{0.650\_ErSi2O7} & \texttt{S} & \texttt{1.298\_BaCdVO(PO4)2} & \texttt{M} & \texttt{1.846\_LaCu3Fe4O12} & \texttt{Lt} \\
        \texttt{0.651\_Er3Cu4Sn4} & \texttt{oS} & \texttt{1.301\_BiMnTeO6} & \texttt{M} & \texttt{1.849\_Na3Ni2SbO6} & \texttt{Lp} \\
        \texttt{0.658\_BaCuTe2O6} & \texttt{M} & \texttt{1.302\_Ba2CoO4} & \texttt{M} & \texttt{1.853\_Ba2YFeO5} & \texttt{M} \\
        \texttt{0.665\_CeMnSbO} & \texttt{oS}* & \texttt{1.304\_ZnMnO3} & \texttt{Lp} & \texttt{1.854\_CoTeMoO6} & \texttt{M} \\
        \texttt{0.667\_LaMnSbO} & \texttt{oS}* & \texttt{1.305\_Mn5Si3} & \texttt{oS} & \texttt{1.855\_Mn3GaC} & \texttt{Lp} \\
        \texttt{0.681\_Ce4Sb3} & \texttt{Lp} & \texttt{1.306\_Na2BaMn(VO4)2} & \texttt{M} & \texttt{1.856\_Mn3GaC} & \texttt{Lp} \\
        \texttt{0.692\_Ba4Ru3O10} & \texttt{oS} & \texttt{1.308\_MnBi2Te4} & \texttt{oS}* & \texttt{1.857\_Li2CoCl4} & \texttt{oS} \\
        \texttt{0.693\_Ba4Ru3O10} & \texttt{oS} & \texttt{1.311\_BaMo(PO4)2} & \texttt{Lp} & \texttt{2.2\_Sr2F2Fe2OS2} & \texttt{Lp} \\
        \texttt{0.694\_Bi2CuO4} & \texttt{oS} & \texttt{1.312\_HoNi2B2C} & \texttt{oS} & \texttt{2.6\_Nd2CuO4} & \texttt{oS}* \\
        \texttt{0.695\_Bi2CuO4} & \texttt{oS} & \texttt{1.314\_NaFeSi2O6} & \texttt{S} & \texttt{2.7\_Sm2CuO4} & \texttt{oS}* \\
        \texttt{0.706\_Tb2Ir3Ga9} & \texttt{oS} & \texttt{1.318\_Sr2Ru0.95Fe0.05O4} & \texttt{Lt} & \texttt{2.13\_UP} & \texttt{Lp} \\
        \texttt{0.708\_CrNb4S8} & \texttt{oS}* & \texttt{1.319\_Sr2Ru0.95Fe0.05O4} & \texttt{Lt} & \texttt{2.14\_NdMg} & \texttt{Lp} \\
        \texttt{0.712\_VNb3S6} & \texttt{oS}* & \texttt{1.320\_Sr2FeWO6} & \texttt{M} & \texttt{2.20\_UAs} & \texttt{Lt} \\
        \texttt{0.714\_Li2Ni(SO4)2} & \texttt{S}* & \texttt{1.321\_Ba2FeWO6} & \texttt{Lp} & \texttt{2.21\_TbOOH} & \texttt{M} \\
        \texttt{0.723\_YbCl3} & \texttt{S} & \texttt{1.334\_Pr2Pd2In} & \texttt{Lp} & \texttt{2.22\_FeTa2O6} & \texttt{Lp} \\
        \texttt{0.724\_BaCoSiO4} & \texttt{M} & \texttt{1.338\_U2Ni2In} & \texttt{oS} & \texttt{2.23\_Sr2CoO2Ag2Se2} & \texttt{S} \\
        \texttt{0.728\_MoP3SiO11} & \texttt{S} & \texttt{1.339\_EuAs3} & \texttt{oS} & \texttt{2.24\_Ba2CoO2Ag2Se2} & \texttt{S} \\
        \texttt{0.733\_AgRuO3} & \texttt{oS}* & \texttt{1.340\_LuMnO3} & \texttt{M} & \texttt{2.30\_CeRh2Si2} & \texttt{Lt} \\
        \texttt{0.740\_Dy3Ga5O12} & \texttt{oS} & \texttt{1.341\_TmMnO3} & \texttt{Lp} & \texttt{2.31\_Mn3ZnN} & \texttt{M} \\
        \texttt{0.741\_Er3Ga5O12} & \texttt{oS} & \texttt{1.345\_NaMnF4} & \texttt{M} & \texttt{2.35\_CrSe} & \texttt{M} \\
        \texttt{0.743\_Ho3Al5O12} & \texttt{oS} & \texttt{1.346\_TlMnF4} & \texttt{oS}* & \texttt{2.36\_TbGe3} & \texttt{Lt} \\
        \texttt{0.744\_Tb3Al5O12} & \texttt{oS} & \texttt{1.347\_CuFeO2} & \texttt{Lt} & \texttt{2.48\_Pr2CuO4} & \texttt{oS}* \\
        \texttt{0.745\_Ho3Ga5O12} & \texttt{oS} & \texttt{1.349\_CoNb3S6} & \texttt{Lp} & \texttt{2.49\_La2O2Fe2OSe2} & \texttt{Lp} \\
        \texttt{0.746\_Tb3Ga5O12} & \texttt{oS} & \texttt{1.354\_EuNiO3} & \texttt{Lp} & \texttt{2.56\_La2O2Fe2OS2} & \texttt{Lp} \\
        \texttt{0.755\_Mn2SeO3F2} & \texttt{oS} & \texttt{1.356\_Ho3Ge4} & \texttt{oS} & \texttt{2.66\_FeSn2} & \texttt{M} \\
        \texttt{0.757\_CeFeO3} & \texttt{oS}* & \texttt{1.361\_DyGe} & \texttt{Lp} & \texttt{2.67\_FeSn2} & \texttt{M} \\
        \texttt{0.758\_CeFeO3} & \texttt{oS} & \texttt{1.363\_TbCu2Si2} & \texttt{Lp} & \texttt{2.68\_FeGe2} & \texttt{M} \\
        \texttt{0.760\_FeOHSO4} & \texttt{oS}* & \texttt{1.364\_HoCu2Si2} & \texttt{Lp} & \texttt{2.71\_HoRh} & \texttt{M} \\
        \texttt{0.761\_SrFe2Se2O} & \texttt{M} & \texttt{1.365\_TbCu2Si2} & \texttt{Lp} & \texttt{2.73\_BaNd2ZnO5} & \texttt{M} \\
        \texttt{0.762\_SrFe2S2O} & \texttt{M} & \texttt{1.366\_HoCu2Si2} & \texttt{Lp} & \texttt{2.77\_Eu2CuO4} & \texttt{oS}* \\
        \texttt{0.766\_YbMnSb2} & \texttt{oS}* & \texttt{1.367\_Pu2O3} & \texttt{Lp} & \texttt{2.78\_Nd2CuO4} & \texttt{oS} \\
        \texttt{0.769\_YbMnBi2} & \texttt{oS} & \texttt{1.368\_Tb2Ni3Si5} & \texttt{oS} & \texttt{2.86\_FeTa2O6} & \texttt{Lp} \\
        \texttt{0.782\_NdScO3} & \texttt{M} & \texttt{1.369\_HFe2Ge2} & \texttt{Lp} & \texttt{2.87\_TbCoGa5} & \texttt{M} \\
        \texttt{0.783\_NdInO3} & \texttt{M} & \texttt{1.370\_Li2CuO2} & \texttt{oS}* & \texttt{2.88\_UNiGa} & \texttt{Lt} \\
        \texttt{0.784\_NdCoO3} & \texttt{oS} & \texttt{1.371\_Nd2NiO4} & \texttt{oS}* & \texttt{2.93\_CoCrO4} & \texttt{M} \\
        \texttt{0.786\_NdVO3} & \texttt{S}* & \texttt{1.374\_HoNiGe} & \texttt{M} & \texttt{2.96\_GdMn2Si2} & \texttt{M} \\
        \texttt{0.787\_YVO3} & \texttt{oS}* & \texttt{1.375\_CeScGe} & \texttt{S} & \texttt{2.99\_TbNiAl} & \texttt{Lt} \\
        \texttt{0.795\_Sr2YRuO6} & \texttt{S}* & \texttt{1.377\_CeScSi} & \texttt{S} & \texttt{2.101\_TbSbTe} & \texttt{Lt} \\
        \texttt{0.798\_MnPd2} & \texttt{oS}* & \texttt{1.379\_ErNiGe} & \texttt{Lp} & \texttt{2.102\_TbSbTe} & \texttt{Lt} \\
        \texttt{0.800\_MnTe} & \texttt{oS}* & \texttt{1.380\_Sr2FeO3Cl} & \texttt{oS}* & \texttt{2.104\_BaNd2ZnS5} & \texttt{M} \\
        \texttt{0.801\_Tl3Fe2S4} & \texttt{oS} & \texttt{1.381\_Sr2FeO3Br} & \texttt{oS}* & \texttt{2.106\_CaCo3V4O12} & \texttt{Lt} \\
        \texttt{0.802\_CuFeS2} & \texttt{oS}* & \texttt{1.382\_Ca2FeO3Cl} & \texttt{oS}* & \texttt{2.107\_DyTe3} & \texttt{Lt} \\
        \texttt{0.803\_NbMnP} & \texttt{M} & \texttt{1.383\_Ca2FeO3Br} & \texttt{oS}* & \texttt{2.116\_Na3Co2SbO6} & \texttt{Lt} \\
        \texttt{0.804\_MoP3SiO11} & \texttt{S} & \texttt{1.384\_USb2} & \texttt{oS} & \texttt{3.1\_TmAgGe} & \texttt{Lt} \\
        \texttt{0.815\_MnNb2O6} & \texttt{oS} & \texttt{1.385\_Sr2FeO3F} & \texttt{oS} & \texttt{3.2\_UO2} & \texttt{oS} \\
        \texttt{0.816\_MnTa2O6} & \texttt{oS} & \texttt{1.386\_Sr2FeO3F} & \texttt{oS}* & \texttt{3.4\_MgCr2O4} & \texttt{M} \\
        \texttt{0.818\_MnTa2O6} & \texttt{M} & \texttt{1.387\_Sr2FeO3F} & \texttt{oS} & \texttt{3.6\_DyCu} & \texttt{oS} \\
        \texttt{0.819\_MnNb2O6} & \texttt{M} & \texttt{1.388\_La2NiO3F2} & \texttt{oS} & \texttt{3.7\_NpBi} & \texttt{oS} \\
        \texttt{0.823\_Sr2MnGaO5} & \texttt{oS}* & \texttt{1.389\_Sr2CoO3Cl} & \texttt{Lp} & \texttt{3.8\_NdZn} & \texttt{oS} \\
        \texttt{0.825\_Ca2MnGaO5} & \texttt{oS} & \texttt{1.397\_Cu3Mg(OD)6Br2} & \texttt{oS} & \texttt{3.9\_NpS} & \texttt{Lt} \\
        \texttt{0.836\_DyFeO3} & \texttt{oS}* & \texttt{1.398\_Pr2CuO4} & \texttt{Lp} & \texttt{3.10\_NpSe} & \texttt{Lt} \\
        \texttt{0.837\_DyFeO3} & \texttt{oS} & \texttt{1.399\_Pr2CuO4} & \texttt{Lp} & \texttt{3.11\_NpTe} & \texttt{Lt} \\
        \texttt{0.838\_DyFeO3} & \texttt{oS} & \texttt{1.400\_TbAg2} & \texttt{Lp} & \texttt{3.12\_USb} & \texttt{oS} \\
        \texttt{0.839\_DyFeO3} & \texttt{oS} & \texttt{1.403\_La2CoO4} & \texttt{oS}* & \texttt{3.13\_CeB6} & \texttt{Lt} \\
        \texttt{0.840\_DyFeO3} & \texttt{oS} & \texttt{1.404\_Sr2CuO2Cl2} & \texttt{Lp} & \texttt{3.16\_Gd2Ti2O7} & \texttt{Lt} \\
        \texttt{0.841\_DyFeO3} & \texttt{oS} & \texttt{1.407\_Nd2CuO4} & \texttt{Lp} & \texttt{3.18\_HoRh} & \texttt{oS} \\
        \texttt{0.842\_DyAlO3} & \texttt{M} & \texttt{1.408\_Nd2CuO4} & \texttt{Lp} & \texttt{3.19\_CoO} & \texttt{Lt} \\
        \texttt{0.854\_Gd2Pt2O7} & \texttt{Lp} & \texttt{1.409\_NaMnO2} & \texttt{Lp} & \texttt{3.21\_TmGa3} & \texttt{oS} \\
        \texttt{0.862\_Eu2Ir2O7} & \texttt{oS}* & \texttt{1.411\_EuMn2P2} & \texttt{oS} & \texttt{3.22\_Eu3PbO} & \texttt{oS} \\
        \texttt{0.881\_CuMnAs} & \texttt{oS} & \texttt{1.413\_Ce3Ni2Ge7} & \texttt{oS} & \texttt{3.24\_CaFe3Ti4O12} & \texttt{Lp} \\
        \texttt{0.896\_NiCrO4} & \texttt{oS}* & \texttt{1.414\_CeNiGe3} & \texttt{oS} & \texttt{3.28\_NiO} & \texttt{Lt} \\
                \bottomrule
\end{longtable*}
\endgroup

\section{\label{appx:stat_ssg_equivalent_sites}Statistics of SSG-equivalent sites in MAGNDATA}

To quantify how much additional symmetry information is captured by SSGs compared with MSGs, we count the number of independent orbits of magnetic sites under four progressively finer symmetry frameworks.
For each magnetic structure in the \textsc{MAGNDATA} database~\cite{MAGNDATA1}, we define the following four quantities.
\begin{enumerate}
\item[(i)] $n_\mathrm{SG}$---the number of orbits of magnetic sites under the family (nonmagnetic) space group $\mathcal{G}$.
      Two magnetic sites belong to the same orbit if and only if they are related by a symmetry operation of $\mathcal{G}$.
\item[(ii)] $n_\mathrm{SG}'$---the number of orbits obtained by further subdividing each $\mathcal{G}$ orbit according to the magnitude of the magnetic moment.
      Sites that belong to the same $\mathcal{G}$ orbit but carry moments of different magnitude are counted as belonging to distinct orbits.
\item[(iii)] $n_\mathrm{SSG}$---the number of orbits of magnetic sites under the spin space group $\mathcal{X}$.
\item[(iv)] $n_\mathrm{MSG}$---the number of orbits of magnetic sites under the magnetic space group $\mathcal{M}$.
\end{enumerate}

Because each successive symmetry framework imposes additional constraints, the four quantities satisfy
\begin{equation}
  n_\mathrm{SG} \le n_\mathrm{SG}' \le n_\mathrm{SSG} \le n_\mathrm{MSG}.
\end{equation}
The first inequality, $n_\mathrm{SG} \le n_\mathrm{SG}'$, holds because the magnitude subdivision can split an existing orbit only into finer classes.
The second inequality, $n_\mathrm{SG}' \le n_\mathrm{SSG}$, holds because sites in the same SSG orbit are necessarily related by an operation of $\mathcal{G}$ (the spatial part of every SSG operation belongs to $\mathcal{G}$) and carry equal-magnitude moments; they, therefore, belong to the same $\mathcal{G}$ orbit and the same magnitude class and, hence, to the same $n_\mathrm{SG}'$ orbit.
The third inequality, $n_\mathrm{SSG} \le n_\mathrm{MSG}$, holds because every MSG operation corresponds to an SSG operation with a specific spin-axis choice, so the MSG is a subgroup of the SSG and its orbits can only be finer.

The case $n_\mathrm{SSG} < n_\mathrm{MSG}$ is of particular interest: It identifies structures in which the SSG recognizes symmetry-equivalent sites (with equal magnetic moments) that the MSG treats as inequivalent.
Such structures directly illustrate the advantage of the SSG framework for enforcing equal-moment constraints on symmetry-related sites.

\begin{table}[t]
\caption{Classification of 1764 nondisordered magnetic structures in \textsc{MAGNDATA} by the number of magnetic-site orbits under four symmetry frameworks. The number in parentheses after each group header indicates the subtotal.}
\label{tab:ssg_equivalent_sites}
\begin{ruledtabular}
\begin{tabular}{llr}
\textrm{Condition} & \textrm{Subcase} & \textrm{Count} \\
\hline
\multicolumn{2}{l}{$n_\mathrm{SSG} < n_\mathrm{MSG}$ (119)} & \\
\multicolumn{2}{l}{$n_\mathrm{SG} < n_\mathrm{SSG} < n_\mathrm{MSG}$ (19)} & \\
  & $n_\mathrm{SG} < n_\mathrm{SG}' = n_\mathrm{SSG}$ & 4 \\
  & $n_\mathrm{SG} < n_\mathrm{SG}' < n_\mathrm{SSG}$ & 0 \\
  & $n_\mathrm{SG} = n_\mathrm{SG}' < n_\mathrm{SSG}$ & 15 \\
$n_\mathrm{SG} = n_\mathrm{SG}' = n_\mathrm{SSG} < n_\mathrm{MSG}$ & & 100 \\
\hline
\multicolumn{2}{l}{$n_\mathrm{SSG} = n_\mathrm{MSG}$ (1645)} & \\
\multicolumn{2}{l}{$n_\mathrm{SG} < n_\mathrm{SSG} = n_\mathrm{MSG}$ (155)} & \\
  & $n_\mathrm{SG} < n_\mathrm{SG}' = n_\mathrm{SSG}$ & 34 \\
  & $n_\mathrm{SG} < n_\mathrm{SG}' < n_\mathrm{SSG}$ & 14 \\
  & $n_\mathrm{SG} = n_\mathrm{SG}' < n_\mathrm{SSG}$ & 107 \\
$n_\mathrm{SG} = n_\mathrm{SG}' = n_\mathrm{SSG} = n_\mathrm{MSG}$ & & 1{,}490 \\
\end{tabular}
\end{ruledtabular}
\end{table}

We analyze the ordered magnetic structures in the \textsc{MAGNDATA} database.
Symmetry identification is performed using the \textsc{moyopy} and \textsc{spinspg} packages~\cite{moyo, Shinohara2024} with \textrm{symprec} $= 1 \times 10^{-4}$ and \textrm{mag\_symprec} $= 1 \times 10^{-2}$, consistent with the parameters used in Sec.~\ref{sec:method_classification}.
A site is classified as magnetic if the magnitude of its magnetic moment exceeds \textrm{mag\_symprec}.
The same \textrm{symprec} value is used across the entire dataset for simplicity.
We note that atomic positions stored in MCIF files do not always preserve the full precision of the reported family space group; a sufficiently large \textrm{symprec} is, therefore, necessary to recover the correct family space group $\mathcal{G}$.
With a tighter \textrm{symprec}, the identified $\mathcal{G}$ (and, hence, the SSG) may be lower than expected, which would reduce the number of structures satisfying $n_\mathrm{SSG} < n_\mathrm{MSG}$.
The results are summarized in Tables~\ref{tab:ssg_equivalent_sites} and \ref{tab:ssg_vs_spin_only_group}; the per-structure data are provided in Supplemental Material~\cite{SM}.
We classify each structure according to the relations among $n_\mathrm{SG}$, $n_\mathrm{SG}'$, $n_\mathrm{SSG}$, and $n_\mathrm{MSG}$.

We note that the \textsc{MAGNDATA} database reflects experimentally reported magnetic structures and is, therefore, biased toward materials and phases that have been studied and published; it does not constitute a representative sample of all magnetic materials in nature.
With this caveat, among the 1764 structures, 119 satisfy $n_\mathrm{SSG} < n_\mathrm{MSG}$, indicating that the SSG identifies symmetry-equivalent magnetic sites that the MSG does not.
This demonstrates the necessity of capturing spin symmetry operations beyond those encoded in the MSG.
Within the $n_\mathrm{SSG} < n_\mathrm{MSG}$ group, the dominant subcase is $n_\mathrm{SG} = n_\mathrm{SG}' = n_\mathrm{SSG} < n_\mathrm{MSG}$ (100 out of 119).
In these structures, sites that are equivalent under the family space group and carry equal-magnitude moments are correctly recognized as equivalent by the SSG but are split into distinct orbits by the MSG.
This shows that equal-magnitude constraints on symmetry-related sites, which are guaranteed by the SSG, are not captured by the MSG alone.
Despite the aforementioned sampling bias, these statistics support the practical usefulness of enforcing the equal-magnitude constraint through the SSG framework.

\begin{table}[t]
\caption{Cross-tabulation of the orbit-splitting condition and the spin-only group type.}
\label{tab:ssg_vs_spin_only_group}
\begin{ruledtabular}
\begin{tabular}{lrr@{\hspace{6pt}}rr}
 & \multicolumn{2}{c}{$n_\mathrm{SSG} < n_\mathrm{MSG}$} & & \\
\cline{2-3}
 & $n_\mathrm{SG} < n_\mathrm{SSG}$ & $n_\mathrm{SG} = n_\mathrm{SSG}$ & $n_\mathrm{SSG} = n_\mathrm{MSG}$ & \textrm{Total} \\
\hline
Collinear    &   4 &  45 & 1014 & 1063 \\
Coplanar     &   6 &  45 &     406 &     457 \\
Noncoplanar  &   9 &  10 &     225 &     244 \\
\hline
Total        &  19 & 100 & 1645 & 1764 \\
\end{tabular}
\end{ruledtabular}
\end{table}

Table~\ref{tab:ssg_vs_spin_only_group} shows that the fraction of structures with $n_\mathrm{SSG} < n_\mathrm{MSG}$ is notably higher for noncollinear cases: 11.2\% for coplanar and 7.8\% for noncoplanar, compared with 4.6\% for collinear structures.
This is expected because noncollinear magnetic orders possess richer spin symmetry operations that are not representable as MSG operations, making the SSG framework particularly advantageous in these cases.

\section{\label{appx:generation_method_comparison}Comparison of symmetry-based methods for magnetic crystal structure generation}

We compare existing symmetry-based approaches for generating candidate magnetic crystal structures.
All methods considered here share the same input: a family (parent) space group $\mathcal{G}$ and a star of a propagation vector $\{\bm{k}_{1}, \dots, \bm{k}_{p}\}$.
From these, each method produces a set of candidate magnetic structures as output.
Note that $\mathcal{G}$ need not coincide with the space group of the nonmagnetic phase; when a magnetic crystal structure has a strictly smaller space group than its nonmagnetic phase, one can try the subgroups of the nonmagnetic space group as family groups, as described in Sec.~\ref{sec:ssg_family_choices}.
Similarly, although the active propagation vectors may not be known beforehand, they can be explored in a brute-force manner, for example, by enumerating vectors with bounded denominators.
These choices of $\mathcal{G}$ and $\{ \bm{k}_{i} \}$ are agnostic to the generation method itself and are, therefore, not attributed to any specific approach below.
Given the same $\mathcal{G}$ and $\{ \bm{k}_{i} \}$, the methods differ in how they exploit group-theoretical constraints, in particular, the symmetry framework employed, the treatment of multidimensional IRs, and whether equal magnetic moments on symmetry-equivalent sites are guaranteed by construction.

\subsection{Representation analysis}

RA~\cite{Bertaut:a05871,BERTAUT1981267,IZYUMOV1979239} decomposes the $3N$-dimensional magnetic degrees of freedom on the Wyckoff positions into IRs of $\mathcal{G}$ (or the little group $\mathcal{G}^{\bm{k}}$ for a nonzero propagation vector $\bm{k}$), where $N$ is the number of magnetic atoms in the unit cell.
Each IR provides basis vectors that serve as candidate magnetic structures.
The key assumption is that the ground state is described by a small number of IRs, which is empirically valid in many cases.
However, for multidimensional IRs, the linear combination of basis vectors is underdetermined, and symmetry-equivalent atoms in $\mathcal{G}$ are not guaranteed to have equal-magnitude moments.

\subsection{Representation analysis with isotropy subgroups}

An extension of RA selects specific linear combinations of multidimensional IR basis vectors by applying the epikernel (isotropy subgroup) principle~\cite{doi:10.1142/0751,E_Ascher_1977,Stokes:vk5013}: One chooses combinations that yield maximal MSGs.
This resolves part of the limitation of RA by restricting to the highest-symmetry orderings within each IR.
However, it still does not guarantee equal-magnitude moments on symmetry-equivalent sites by construction.

Zhou \textit{et al.}~\cite{Zhou2026arXiv} follow essentially the same MSG and IR framework (note that it is a preprint; the details described here are based on the version available at the time of writing).
They decompose the $3N$-dimensional magnetic degrees of freedom into IRs of the parent crystallographic space group and enumerate maximal-symmetry MSGs for each IR.
For repeated equivalent IRs, they retain only those linear combinations whose resulting magnetic moments align with selected high-symmetry crystallographic directions.
This practically useful finite-enumeration rule could, in principle, be incorporated into other symmetry-based generation schemes as well.

\subsection{Cluster multipole theory}

CMP theory~\cite{Suzuki2017,Suzuki2019,Yanagi2023} generates a symmetry-adapted basis for magnetic structures by constructing magnetic and magnetic-toroidal multipoles from vector spherical harmonics on a virtual atomic cluster, mapping them to the crystal, and orthonormalizing the resulting magnetic structures.
This provides a systematic and physically interpretable basis, often yielding practically reasonable trial magnetic structures.
However, the symmetry framework itself is still limited to the action of magnetic symmetry operations on magnetic moments.
In this sense, CMP shares the same symmetry framework as conventional RA: It replaces the usual projection-operator basis construction by a multipole-based one but does not introduce a spin symmetry operation or guarantee equal moments on symmetry-equivalent sites by construction.
Therefore, CMP would be best regarded as a physically motivated symmetry-adapted basis construction and classification scheme rather than as a distinct symmetry framework for exhaustive magnetic-structure generation.
A more natural route to a complete generation scheme is to generalize RA with isotropy subgroups from MSGs to spin space groups, which is the approach taken in this work.

\subsection{Spin space group approaches}

SSG approaches describe magnetic structures as subgroups of $\mathcal{G} \times \mathrm{O}(3)$, where spatial and spin operations are decoupled.
Unlike MSG-based approaches, they directly capture spin configurations in the absence of SOC and can distinguish magnetic structures that share the same MSG but differ in their spin space group.

Jiang \textit{et al.}~\cite{Jiang2024PRX} establishes the basic enumeration framework by systematically listing SSGs up to $12\times$ unit cells.
Section~6D in their work also extends the standard MSG-based neutron-diffraction refinement procedure to the SSG setting: For a given SSG and Wyckoff position, the site-symmetry group constrains the symmetry-allowed magnetic moments, thereby reducing the number of free fitting parameters compared with the corresponding MSG description.
However, as the authors themselves note, this refinement-oriented use of SSGs does not by itself resolve the global spin-axis problem.
Because spin and real space are decoupled in an SSG, the symmetry-allowed magnetic moments are determined only up to an overall spin-space rotation.
Accordingly, the spin axis or spin plane must still be fixed externally, for example, from experimental input.
Thus, Jiang \textit{et al.}~\cite{Jiang2024PRX} provide a powerful SSG classification and constraint framework for refinement but not yet a symmetry-complete generation scheme for magnetic structures.

Our method builds on the SSG framework in a generative direction.
We generate SSA structures as totally symmetric IRs of each SSG, which guarantees equal moments on symmetry-equivalent Wyckoff sites by construction.
The oriented SSG concept~\cite{Liu2025arXiv} then fixes the global spin orientation by descending to compatible maximal MSGs, yielding oriented SSA structures.

Li \textit{et al.}~\cite{Li2025arXiv} adopt a closely related hierarchical SSG-to-MSG work flow.
They first enumerate SSG-allowed magnetic geometries in the SOC-free limit, rank them by nonrelativistic DFT, and then enumerate MSG-allowed magnetic orientations compatible with the selected SSG geometry and rank them by relativistic DFT calculations including SOC.
Thus, at the level of overall work flow, their method is conceptually very close to ours.
At the same time, the currently public description does not elaborate on all implementation details.
Their SSG enumeration appears to rely heavily on existing crystallographic subgroup infrastructure, in particular, the subgroup data of $\mathcal{G}$ obtained from the \textsc{CELLSUB} program on the Bilbao Crystallographic Server \cite{AroyoPerezMatoCapillasKroumovaIvantchevMadariagaKirovWondratschek}.
By contrast, the subsequent enumeration of MSG descendants from a given SSG is described at a high level in the public preprint.
Since this SSG-to-MSG step is nontrivial, a detailed comparison with our oriented-SSG construction awaits further details from the authors.

\section{Complete list of SDFT calculations}
\label{appx:sdft_list}
\begin{table*}[t]
\centering
\caption{Correctly and incorrectly predicted magnetic structures by SDFT at the SSA level.}
\label{tab:sdft_ssa_list}
\small
\begin{tabularx}{\textwidth}{>{\raggedright\arraybackslash}p{0.09\textwidth} X}
\hline
Category & Materials \\
\hline
Correctly predicted&
\texttt{0.1\_LaMnO3}, \texttt{0.3\_Ca3LiOsO6}, \texttt{0.15\_MnF2}, \texttt{0.18\_BaMn2As2}, \texttt{0.19\_MnTiO3}, \texttt{0.20\_MnTe2}, \texttt{0.21\_PbNiO3}, \texttt{0.23\_Ca3Mn2O7}, \texttt{0.24\_LiMnPO4}, \texttt{0.25\_NaOsO3}, \texttt{0.56\_Ba2CoGe2O7}, \texttt{0.58\_CoAl2O4}, \texttt{0.59\_Cr2O3}, \texttt{0.65\_Fe2O3-alpha}, \texttt{0.72\_CaMnBi2}, \texttt{0.73\_SrMnBi2}, \texttt{0.74\_Mn3Cu0.5Ge0.5N}, \texttt{0.75\_Cr2WO6}, \texttt{0.76\_Cr2TeO6}, \texttt{0.79\_CaIrO3}, \texttt{0.83\_LiFeP2O7}, \texttt{0.87\_NaFePO4}, \texttt{0.89\_BaMn2Bi2}, \texttt{0.92\_CaMn2Sb2}, \texttt{0.95\_LiFePO4}, \texttt{0.108\_Mn3Ir}, \texttt{0.109\_Mn3Pt}, \texttt{0.112\_FeBO3}, \texttt{0.113\_NiCO3}, \texttt{0.114\_CoCO3}, \texttt{0.115\_MnCO3}, \texttt{0.125\_MnGeO3}, \texttt{0.128\_FeSO4F}, \texttt{0.131\_Mn(N(CN2))2}, \texttt{0.137\_Cu2V2O7}, \texttt{0.142\_Fe2TeO6}, \texttt{0.148\_La2LiRuO6}, \texttt{0.150\_NiS2}, \texttt{0.156\_CaMnGe2O6}, \texttt{0.163\_MnPS3}, \texttt{0.177\_Mn3GaN}, \texttt{0.178\_CoF2}, \texttt{0.180\_MnPSe3}, \texttt{0.186\_CeMnAsO}, \texttt{0.193\_LiCoPO4}, \texttt{0.199\_Mn3Sn}, \texttt{0.209\_TlFe1.6Se2}, \texttt{0.212\_Sr2Mn3As2O2}, \texttt{0.215\_BaNi2P2O8}, \texttt{0.222\_CuMnAs}, \texttt{0.229\_Ba2MnSi2O7}, \texttt{0.239\_Ca3LiRuO6}, \texttt{0.267\_YbMnBi2}, \texttt{0.273\_Mn3ZnN}, \texttt{0.277\_MgMnO3}, \texttt{0.279\_Mn3As}, \texttt{0.284\_KOsO4}, \texttt{0.301\_Sr2CoTeO6}, \texttt{0.303\_BaCrF5}, \texttt{0.307\_ScCrO3}, \texttt{0.308\_InCrO3}, \texttt{0.309\_TlCrO3}, \texttt{0.334\_CoF3}, \texttt{0.335\_FeF3}, \texttt{0.336\_NdFeO3}, \texttt{0.348\_Bi2CuO4}, \texttt{0.351\_TbFeO3}, \texttt{0.354\_TbCrO3}, \texttt{0.361\_Sr3LiRuO6}, \texttt{0.364\_SrCr2As2}, \texttt{0.365\_BaCr2As2}, \texttt{0.377\_Mn3Ge}, \texttt{0.379\_SmFeO3}, \texttt{0.399\_FeOOH}, \texttt{0.404\_Sr3NaRuO6}, \texttt{0.416\_LaCrO3}, \texttt{0.420\_Sr2LuRuO6}, \texttt{0.421\_EuMnSb2}, \texttt{0.426\_EuMnBi2}, \texttt{0.433\_KMnF3}, \texttt{0.434\_K2ReI6}, \texttt{0.462\_MnAl2O4}, \texttt{0.463\_Co3O4}, \texttt{0.464\_BaMn2P2}, \texttt{0.465\_HoCr2Si2}, \texttt{0.466\_ThCr2Si2}, \texttt{0.470\_BaMn2Sb2}, \texttt{0.471\_Ba2Mn3Sb2O2}, \texttt{0.472\_LaMn2Si2}, \texttt{0.474\_EuMn2Ge2}, \texttt{0.482\_SrMn2As2}, \texttt{0.483\_YbMn2Sb2}, \texttt{0.486\_ErCr2Si2}, \texttt{0.501\_LiFe2F6}, \texttt{0.504\_NaCrSi2O6}, \texttt{0.518\_TbCr2Si2}, \texttt{0.528\_CrSb}, \texttt{0.575\_ZnFeF5(H2O)2}, \texttt{0.582\_Fe3F8(H2O)2}, \texttt{0.586\_YCrO3}, \texttt{0.587\_TmCrO3}, \texttt{0.591\_ErCrO3}, \texttt{0.592\_DyCrO3}, \texttt{0.598\_AlCr2}, \texttt{0.599\_CaMnSi}, \texttt{0.601\_CaMnGe}, \texttt{0.603\_CaMn2Ge2}, \texttt{0.605\_BaMn2Ge2}, \texttt{0.611\_BaMnSb2}, \texttt{0.617\_KMnSb}, \texttt{0.618\_KMnBi}, \texttt{0.619\_LaMnAsO}, \texttt{0.620\_NdMnAsO}, \texttt{0.626\_NaMnP}, \texttt{0.629\_NaMnAs}, \texttt{0.631\_NaMnSb}, \texttt{0.633\_KFeS2}, \texttt{0.634\_NaMnBi}, \texttt{0.636\_RbFeS2}, \texttt{0.637\_KFeSe2}, \texttt{0.638\_RbFeSe2}, \texttt{0.639\_Mn2Au}, \texttt{0.665\_CeMnSbO}, \texttt{0.667\_LaMnSbO}, \texttt{0.712\_VNb3S6}, \texttt{0.714\_Li2Ni(SO4)2}, \texttt{0.757\_CeFeO3}, \texttt{0.760\_FeOHSO4}, \texttt{0.766\_YbMnSb2}, \texttt{0.795\_Sr2YRuO6}, \texttt{0.798\_MnPd2}, \texttt{0.800\_MnTe}, \texttt{0.802\_CuFeS2}, \texttt{0.823\_Sr2MnGaO5}, \texttt{0.836\_DyFeO3}, \texttt{0.896\_NiCrO4}, \texttt{0.917\_Sr2ScOsO6}, \texttt{0.920\_ThMnPN}, \texttt{0.922\_ThMnAsN}, \texttt{0.934\_Sr2NiTeO6}, \texttt{0.936\_Sr2MnTeO6}, \texttt{0.991\_HoFeO3}, \texttt{0.1018\_SrMnO3}, \texttt{0.1021\_TmFeO3}, \texttt{0.1036\_CoTa4Se8}, \texttt{0.1059\_PrBaMn2O6}, \texttt{0.1065\_MnGeN2}, \texttt{0.1066\_MnSiN2}, \texttt{0.1085\_PrFeO3}, \texttt{0.1091\_La2O3Mn2Se2}, \texttt{0.1104\_GdCrO3}, \texttt{0.1111\_MnSb2O4}, \texttt{0.1112\_CoNb4Se8}, \texttt{1.16\_BaFe2As2}, \texttt{1.17\_CoV2O6-alpha}, \texttt{1.23\_La2CuO4}, \texttt{1.42\_La2NiO4}, \texttt{1.52\_CaFe2As2}, \texttt{1.65\_SrFeO2}, \texttt{1.97\_Li2MnO3}, \texttt{1.125\_LaFeAsO}, \texttt{1.146\_LaCrAsO}, \texttt{1.159\_Li2Ni(WO4)2}, \texttt{1.186\_SrRu2O6}, \texttt{1.194\_NiWO4}, \texttt{1.219\_CuF2}, \texttt{1.230\_NiPS3}, \texttt{1.237\_VCl2}, \texttt{1.238\_VBr2}, \texttt{1.244\_CrCl3}, \texttt{1.252\_CaCo2P2}, \texttt{1.253\_CeCo2P2}, \texttt{1.256\_BaNi2V2O8}, \texttt{1.278\_Cu(NCS)2}, \texttt{1.287\_V2O3}, \texttt{1.371\_Nd2NiO4}, \texttt{1.382\_Ca2FeO3Cl}, \texttt{1.386\_Sr2FeO3F}, \texttt{1.449\_Li2CuW2O8}, \texttt{1.452\_FeSn}, \texttt{1.453\_EuMn2Si2}, \texttt{1.458\_CsCo2Se2}, \texttt{1.468\_TbMn2Si2}, \texttt{1.469\_YMn2Si2}, \texttt{1.488\_CeMn2Si2}, \texttt{1.491\_PrMn2Si2}, \texttt{1.493\_NdMn2Si2}, \texttt{1.496\_YMn2Ge2}, \texttt{1.508\_Mn2AlB2}, \texttt{1.520\_NiSO4}, \texttt{1.521\_FeSO4}, \texttt{1.522\_CrVO4}, \texttt{1.526\_LiCoF4}, \texttt{1.538\_Ba2MnTeO6}, \texttt{1.539\_KMnP}, \texttt{1.541\_RbMnP}, \texttt{1.543\_RbMnAs}, \texttt{1.545\_RbMnBi}, \texttt{1.546\_CsMnBi}, \texttt{1.547\_CsMnP}, \texttt{1.550\_LiMnAs}, \texttt{1.553\_KMnAs}, \texttt{1.556\_FeSn2}, \texttt{1.557\_FeGe2}, \texttt{1.588\_NdFeAsO}, \texttt{1.629\_FeGe}, \texttt{1.630\_LuMn6Sn6}, \texttt{1.631\_YMn6Ge6}, \texttt{1.636\_ErMn2Si2}, \texttt{1.638\_ErMn2Ge2}, \texttt{1.642\_TlFeS2}, \texttt{1.653\_FeWO4}, \texttt{1.678\_CrN}, \texttt{1.689\_LuMn2Ge2}, \texttt{1.690\_TmMn2Ge2}, \texttt{1.694\_TbMn2Ge2}, \texttt{1.767\_Li2CoCl4}, \texttt{1.778\_ThCr2Si2C}, \texttt{1.799\_MnO}, \texttt{1.802\_CrSBr}, \texttt{1.809\_BiCoO3}, \texttt{1.838\_LaMn2Au4}, \texttt{2.6\_Nd2CuO4}, \texttt{2.7\_Sm2CuO4}, \texttt{2.48\_Pr2CuO4} \\
\hline
Incorrectly predicted&
\texttt{0.2\_Cd2Os2O7}, \texttt{0.116\_FeCO3}, \texttt{0.211\_Ca2MnO4}, \texttt{0.264\_Fe3(PO4)2}, \texttt{0.285\_KRuO4}, \texttt{0.401\_Sr4Fe4O11}, \texttt{0.461\_CoRh2O4}, \texttt{0.499\_UCr2Si2C}, \texttt{0.513\_YRuO3}, \texttt{0.607\_RuO2}, \texttt{0.608\_PrMnO3}, \texttt{0.609\_NdMnO3}, \texttt{0.708\_CrNb4S8}, \texttt{0.733\_AgRuO3}, \texttt{0.786\_NdVO3}, \texttt{0.787\_YVO3}, \texttt{0.862\_Eu2Ir2O7}, \texttt{0.945\_Yb2Ir2O7}, \texttt{0.966\_V2WO6}, \texttt{0.979\_TmVO3}, \texttt{0.984\_LuVO3}, \texttt{0.1023\_Cr2MoO6}, \texttt{1.5\_YBa2Cu3O6+d}, \texttt{1.18\_MnS2}, \texttt{1.37\_VOCl}, \texttt{1.57\_CuMnO2}, \texttt{1.69\_CoO}, \texttt{1.70\_CoV2O6}, \texttt{1.110\_ScMn6Ge6}, \texttt{1.183\_FePS3}, \texttt{1.264\_CoPS3}, \texttt{1.281\_YBaCuFeO5}, \texttt{1.346\_TlMnF4}, \texttt{1.380\_Sr2FeO3Cl}, \texttt{1.381\_Sr2FeO3Br}, \texttt{1.383\_Ca2FeO3Br}, \texttt{1.403\_La2CoO4}, \texttt{1.420\_YBa2Cu3O6}, \texttt{1.461\_Sr2Cr3As2O2}, \texttt{1.580\_NiTiO3}, \texttt{1.585\_PrFeAsO}, \texttt{1.702\_YBaCo2O5}, \texttt{1.741\_KNiAsO4}, \texttt{1.782\_FeBr3}, \texttt{1.830\_FeBi4S7}, \texttt{2.77\_Eu2CuO4} \\
\hline
\end{tabularx}
\end{table*}

\begin{table*}[t]
\centering
\caption{Correctly and incorrectly predicted magnetic structures by SDFT at the oriented-SSA level.}
\label{tab:sdft_ossa_list}
\small
\begin{tabularx}{\textwidth}{>{\raggedright\arraybackslash}p{0.09\textwidth} X}
\hline
Category & Materials \\
\hline
Correctly predicted&
\texttt{0.1\_LaMnO3}, \texttt{0.2\_Cd2Os2O7}, \texttt{0.3\_Ca3LiOsO6}, \texttt{0.15\_MnF2}, \texttt{0.18\_BaMn2As2}, \texttt{0.19\_MnTiO3}, \texttt{0.20\_MnTe2}, \texttt{0.21\_PbNiO3}, \texttt{0.23\_Ca3Mn2O7}, \texttt{0.24\_LiMnPO4}, \texttt{0.25\_NaOsO3}, \texttt{0.56\_Ba2CoGe2O7}, \texttt{0.58\_CoAl2O4}, \texttt{0.59\_Cr2O3}, \texttt{0.65\_Fe2O3-alpha}, \texttt{0.72\_CaMnBi2}, \texttt{0.73\_SrMnBi2}, \texttt{0.76\_Cr2TeO6}, \texttt{0.79\_CaIrO3}, \texttt{0.87\_NaFePO4}, \texttt{0.89\_BaMn2Bi2}, \texttt{0.92\_CaMn2Sb2}, \texttt{0.95\_LiFePO4}, \texttt{0.125\_MnGeO3}, \texttt{0.128\_FeSO4F}, \texttt{0.131\_Mn(N(CN2))2}, \texttt{0.137\_Cu2V2O7}, \texttt{0.142\_Fe2TeO6}, \texttt{0.150\_NiS2}, \texttt{0.177\_Mn3GaN}, \texttt{0.186\_CeMnAsO}, \texttt{0.193\_LiCoPO4}, \texttt{0.199\_Mn3Sn}, \texttt{0.209\_TlFe1.6Se2}, \texttt{0.212\_Sr2Mn3As2O2}, \texttt{0.222\_CuMnAs}, \texttt{0.239\_Ca3LiRuO6}, \texttt{0.267\_YbMnBi2}, \texttt{0.273\_Mn3ZnN}, \texttt{0.277\_MgMnO3}, \texttt{0.284\_KOsO4}, \texttt{0.303\_BaCrF5}, \texttt{0.307\_ScCrO3}, \texttt{0.308\_InCrO3}, \texttt{0.309\_TlCrO3}, \texttt{0.334\_CoF3}, \texttt{0.335\_FeF3}, \texttt{0.336\_NdFeO3}, \texttt{0.348\_Bi2CuO4}, \texttt{0.351\_TbFeO3}, \texttt{0.354\_TbCrO3}, \texttt{0.361\_Sr3LiRuO6}, \texttt{0.364\_SrCr2As2}, \texttt{0.365\_BaCr2As2}, \texttt{0.379\_SmFeO3}, \texttt{0.401\_Sr4Fe4O11}, \texttt{0.404\_Sr3NaRuO6}, \texttt{0.416\_LaCrO3}, \texttt{0.421\_EuMnSb2}, \texttt{0.426\_EuMnBi2}, \texttt{0.433\_KMnF3}, \texttt{0.462\_MnAl2O4}, \texttt{0.463\_Co3O4}, \texttt{0.464\_BaMn2P2}, \texttt{0.465\_HoCr2Si2}, \texttt{0.470\_BaMn2Sb2}, \texttt{0.472\_LaMn2Si2}, \texttt{0.474\_EuMn2Ge2}, \texttt{0.482\_SrMn2As2}, \texttt{0.486\_ErCr2Si2}, \texttt{0.501\_LiFe2F6}, \texttt{0.518\_TbCr2Si2}, \texttt{0.528\_CrSb}, \texttt{0.575\_ZnFeF5(H2O)2}, \texttt{0.582\_Fe3F8(H2O)2}, \texttt{0.586\_YCrO3}, \texttt{0.587\_TmCrO3}, \texttt{0.591\_ErCrO3}, \texttt{0.592\_DyCrO3}, \texttt{0.599\_CaMnSi}, \texttt{0.603\_CaMn2Ge2}, \texttt{0.605\_BaMn2Ge2}, \texttt{0.611\_BaMnSb2}, \texttt{0.617\_KMnSb}, \texttt{0.618\_KMnBi}, \texttt{0.619\_LaMnAsO}, \texttt{0.620\_NdMnAsO}, \texttt{0.626\_NaMnP}, \texttt{0.629\_NaMnAs}, \texttt{0.631\_NaMnSb}, \texttt{0.634\_NaMnBi}, \texttt{0.637\_KFeSe2}, \texttt{0.638\_RbFeSe2}, \texttt{0.639\_Mn2Au}, \texttt{0.665\_CeMnSbO}, \texttt{0.667\_LaMnSbO}, \texttt{0.712\_VNb3S6}, \texttt{0.757\_CeFeO3}, \texttt{0.760\_FeOHSO4}, \texttt{0.766\_YbMnSb2}, \texttt{0.798\_MnPd2}, \texttt{0.802\_CuFeS2}, \texttt{0.823\_Sr2MnGaO5}, \texttt{0.836\_DyFeO3}, \texttt{0.896\_NiCrO4}, \texttt{0.920\_ThMnPN}, \texttt{0.922\_ThMnAsN}, \texttt{0.991\_HoFeO3}, \texttt{0.1018\_SrMnO3}, \texttt{0.1021\_TmFeO3}, \texttt{0.1036\_CoTa4Se8}, \texttt{0.1059\_PrBaMn2O6}, \texttt{0.1065\_MnGeN2}, \texttt{0.1066\_MnSiN2}, \texttt{0.1085\_PrFeO3}, \texttt{0.1091\_La2O3Mn2Se2}, \texttt{0.1104\_GdCrO3}, \texttt{0.1111\_MnSb2O4}, \texttt{0.1112\_CoNb4Se8}, \texttt{1.16\_BaFe2As2}, \texttt{1.23\_La2CuO4}, \texttt{1.42\_La2NiO4}, \texttt{1.65\_SrFeO2}, \texttt{1.186\_SrRu2O6}, \texttt{1.237\_VCl2}, \texttt{1.238\_VBr2}, \texttt{1.241\_FeCl2}, \texttt{1.250\_KNiF3}, \texttt{1.252\_CaCo2P2}, \texttt{1.253\_CeCo2P2}, \texttt{1.370\_Li2CuO2}, \texttt{1.371\_Nd2NiO4}, \texttt{1.382\_Ca2FeO3Cl}, \texttt{1.386\_Sr2FeO3F}, \texttt{1.452\_FeSn}, \texttt{1.453\_EuMn2Si2}, \texttt{1.458\_CsCo2Se2}, \texttt{1.468\_TbMn2Si2}, \texttt{1.469\_YMn2Si2}, \texttt{1.488\_CeMn2Si2}, \texttt{1.491\_PrMn2Si2}, \texttt{1.493\_NdMn2Si2}, \texttt{1.496\_YMn2Ge2}, \texttt{1.508\_Mn2AlB2}, \texttt{1.520\_NiSO4}, \texttt{1.521\_FeSO4}, \texttt{1.539\_KMnP}, \texttt{1.541\_RbMnP}, \texttt{1.543\_RbMnAs}, \texttt{1.545\_RbMnBi}, \texttt{1.546\_CsMnBi}, \texttt{1.547\_CsMnP}, \texttt{1.550\_LiMnAs}, \texttt{1.553\_KMnAs}, \texttt{1.556\_FeSn2}, \texttt{1.557\_FeGe2}, \texttt{1.629\_FeGe}, \texttt{1.631\_YMn6Ge6}, \texttt{1.636\_ErMn2Si2}, \texttt{1.638\_ErMn2Ge2}, \texttt{1.689\_LuMn2Ge2}, \texttt{1.690\_TmMn2Ge2}, \texttt{1.694\_TbMn2Ge2}, \texttt{1.767\_Li2CoCl4}, \texttt{1.778\_ThCr2Si2C}, \texttt{1.802\_CrSBr}, \texttt{1.803\_LiCrTe2}, \texttt{1.809\_BiCoO3}, \texttt{1.838\_LaMn2Au4}, \texttt{2.6\_Nd2CuO4}, \texttt{2.7\_Sm2CuO4}, \texttt{2.48\_Pr2CuO4}\\
\hline
Incorrectly predicted&
\texttt{0.74\_Mn3Cu0.5Ge0.5N}, \texttt{0.75\_Cr2WO6}, \texttt{0.108\_Mn3Ir}, \texttt{0.109\_Mn3Pt}, \texttt{0.112\_FeBO3}, \texttt{0.116\_FeCO3}, \texttt{0.178\_CoF2}, \texttt{0.211\_Ca2MnO4}, \texttt{0.229\_Ba2MnSi2O7}, \texttt{0.279\_Mn3As}, \texttt{0.285\_KRuO4}, \texttt{0.377\_Mn3Ge}, \texttt{0.399\_FeOOH}, \texttt{0.461\_CoRh2O4}, \texttt{0.466\_ThCr2Si2}, \texttt{0.471\_Ba2Mn3Sb2O2}, \texttt{0.499\_UCr2Si2C}, \texttt{0.513\_YRuO3}, \texttt{0.607\_RuO2}, \texttt{0.608\_PrMnO3}, \texttt{0.708\_CrNb4S8}, \texttt{0.733\_AgRuO3}, \texttt{0.787\_YVO3}, \texttt{0.800\_MnTe}, \texttt{0.862\_Eu2Ir2O7}, \texttt{0.945\_Yb2Ir2O7}, \texttt{0.966\_V2WO6}, \texttt{0.979\_TmVO3}, \texttt{0.984\_LuVO3}, \texttt{0.1023\_Cr2MoO6}, \texttt{1.5\_YBa2Cu3O6+d}, \texttt{1.18\_MnS2}, \texttt{1.52\_CaFe2As2}, \texttt{1.110\_ScMn6Ge6}, \texttt{1.125\_LaFeAsO}, \texttt{1.146\_LaCrAsO}, \texttt{1.242\_FeBr2}, \texttt{1.281\_YBaCuFeO5}, \texttt{1.308\_MnBi2Te4}, \texttt{1.346\_TlMnF4}, \texttt{1.380\_Sr2FeO3Cl}, \texttt{1.381\_Sr2FeO3Br}, \texttt{1.383\_Ca2FeO3Br}, \texttt{1.403\_La2CoO4}, \texttt{1.420\_YBa2Cu3O6}, \texttt{1.461\_Sr2Cr3As2O2}, \texttt{1.585\_PrFeAsO}, \texttt{1.588\_NdFeAsO}, \texttt{1.630\_LuMn6Sn6}, \texttt{1.678\_CrN}, \texttt{1.702\_YBaCo2O5}, \texttt{1.782\_FeBr3}, \texttt{1.830\_FeBi4S7}, \texttt{2.77\_Eu2CuO4} \\
\hline
\end{tabularx}
\end{table*}

In Tables~\ref{tab:sdft_ssa_list} and \ref{tab:sdft_ossa_list}, we present the complete lists of the correctly and incorrectly predicted materials. In this study, following the criterion described in Sec.~\ref{sec:predictive}, calculations whose converged magnetic structure deviates by more than 10\% from the initial structure are excluded from the energy comparison. For completeness, we summarize how the classification changes when all converged calculations are retained in the energy comparison. At the SSA level, \texttt{0.401\_Sr4Fe4O11} and \texttt{0.607\_RuO2} change from incorrect to correct, whereas \texttt{0.25\_NaOsO3}, \texttt{0.212\_Sr2Mn3As2O2}, \texttt{0.471\_Ba2Mn3Sb2O2}, \texttt{1.146\_LaCrAsO}, \texttt{2.6\_Nd2CuO4}, \texttt{2.7\_Sm2CuO4}, and \texttt{2.48\_Pr2CuO4} change from correct to incorrect, giving 220 out of 271 materials (approximately 81\%). At the oriented-SSA level, \texttt{0.461\_CoRh2O4}, \texttt{0.499\_UCr2Si2C}, \texttt{0.607\_RuO2}, \texttt{1.5\_YBa2Cu3O6+d}, and \texttt{1.420\_YBa2Cu3O6} change from incorrect to correct, whereas \texttt{0.79\_CaIrO3}, \texttt{0.212\_Sr2Mn3As2O2}, \texttt{2.6\_Nd2CuO4}, \texttt{2.7\_Sm2CuO4}, and \texttt{2.48\_Pr2CuO4} change from correct to incorrect, giving 172 out of 226 materials (approximately 76\%). Thus, although the classification of a small number of difficult cases depends on the precise judgment criterion, the overall statistical performance remains essentially unchanged. We also note that the filtered criterion is insensitive to the threshold used to detect the change of the magnetic structure: Varying the overlap threshold between 0.80 and 0.90 leaves all the numbers unchanged.

\section{Details of \texorpdfstring{SDFT$+U$}{SDFT+U} calculations}
\label{appx:sdft_u}
Magnetic materials are often correlated systems, and the SDFT+$U$ method is frequently used to improve the description of localized $d$ electrons. To assess the sensitivity of the results, we examine SDFT+$U$ calculations for the same dataset as that used in Sec.~\ref{sec:sdft}. The computational setup was the same as that in Sec.~\ref{sec:method_calc}, except that a Hubbard correction of $U=3$~eV was applied to the $d$ orbitals of the transition-metal elements. In the following, we discuss the results at the SSA level.

With the inclusion of $U$, the predicted magnetic structure agrees with experiment for 219 out of 271 materials, which is nearly unchanged from the 225 out of 271 obtained without $U$. The overlap between the two correctly predicted sets is 203 materials, indicating that the SDFT and SDFT+$U$ predictions are consistent for more than 90\% of the dataset. This result shows that, for a large fraction of candidate magnetic structures, the energetic ordering is not strongly affected by the inclusion of $U$. The absence of a systematic improvement upon introducing $U$ is also consistent with the previous CMP-based study by Ref.~\cite{Mita2021}.

To examine the material dependence more closely, we analyze the compounds for which the prediction changes upon introducing $U$. The corresponding list is summarized in Table~\ref{tab:sdft_vs_sdftu}. Among the materials that are correctly predicted by SDFT but not by SDFT+$U$, the transition elements are distributed as follows: Cr: 2, Mn: 13, Fe: 2, Co: 1, Cu: 3, and Re: 1, giving 22 materials in total. Conversely, among the materials that are correctly predicted by SDFT+$U$ but not by SDFT, the corresponding distribution is V: 4, Mn: 1, Fe: 3, Co: 3, Ru: 3, Os: 1, and Ir: 1, giving 16 materials in total. These results suggest that, at least for Mn-based magnets in the present dataset, applying $+U$ does not improve the prediction and may even worsen it. By contrast, somewhat better results are obtained with $+U$ for V- and 4$d$/5$d$-based compounds, for which plain SDFT tends to yield smaller magnetic moments. However, this tendency should be interpreted with some caution. Since the present dataset consists entirely of magnetic materials, the apparent improvement for V- and 4$d$/5$d$-containing compounds does not necessarily imply that SDFT+$U$ provides a uniformly better description in realistic high-throughput applications. Indeed, compounds containing these elements are often nonmagnetic, and applying $U$ may incorrectly overstabilize magnetic solutions. For practical high-throughput applications, calculations without $U$ appear to be a reasonable default choice. If one wishes to reduce the risk of missing possible magnetic structures, supplementary SDFT+$U$ calculations may be useful, in particular, for V- and 4$d$/5$d$-based systems.

\begin{table*}[t]
\centering
\caption{Materials that are correctly predicted at the SSA level by only one of the SDFT and SDFT$+U$ methods.}
\label{tab:sdft_vs_sdftu}
\small
\begin{tabularx}{\textwidth}{>{\raggedright\arraybackslash}p{0.27\textwidth} X}
\hline
Category & Materials \\
\hline
Correctly predicted by SDFT+$U$ but not by SDFT &
\texttt{0.2\_Cd2Os2O7}, \texttt{0.116\_FeCO3}, \texttt{0.285\_KRuO4}, \texttt{0.461\_CoRh2O4}, \texttt{0.607\_RuO2}, \texttt{0.609\_NdMnO3}, \texttt{0.733\_AgRuO3}, \texttt{0.787\_YVO3}, \texttt{0.945\_Yb2Ir2O7}, \texttt{0.966\_V2WO6}, \texttt{0.979\_TmVO3}, \texttt{1.37\_VOCl}, \texttt{1.381\_Sr2FeO3Br}, \texttt{1.403\_La2CoO4}, \texttt{1.702\_YBaCo2O5}, \texttt{1.782\_FeBr3} \\
\hline
Correctly predicted by SDFT but not by SDFT$+U$ &
\texttt{0.1\_LaMnO3}, \texttt{0.193\_LiCoPO4}, \texttt{0.277\_MgMnO3}, \texttt{0.434\_K2ReI6}, \texttt{0.504\_NaCrSi2O6}, \texttt{0.582\_Fe3F8(H2O)2}, \texttt{1.97\_Li2MnO3}, \texttt{1.382\_Ca2FeO3Cl}, \texttt{1.453\_EuMn2Si2}, \texttt{1.468\_TbMn2Si2}, \texttt{1.488\_CeMn2Si2}, \texttt{1.491\_PrMn2Si2}, \texttt{1.493\_NdMn2Si2}, \texttt{1.631\_YMn6Ge6}, \texttt{1.638\_ErMn2Ge2}, \texttt{1.689\_LuMn2Ge2}, \texttt{1.690\_TmMn2Ge2}, \texttt{1.694\_TbMn2Ge2}, \texttt{1.778\_ThCr2Si2C}, \texttt{2.6\_Nd2CuO4}, \texttt{2.7\_Sm2CuO4}, \texttt{2.48\_Pr2CuO4} \\
\hline
\end{tabularx}
\end{table*}

\section{Details of the calculations with fully self-consistent treatment of SOC}
\label{appx:sdft_soc}
In this study, we employ fixed-charge calculations to compare the energies within the oriented SSA manifold. This is a practically well-justified approach for noncollinear magnetic structures with SOC, because such systems generally exhibit highly complex energy landscapes with many locally metastable electronic solutions. It is also consistent with the recommendation in the VASP documentation. At the same time, however, it is valuable to benchmark this approach systematically against fully self-consistent SOC calculations. To this end, we performed fully self-consistent SOC calculations for all 2370 oriented-SSG structures considered in the main text. The computational settings were essentially the same as those used for the SSA-level calculations in Sec.~\ref{sec:method_calc}, except that a looser convergence criterion, \texttt{EDIFF}=$10^{-6}$, was adopted instead of the tighter threshold used in the fixed-charge calculations.

We found that the convergence behavior deteriorates substantially in the fully self-consistent treatment. Even when the convergence criterion is relaxed to $10^{-6}$, the number of unconverged calculations increases from only one in the original work flow to 61. This makes practical applications of the method, such as the construction of spin models via the energy-mapping method, difficult in practice. The average energy difference $\bar{\Delta}$ among oriented-SSA structures increases from 290~$\mu$eV per magnetic atom in the fixed-charge calculations to 670~$\mu$eV per magnetic atom. Note that this increase may not be entirely attributable to enhanced SOC effects, since some calculations may converge to different metastable solutions irrespective of whether SOC is included. Nevertheless, to account for the larger energy scale, we increased the energy tolerance for judging agreement between the oriented-SSA and experimental structures from 30 to 60~$\mu$eV per magnetic atom. Despite this relaxed tolerance, the final predictive performance deteriorates, decreasing from 172 out of 226 materials in the original work flow to 147 out of 226 materials in the fully self-consistent treatment, owing to the numerical difficulties in the fully self-consistent SOC calculations. These results clearly demonstrate the practical advantage of our approach: It is numerically much more stable than a fully self-consistent treatment and also provides better predictive performance.

\bibliography{references}

\end{document}